\documentclass[a4paper,UKenglish,cleveref, authorcolumns, thm-restate]{tgdk-v2021}
\pdfoutput=1 %uncomment to ensure pdflatex processing (mandatatory e.g. to submit to arXiv)
\hideTGDK  %uncomment to remove references to TGDK series (logo, DOI, ...), e.g. when preparing a pre-final version to be uploaded to arXiv or another public repository

\usepackage{amssymb}
\usepackage{amsmath}
\usepackage{amsthm}
\usepackage{stmaryrd}
\usepackage{fancyvrb}
\usepackage{xspace}
\usepackage{appendix}
\AtBeginEnvironment{appendices}{\crefalias{section}{appendix}}

\usepackage{subcaption}

\usepackage{graphicx}
\usepackage[
 svgnames,
 table,
 ]{xcolor}

\usepackage{listings}
\usepackage{bussproofs}
\usepackage{tikz}

\usepackage{xspace,cite}
\usepackage{pdfpages}
\newcommand{\etal}{et al.\xspace}
\usetikzlibrary{arrows,shapes,trees,positioning,decorations,shapes}
\usetikzlibrary{arrows.meta}

\usepackage[skins]{tcolorbox}
\tcbuselibrary{listings}
\DeclareOldFontCommand{\rm}{\normalfont\rmfamily}{\mathrm}
\DeclareOldFontCommand{\sf}{\normalfont\sffamily}{\mathsf}
\DeclareOldFontCommand{\tt}{\normalfont\ttfamily}{\mathtt}
\DeclareOldFontCommand{\bf}{\normalfont\bfseries}{\mathbf}
\DeclareOldFontCommand{\it}{\normalfont\itshape}{\mathit}
\DeclareOldFontCommand{\sl}{\normalfont\slshape}{\@nomath\sl}
\DeclareOldFontCommand{\sc}{\normalfont\scshape}{\@nomath\sc}
\DeclareRobustCommand*\cal{\@fontswitch\relax\mathcal}
\DeclareRobustCommand*\mit{\@fontswitch\relax\mathnormal}

\colorlet{keywordcolor}{blue!50!black}
\colorlet{commentcolor}{green!60!black}
\colorlet{typecolor}{black}
 
\newcommand{\sourcefont}{\ttfamily\small}
\newcommand{\commentfont}{\slshape\rmfamily\color{commentcolor}}

\lstdefinelanguage{SMOL}{
        keywords={links,domain,hidden,construct, access,
                  member, validate,rule},
        keywordstyle=\color{keywordcolor}\bfseries\sffamily,
        morekeywords=[2]{abstract,for,in,out,destroy,simulate,
                         new,end,od,do,class,if,then,else,while, return,
                         skip, main, print, null,this,foreach,extends},
        keywordstyle=[2]\color{typecolor}\bfseries\sffamily,
        sensitive=true,
        comment=[l]{//},
        morecomment=[s]{/*}{*/},
        morestring=[b]"
}

\lstdefinestyle{code}{
        basicstyle=\sourcefont\upshape,
        keywordstyle=\color{keywordcolor}\bfseries\sffamily,
        commentstyle=\commentfont,
        columns=fullflexible,
        mathescape=true,
        escapechar={\#},
        keepspaces=true,
        showstringspaces=false,
        aboveskip=8pt, % default is \medskipamount
        numbers=left,
        stepnumber=1, 
        numberstyle=\ttfamily\scriptsize\color{gray},
        numbersep=4pt,
        xleftmargin=1.5em,
        xrightmargin=1.5em,
        framexleftmargin=1.2em,
        framexrightmargin=1em,
        framextopmargin=0.5ex,
        breaklines=true,
        breakindent=3pt,
}

\lstdefinestyle{smol}{
        style=code,
        language=SMOL,
}

\newcommand{\code}[2][]{\lstinline[style=code,basicstyle=\ttfamily\upshape,#1]{#2}}
\newcommand{\smol}[2][]{\code[style=smol,#1]{#2}}

\lstnewenvironment{smolcode}[1][]{
        \minipage{0.92\textwidth}
        \lstset{style=smol,
        framerule=1pt,
        backgroundcolor=\color{white},
        rulecolor=\color{gray!50},
        frame=tblr,
        xleftmargin=0cm,
        xrightmargin=0cm,
        captionpos=b,
        #1}
}{\endminipage}

\newtcblisting{smolframe}[1][]{%
enhanced,colback=white,colframe=black,coltitle=black,
sharp corners,boxrule=0.4pt,
fonttitle=\itshape,
attach boxed title to top left={yshift=-0.3\baselineskip-0.4pt,xshift=2mm},
boxed title style={tile,size=minimal,left=0.5mm,right=0.5mm,
colback=white,before upper=\strut},
listing only, listing options={style=smol,framerule=1pt,
        backgroundcolor=\color{white},
        frame=none,
        xleftmargin=0cm,
        xrightmargin=0cm,#1},
top=-5pt,
bottom=-5pt,
title={\normalfont\texttt{SMOL}}
}

\lstdefinelanguage{RDF}{
        keywords={},
        keywordstyle=\color{keywordcolor}\bfseries\sffamily,
        morekeywords=[2]{link, load, anchor, retrieve, extends},
        keywordstyle=[2]\color{typecolor},
        sensitive=true,
        comment=[l]{//},
        morecomment=[s]{/*}{*/},
        morestring=[b]"
}

\lstdefinelanguage{OWL}{ keywords={Class, SubClassOf,SubPropertyOf, and, some,
    DataProperty, Domain, Range, Characteristics,,EquivalentTo,DisjointWith,DisjointUnionOf, ObjectProperty, not, exactly, only, Individual, Facts, Types, InverseFunctional, Functional,AllDisjointClasses},
  keywordstyle=\bfseries\sffamily,
        sensitive=true,
        comment=[l]{//},
        morecomment=[s]{/*}{*/},
        morestring=[b]"
}

\lstdefinelanguage{SPARQL}{ keywords={SELECT, ASK, FILTER, NOT, EXISTS, WHERE},
  keywordstyle=\bfseries\sffamily,
        keywordstyle=[2]\color{keywordcolor}\bfseries\sffamily,
        sensitive=true,
        comment=[l]{//},
        morecomment=[s]{/*}{*/},
        morestring=[b]"
}

\lstdefinestyle{owl}{
        style=code,
        language=OWL,
}

\lstdefinestyle{sparql}{
        style=code,
        language=SPARQL,
}

\lstdefinestyle{rdf}{
        style=code,
        language=RDF,
}

\newtcblisting{rdfframe}[1][]{%
enhanced,colback=white,colframe=black,coltitle=black,
sharp corners,boxrule=0.4pt,
fonttitle=\itshape,
attach boxed title to top left={yshift=-0.3\baselineskip-0.4pt,xshift=2mm},
boxed title style={tile,size=minimal,left=0.5mm,right=0.5mm,
colback=white,before upper=\strut},
listing only, listing options={style=rdf,framerule=1pt,
        backgroundcolor=\color{white},
        frame=none,
        xleftmargin=0cm,
        xrightmargin=0cm},
top=-5pt,
bottom=-5pt,
title={\normalfont\texttt{RDF}},#1
}

\newtcblisting{owlframe}[1][]{%
enhanced,colback=white,%colframe=black,
coltitle=black,
sharp corners,boxrule=0.4pt,
fonttitle=\itshape,
attach boxed title to top left={yshift=-0.3\baselineskip-0.4pt,xshift=2mm},
boxed title style={tile,size=minimal,left=0.5mm,right=0.5mm,
colback=white,before upper=\strut},
listing only, listing options={style=owl,framerule=1pt,
        backgroundcolor=\color{white},
        frame=none,
        xleftmargin=0cm,
        xrightmargin=0cm},
top=-5pt,
bottom=-5pt,
title={\normalfont\texttt{OWL}},#1
}

\newtcblisting{sparqlframe}[1][]{%
enhanced,colback=white,colframe=black,coltitle=black,
sharp corners,boxrule=0.4pt,
fonttitle=\itshape,
attach boxed title to top left={yshift=-0.3\baselineskip-0.4pt,xshift=2mm},
boxed title style={tile,size=minimal,left=0.5mm,right=0.5mm,
colback=white,before upper=\strut},
listing only, listing options={style=sparql,framerule=1pt,
        backgroundcolor=\color{white},
        frame=none,
        xleftmargin=0cm,
        xrightmargin=0cm,#1},
top=-5pt,
bottom=-5pt,
title={\normalfont\texttt{SPARQL}}
}

\newtheorem{assumption}{Assumption}
\makeatletter
\newsavebox{\@brx}
\newcommand{\llangle}[1][]{\savebox{\@brx}{\(\m@th{#1\langle}\)}%
  \mathopen{\copy\@brx\kern-0.5\wd\@brx\usebox{\@brx}}}
  \newcommand{\rrangle}[1][]{\savebox{\@brx}{\(\m@th{#1\rangle}\)}%
    \mathclose{\copy\@brx\kern-0.5\wd\@brx\usebox{\@brx}}}
    \makeatother

\let\temp\phi
\let\phi\varphi
\let\varphi\temp

\makeatletter
\newcommand{\xRightarrow}[2][]{\ext@arrow 0359\Rightarrowfill@{#1}{#2}}
\makeatother

\newcommand{\SMOL}{\ensuremath{\mathtt{SMOL}}\xspace}

\newcommand{\xsmol}[1]{\text{\smol{#1}}}
\newcommand{\sem}[1]{\ensuremath{ \llbracket #1 \rrbracket \xspace}}

\newcommand{\many}[1]{\overline{#1}}

\newcommand{\newrev}[1]{\textcolor{black}{#1}}

\newcommand{\rulename}[1]{\textbf{\upshape\scriptsize(\textsf{#1})}}

\newcommand{\TTTINFER}[6][]{
\begin{prooftree}
\AxiomC{\ensuremath{#3}}
\noLine
\UnaryInfC{\ensuremath{#4}}
\noLine
\UnaryInfC{\ensuremath{#5}}
\LeftLabel{\rulename{#2}}
\RightLabel{#1}
\UnaryInfC{\ensuremath{#6}}
\end{prooftree}
}

\newcommand{\TTINFER}[5][]{
\begin{prooftree}
\AxiomC{\ensuremath{#3}}
\noLine
\UnaryInfC{\ensuremath{#4}}
\LeftLabel{\rulename{#2}}
\RightLabel{#1}
\UnaryInfC{\ensuremath{#5}}
\end{prooftree}
}

\newcommand{\TINFER}[4][]{
\begin{prooftree}
\AxiomC{\ensuremath{#3}}
\LeftLabel{\rulename{#2}}
\RightLabel{#1}
\UnaryInfC{\ensuremath{#4}}
\end{prooftree}
}

\newcommand{\ssep}{\ | \ }

\newcommand{\Prgm}{\ensuremath{\mathsf{Prog}}\xspace}
\newcommand{\conf}{\ensuremath{\mathsf{conf}}\xspace}
\newcommand{\classtable}{\ensuremath{\mathsf{CT}}\xspace}
\newcommand{\dll}{{\ensuremath{\mathsf{street}}}\xspace}
\newcommand{\userkb}{\ensuremath{{\mathcal{K}_\mathbf{domain}}}\xspace}
\newcommand{\smolkb}{\ensuremath{{\mathcal{K}_\SMOL}}\xspace}

\newcommand{\userpre}[1]{\ensuremath{{#1}^\mathbf{domain}}}
\newcommand{\progpre}[1]{\ensuremath{{#1}^\mathbf{prog}}}
\newcommand{\runpre}[1]{\ensuremath{{#1}^\mathbf{run}}}
\newcommand{\smolpre}[1]{\ensuremath{{#1}^\SMOL}}

\title{Semantically~Reflected~Programs} 

\author{Eduard Kamburjan}{IT University of Copenhagen, Denmark\newline University of Oslo, Norway}{eduard.kamburjan@itu.dk}{0000-0002-0996-2543}{}
\author{Vidar Norstein Klungre}{University of Oslo, Norway}{}{0000-0003-1925-5911}{}
\author{Yuanwei Qu}{University of Oslo, Norway}{quy@ifi.uio.no}{0000-0003-3220-2101}{}
\author{Rudolf Schlatte}{University of Oslo, Norway}{rudi@ifi.uio.no}{0000-0001-5601-5517}{}
\author{Egor V. Kostylev}{University of Oslo, Norway}{egork@ifi.uio.no}{0000-0002-8886-6129}{}
\author{Martin Giese}{University of Oslo, Norway}{martingi@ifi.uio.no}{0000-0002-2058-2728}{}
\author{{Einar Broch} Johnsen}{University of Oslo, Norway}{einarj@ifi.uio.no}{0000-0001-5382-3949}{}

\authorrunning{E. Kamburjan et al.} %TODO mandatory. First: Use abbreviated first/middle names. Second (only in severe cases): Use first author plus 'et al.'

\Copyright{Eduard Kamburjan, Vidar Norstein Klungre, Yuanwei Qu, Rudolf Schlatte, Egor V. Kostylev, Martin Giese, Einar Broch Johnsen} %TODO mandatory, please use full first names. TGDK license is "CC-BY";  http://creativecommons.org/licenses/by/4.0/

\ccsdesc[500]{Software and its engineering~Object oriented languages}
\ccsdesc[500]{Computing methodologies~Knowledge representation and reasoning}
\ccsdesc[100]{Computing methodologies~Modeling and simulation}

\keywords{Knowledge Graphs, Ontologies, Object-Oriented Modelling, \newrev{Imperative} Programming Languages, Reflection, Type Safety} %TODO mandatory; please add comma-separated list of keywords

\category{} %optional, e.g. invited paper

\relatedversion{} %optional, e.g. full version hosted on arXiv, HAL, or other respository/website
\supplement{All supplementary resources are available in the github repository linked below.}%optional, e.g. related research data, source code, ... hosted on a repository like zenodo, figshare, GitHub, ...
\supplementdetails[]{Software}{https://github.com/smolang/SemanticObjects/tree/prepare-1.0} %linktext, cite, and subcategory are optional

\acknowledgements{We thank the anonymous reviewers for their constructive feedback, in particular for the suggestion of the ABox-based algorithm for typing described in Sec.~5.2.2.}%optional

\nolinenumbers %uncomment to disable line numbering

\SeriesVolume{4}
\SeriesIssue{1}
\SectionAreaEditor{}

\DateSubmission{}
\DateAcceptance{}
\DatePublished{}

\ArticleNo{3}
\begin{document}

\maketitle

\begin{abstract}
  This paper addresses the dichotomy between the formalization of
  structural and the formalization of \emph{\newrev{executable}} behavioral knowledge by means of
  semantically lifted programs, which explore an intuitive connection
  between \newrev{imperative} programs and knowledge graphs.
  While knowledge graphs and ontologies are eminently useful to
  represent formal knowledge about a system's individuals and
  universals, programming languages are designed to describe the
  system's evolution.
  To address this dichotomy, we introduce a \emph{semantic lifting} of
  the program states
  % runtime configurations
  of an executing progam into a knowledge graph, for an
  object-oriented programming language.  The resulting graph is
  exposed as a \emph{semantic reflection} layer within the programming
  language, allowing programmers to leverage knowledge of the
  application domain in their programs \newrev{during execution}.
  In this paper, we formalize semantic lifting and semantic reflection
  for a small \newrev{imperative} programming language, SMOL, explain the operational
  aspects of the language, and consider type correctness and
  virtualization for runtime program queries through the semantic
  reflection layer.  We illustrate semantic lifting and semantic
  reflection through a case study of geological modeling and discuss
  different applications of the technique.  The language
  implementation is open source and available online.
\end{abstract}

\section{Introduction}\label{sec:intro}
There is a dichotomy between the formalization of structural and the
formalization of behavioral knowledge, which can be expressed through knowledge graphs and programming languages, respectively. We address this dichotomy by introducing a semantic lifting from program states to description logic (DL) ontologies that enables \newrev{imperative} programs to exploit a semantic view of their own state during execution. This way, structural knowledge can be used from within behavioral knowledge. \looseness=-1

Knowledge graphs and ontologies are eminently useful representations of formal knowledge about the individuals and universals of systems. Among others, they give us decidable reasoning, easy avenues for negotiating domain knowledge with non-technical stakeholders, ‘native’ ways of integrating information sources, and, not least, a wealth of well established standards.  However, they are less suitable for the representation of change, and in particular dynamic behavior. Although concepts of change have been investigated ontologically~\cite{ZambGuiz2010,ZambGuiz2014}, and time stamped sensor readings can be represented in RDF~\cite{SensorNetworkOntology}, the essence of state change remains external to description logic-based knowledge representation, and \emph{how} states change is not readily expressed.

In contrast, programming languages are specifically designed to describe \emph{behavior}, i.e., the evolution of systems.  The most common use of programming languages is to specify programs to be executed, but the use of programming languages for behavioral modeling for simulation and analysis is also well established~\cite{SpecSharp,JohnsenHSSS10}.  In fact, the object-oriented programming paradigm emerged from discrete event simulation languages as a more natural way of representing the interaction between different entities~\cite{dahl04birth}. However, the systems specified by programming languages are rarely pure models, but contain additional implementation-driven structure that interferes with domain modeling and may even become the dominant view of a system, especially when independently developed models need to be integrated.

It is natural to ask for a formalism that combines the advantages of semantic technologies for the representation of states with the elegance and maturity of programming languages to describe the \newrev{execution and} evolution of states.  Different approaches have been proposed that attempt such a combination.  For example, one can try to express program behavior in terms of actions on a description logic interpretation~\cite{ZarriessC15} or a DL ontology~\cite{Calvanese11}. A recent approach~\cite{DubslaffKT21,DubslaffKT20} has combined a guarded command language with DL reasoning to enable probabilistic model checking over the combination.  A combination of RDF and rewriting theories in Maude has also been investigated~\cite{DinKPSYO19,Yu2021}.  These approaches are all quite far from current state-of-the-art programming paradigms, and come with their own set of technical challenges. \newrev{Other approaches, such as Owlready~\cite{DBLP:journals/artmed/Lamy17} or libraries operating directly on the OWL metamodel such as the OWL-API~\cite{DBLP:conf/semweb/HorridgeB08}, give up the separation of concerns between ontological modeling and data modeling, a problem known as the semantic gap or impedance mismatch~\cite{DBLP:journals/ijseke/BasetS18,DBLP:conf/sigmod/CopelandM84,lazy}. The semantic gap leads to a lack of tool support and prohibits the use of patterns aiming at code reuse and behavior in the program directly.}

\newrev{While processes and other kinds of behavior can be described by ontologies~\cite{DBLP:journals/tgdk/HarthKRCKG24}, they follow different patterns and aims than programs, which describe \emph{executable behavior} and are thus concerned with code reuse. For example, behavioral subtyping~\cite{DBLP:journals/toplas/LiskovW94} is a crucial pattern that expresses that from the perspective of a caller, the contract of a method $m$ in a class $C$ has also to hold for $m$ in all subclasses of $C$. This pattern is not decidable (as it compares two Turing complete methods), and is not useful in a purely conceptual setting where execution and code reuse are not present or are not presented.}

We propose a connection between programs and knowledge graphs that integrates both kinds of knowledge: we develop a semantic lifting that maps from program states in an object-oriented programming language to an RDF graph, including the running program's objects, fields, and call stack.  Abstraction is supported in the mapping by integrating computations in the lifting process, thereby allowing, e.g., implementation-specific structure to be ignored by the mapping. The RDF graph can be exposed within the programming language, which adds a \emph{semantic reflection} layer to programs.  This reflection layer enables \emph{semantic programming} where the semantic view of the state can be exploited by the program; in particular, formalized knowledge of the application domain can be used within the program by querying for objects using domain knowledge.

In this paper, we focus on the essence of semantic lifting and semantic reflection: the paper formalizes semantic lifting of object-oriented program states and semantic reflection for a small programming language \SMOL (short for Semantic Micro Object Language) and explains both the operational aspects of the language and the mapping between states and RDF graphs; further, we discuss type correctness and virtualization for queries on semantically lifted program states from within the programs.  An important aspect of this work lies in the intricate relationship between object-oriented typing%
%and that of RDF and its extension RDFS. 
\newrev{, and class membership and subclassing in RDF(s).}

\subsection*{Contributions}
This paper, which builds on work published at ESWC \cite{DBLP:conf/esws/KamburjanKSJG21}, reports on a strand of research on semantically lifted programs. Compared to the previous paper, this paper features a reworked presentation of \SMOL based on our experiences with several case studies and applications --- including the removal of features that proved to be less useful in practice.

This paper includes the following technical improvements to semantic lifting and semantic reflection, compared to the original publications on semantic lifting~\cite{DBLP:conf/esws/KamburjanKSJG21} and its type system~\cite{DBLP:conf/dlog/KamburjanK21}:
\begin{enumerate}
\item a new \emph{semantic pointer mechanism} that explicitly connects   the program knowledge graph with a domain knowledge graph;
\item The \emph{ontology} of the lifting has been remodeled, compared   to the previous work~\cite{DBLP:conf/esws/KamburjanKSJG21};
\item a \emph{full formalization} of the type system, including a new   result that shows that all reachable states are semantically lifted   to \emph{consistent} knowledge graphs; and
\item a discussion of the \emph{virtualization} of semantically   lifted program states.
\end{enumerate}
We furthermore discuss several published case studies and applications of semantic lifting outside the \SMOL language.

\subsection*{Paper Overview}
\Cref{sec:overview} gives a general overview of semantic lifting and reflection by means of a motivating example.  \Cref{sec:lang} introduces \SMOL, a small object-oriented language and \cref{sec:graph} details its semantic lifting mechanism. \Cref{sec:internal} explains semantic reflection in \SMOL and type safety for queries through the semantic reflection layer. We discuss the implementation of \SMOL and describe how our work with applications influenced the language design in \cref{sec:eval}.  Related work is reviewed in \cref{sec:related} and \cref{sec:conclusion} concludes the paper.

\subsection*{A Note on Notation}
We assume a general familiarity with the standard Semantic Web stack of RDF, OWL, SPARQL and SHACL; for an introduction, see, e.g., Hitzler et al.~\cite{Hitzler2010} and the online documentation.\footnote{\url{https://www.w3.org/TR/rdf11-primer/}}

Some notions, most prominently ``class'' and ``object'', denote different entities in program semantics and knowledge representation. In cases where the exact meaning is not clear from the immediate context, we use ``concept'', ``individual'' and ``node'' for the knowledge representation entities and ``class'', ``instance'' and ``runtime object'' for the program semantics entities.

In this paper, we use DL syntax for axioms in the program semantics and RDF Turtle syntax in examples.  Given a SPARQL query $\mathsf{Q}$, an entailment regime $\mathsf{er}$ and an knowledge graph $\mathcal{K}$, the function $\mathsf{Ans}_\mathsf{er}(\mathcal{K},\mathsf{Q})$ returns the result set, $\mathsf{Sha}(\mathcal{K},\mathsf{shacl})$ returns a Boolean depending on whether $\mathcal{K}$ conforms to the SHACL shape $\mathsf{shacl}$, and $\mathsf{Mem}(\mathcal{K},\mathsf{owl})$ returns all members of the OWL concept $\mathsf{owl}$ in $\mathcal{K}$. Query containment for queries $\mathsf{Q}_1$ and $\mathsf{Q}_2$ under an entailment regime $\mathsf{er}$ and a knowledge graph $\mathcal{K}$ is denoted $\mathsf{Q}_1 \subseteq^\mathcal{K}_\mathsf{er} \mathsf{Q}_2$.

% \paragraph*{ToDos}
% Before detailed work, we need to restructure the whole thing.
% \begin{itemize}
%     \item Section 2: Fill in Gaps
%     \item Section 3: Ok
%     \item Section 4: Needs more explanation, especially of the semantic pointers. Simplification now that we ignore AST and function stack.
%     \item Section 5: Remove remains of CSSA. Shorten and add second theorem, maybe just add usual type safety?
%     \item Section 6: We only have 1 RQ now that CSSA is removed, some thought required here.
%     \item Section 7: Move all discussions about alternative here, add structure.
%     \item Section 8: Add papers and points discovered/published since last pass
%     \item Section 9: Ok
% \end{itemize}

%%% Local Variables: 
%%% mode: latex
%%% TeX-master: "../main"
%%% End: 

\section{Motivating Example}\label{sec:overview}
We introduce the techniques of semantic lifting and semantic
reflection through a motivating example to illustrate how these
techniques allow us to combine domain knowledge for static modeling
and programming for dynamic modeling. We consider an example based on
a simulator for geological processes, developed in \SMOL by Yu et
al.~\cite{geo2}, to show how complex domain knowledge expressed in an
ontology can be integrated into a program.

Let us implement a program that simulates geological processes in a
system that captures the deposition and erosion of geological layers
in petroleum geoscience, as well as the transformation of organic
matter inside these layers to petroleum.  The program needs to access
domain knowledge about conditions that trigger such transformations in
order to perform a meaningful simulation.  Whereas Yu et
al.~\cite{geo2} considered a realistic ontology for this domain, our
ontology will be simplified to focus on the interactions between the
program and the ontology.

A \emph{petroleum system} in the energy industry describes the
different entities that relate to hydrocarbon production and
storage~\cite{peters_petroleum_1994}.  We focus on the
physical-geological components and processes that are involved in the
formation of hydrocarbon accumulation, which can be separated into
three classes: \emph{physical-geological components}, the different
geological layers and their types of rocks and properties; \emph{thermal
  transformations}, the processes describing transformation and
accumulation of hydrocarbons within these layers; and
\emph{compaction}, the change of physical properties in the rock
during its burial.  We consider stacks of layers; i.e., the geological
layers are layered upon each other.

We distinguish between \emph{source rock}, that can generate
petroleum, and \emph{reservoir rock}, that can store it. Each layer
has one type of homogeneous rock as material, where we model
\emph{shale}, \emph{limestone}, and \emph{sandstone}.  In our model,
each layer of rock has homogeneous rock properties such as grain size,
porosity, and permeability.

Given a description of the state of a geological system, different
geological processes can affect the layers. Let us consider
\emph{cooking}, which transforms \emph{kerogen} in a source rock into
petroleum. Kerogen refers to a collection of large and complex,
insoluble molecules that are dehydrated from fresh organic matter after
burial and compaction by overlying at least 100~$m$ sediments
\cite{bjorlykke_source_2010}.\looseness=-1

Temperature plays a key role during kerogen’s thermal transformation,
although other factors such as pressure, time, and mineral type also
play a role.
% influence this transformation.
We concentrate on the North Sea and the Norwegian Sea, where the
general gradient is about 30 $^\circ C$ increase in temperature for each
kilometer depth~\cite{nathenson_geothermal_1988,
  bjorlykke_source_2010}.  Cooking of oil starts at 60
$^\circ C$~\cite{bjorlykke_source_2010}.
% , while burning starts at 150 C~\cite{bjorlykke_source_2010}.

% \Cref{fig:models} shows interactions of dynamic and static models.
The static models, i.e., knowledge graphs and ontologies, are used to
model the structure of the domain and the current state of the
geological layers.
% units.
The dynamic models, i.e., programs, are used to describe the processes
that transfer the system between states. At their interaction, we must
be able to interpret the program state in the static model and
retrieve information from it to determine the triggering layers for
the processes.\looseness=-1

% \begin{figure}[t]
%     \centering
%     \includegraphics[width=0.5\linewidth]{figures/geo.pdf}
%     \caption{Static and dynamic models.}
%     \label{fig:models}
% \end{figure}

%%% Local Variables: 
%%% mode: latex
%%% TeX-master: "../main"
%%% End: 

\subsection{An Ontology for the Static Model}\label{ssec:trigger}

The concepts of layers, their properties and their relation to each
other can be described in an ontology. The ontology does not describe
processes, but rather describes \emph{triggers}: A layer is a trigger
if it fulfills the conditions to trigger some geological process.  For
example, a layer is a \emph{cooking trigger}, if it \emph{(a)}
contains uncooked kerogen, \emph{(b)} is below a certain minimal depth
and \emph{(c)} is above a certain maximal depth.

The basic geological notions that we need for our simulator are
organic matter, rocks and layers. Organic matter is kerogen,
oil or gas. These notions are represented as follows.

\begin{owlframe}[]
domain:Oil SubClassOf domain:OrganicMatter
domain:Gas SubClassOf domain:OrganicMatter
domain:Kerogen SubClassOf domain:OrganicMatter
\end{owlframe}

% \begin{align*}
% \mathtt{domain}&\mathtt{:\!Oil}~\mathbf{SubClassOf}~\mathtt{domain\!:\!OrganicMatter}\\
% \mathtt{domain}&\mathtt{:\!Gas}~\mathbf{SubClassOf}~\mathtt{domain\!:\!OrganicMatter}\\
% \mathtt{domain}&\mathtt{:\!Kerogen}~\mathbf{SubClassOf}~\mathtt{domain\!:\!OrganicMatter}
% \end{align*}

We here focus on two types of rocks, \emph{shale} and
\emph{sandstone}, among the different rocks and layers.  A layer
consists of one kind of rock and may contain organic matter. We model
stratigraphic layers that are stacked on each other.

\begin{owlframe}
domain:SiliciclasticRock SubClassOf domain:Rock
domain:Shale SubClassOf domain:SiliciclasticRock
domain:Sandstone SubClassOf domain:SiliciclasticRock
domain:StratigraphicLayer SubClassOf domain:constitutedBy exactly 1 domain:Rock
domain:StratigraphicLayer SubClassOf 
         domain:constitutedBy only domain:SiliciclasticRock
\end{owlframe}

\medskip

In addition to the geological notions, we model \emph{triggers}. A
trigger is a stratigraphic layer that enables some process.  We focus
on the trigger for the \emph{cooking} process here; in general, any
layer can be in a state that triggers a process.  Thus, a trigger is a
layer, expressed using the following axiom:\looseness=-1
\begin{owlframe}
domain:Trigger SubClassOf domain:StratigraphicLayer
\end{owlframe}

A layer can trigger the cooking process if it contains kerogen and is
below 2000~$m$ depth but above 5000~$m$. We do not describe the
cooking process itself, i.e., what happens to the kerogen during or
after cooking, in the ontology.

\begin{owlframe}
domain:CookingTrigger EquivalentTo domain:Trigger
  and (domain:constitutedBy some (domain:contains some domain:Kerogen))
  and (domain:depth some xsd:integer[$\geq$ 2000, $\leq$ 5000])  
\end{owlframe}

\subsection{A Program for the Dynamic Model}\label{ssec:simulate}
\begin{figure}[t]
%\begin{subfigure}{\textwidth}
\begin{smolframe}
abstract class GeoLayer(domain Int thickness,  domain depth,
                      hidden GeoLayer above, hidden GeoLayer below) 
  Unit update()
    Int res = 0;
    if(this.above != null) then 
      res = this.above.depth + this.above.thickness; 
    end
    this.depth = res;
  end
  Boolean canPropagate() return False; end
  Unit migrate() skip; end
end
  
class Bedrock extends GeoLayer() end 

class Shale extends GeoLayer(hidden Int kerogen) 
  links(this.kerogen == 1 || this.kerogen == 2) 
    "a domain:Stratigraphic_Layer; 
     domain:constitutedBy [a domain:Shale]; 
     domain:constitutedBy [domain:contains [a domain:Kerogen]].";
  links "a domain:Stratigraphic_Layer; 
          domain:constitutedBy [a domain:Shale].";
  Unit cook() this.kerogen = this.kerogen +1; end
end
\end{smolframe}
% \caption{\label{fig:smolshale}Geological shale layers in the simulator.}
% \end{subfigure}
% \caption{\label{fig:overviewtotal}Excerpts of a semantically reflective simulator.}
\caption{\label{fig:smollayer}Geological layers in the simulator.}
\end{figure}

The \SMOL program uses the ontology developed in \cref{ssec:trigger}
to simulate geological process.  The program's input is a
\emph{geological scenario}, which is a sequence of deposition and
erosion events, and its output is the final state of the system.  The
program's internal structure mirrors the structure of the domain, so
its central data structure is a stack of geological layer objects.

Observe that these geological layers play a dual role as both
computational and domain-specific
artifacts~\cite{DBLP:conf/birthday/KamburjanF20}.  On one hand, they
implement behavior like migration of hydrocarbons or perform
computations like their current depth. On the other hand, they relate
to the domain knowledge encoded in the above axioms.  Let us first
examine the classes in \cref{fig:smollayer}. They model generic
geological layers as class \smol{GeoLayer}, with a state that includes
a given thickness, depth and neighboring layers, and methods to
manipulate the state. The \smol{Bedrock} class describes the lowest
layer of rock that we consider in our scenarios. The \smol{Shale}
class specializes \smol{GeoLayer} to a layer that contains only
shale. This class has a field \smol{kerogen} that contains the status
of kerogen within the modeled layer. If this field has value 1 or 2,
the layer contains kerogen, if the field has value 0, the layer has no
kerogen, if the field has any other value, the layer contains
overcooked kerogen.

Let us now examine the semantic lifting of a \smol{Shale} object, for
the moment ignoring
% but ignore
the \smol{links} clause, and the \smol{domain} and \smol{hidden}
modifiers of the class definition (see \cref{fig:smollayer}). For this
example, we consider an object created with the following statements.

\begin{smolframe}
Bedrock bed = 
  new Bedrock(/*thickness:*/100, /*depth:*/100, /*above:*/null, /*below:*/null);
Shale sh = 
  new Shale(/*kerogen:*/1,/*thickness:*/100,/*depth:*/0,/*above:*/null,/*below:*/bed);
bed.above = sh;
\end{smolframe}

\Cref{fig:ex:rdf1} shows an excerpt of the resulting semantic lifting
(ignoring the modifiers and special clauses, and the class table). It
is a serialization in RDF, outlined for a node \texttt{run:obj1} for
the shale object and a node \texttt{run:obj2} for the bedrock object.

\begin{figure}[bht]
%\begin{subfigure}{\textwidth}
\begin{rdfframe}
prog:GeoLayer a owl:Class;
prog:Shale owl:subClassOf prog:GeoLayer.
prog:Bedrock owl:subClassOf prog:GeoLayer.
run:obj1 a prog:Shale;
         prog:kerogen 1;  prog:thickness 100; prog:depth 0;
         prog:above smol:null; prog:below run:obj2.
run:obj2 a prog:Bedrock;
         prog:thickness 100; prog:depth 100;
         prog:above run:obj1; prog:below smol:null.
\end{rdfframe}
\caption{Excerpt of the lifting without modifiers and linking
  clause.\label{fig:ex:rdf1}}
% \end{figure}

% \begin{figure}[thb]
\begin{rdfframe}[]
prog:GeoLayer a owl:Class;
prog:Shale owl:subClassOf prog:GeoLayer.
prog:Bedrock owl:subClassOf prog:GeoLayer.
run:obj1 a prog:Shale; smol:links run:l1.
run:l1 domain:thickness 100; domain:depth 0; 
       a domain:Stratigraphic_Layer; domain:constitutedBy [a domain:Shale];
       domain:constitutedBy [domain:contains [a domain:Kerogen]].
\end{rdfframe}
\caption{Excerpt of the lifting with modifiers and linking
  clause.\label{fig:ex:rdf2}}
\end{figure}

Observe that the semantic lifting of objects,
% Already this serialization,
without any connection to the domain ontology, is already useful. For
example, we can use SHACL to formulate the restriction that (a) there
is only one object acting as bedrock and (b) a bedrock object is the
lowest one. In other words, semantic technologies can be used as a
specification language for object-oriented programs\newrev{, with
  built-in support for logical inference}.
% Similarly,
\newrev{Reflective features have been explored in object-oriented
  programming languages such as CLOS, Java and Smalltalk
  \cite{bracha04oopsla,kiczales91mop}, to support, e.g., introspection
  by implementing methods that reveal structural aspects of the
  program such as the class of a given object. Semantic lifting also
  enables introspection, but it happens outside of the runtime
  system:} we can use SPARQL to retrieve objects that satisfy
particular \newrev{logical} properties, without the need to manually
traverse the state using a debugger. We refer to this way of using the
lifted state, which is external to the program semantics, as
\emph{semantic state access}.

The \smol{Shale} object is lifted as a node of class
\texttt{prog:Shale}. This class is \emph{not} part of the domain
ontology. In fact, this node is not part of the geological domain at
all: If it were, the node would have the properties of the
\texttt{domain:StratigraphicLayer} class and be restricted by the
axioms governing the domain ontology.  Such a design would be problematic
% This would be a doubtful choice, as
because this would restrict the program with constraints not concerned
with computational structures and merge the domain model with the
computational model.

We want to preserve the separation of concerns between these two
modeling paradigms, and instead \emph{link} the lifted state to the
domain. For each \SMOL object, two nodes are generated: one representing
the object itself (the above \texttt{run:obj1}) and one node
representing an entity in the domain to which the object is
linked. These two objects are connected using a special relation
\texttt{smol:links}.\looseness=-1

Semantic lifting serializes the program state, and provides a way to
specify how domain objects link to the program state. In the \SMOL
code of \cref{fig:smollayer}, these are the modifiers and the
\smol{links} clause.  The \smol{hidden} modifiers prohibit a field
from being lifted.  
\newrev{This allows us to control the knowledge graph: As parts of the program that are unrelated to the operations to the lifted state are removed, the resulting graph is (a) smaller and (b) more focused.}
% This allows us to control the size of the
% knowledge graph if some part of the program is unrelated to the
% operations performed on the lifted state.  
In contrast, the
\smol{domain} modifier moves information from the computational object
to the linked domain node. In the example above, this will attach the
edge lifting field \smol{depth} not to the object \texttt{run:obj1},
but to its linked node.

The \smol{links} clause is a general way to annotate information to
the linked object. The clause in the \smol{Shale} class (see
\cref{fig:smollayer}) expresses that every node linked to the lifting
of a \smol{Shale} object is a stratigraphic layer constituted by
shale. The class has two \smol{links} clauses. The first is
conditional --- if the expression \smol{this.kerogen == 1 ||
  this.kerogen == 2} evaluates to true, then the linked object
contains kerogen, otherwise the unconditional clause is used and the
linked object does not contain kerogen. This way, the semantic lifting
precisely captures the meaning of the \smol{kerogen} field in terms of
the domain ontology. The above \smol{Shale} object is, when these
features are considered, lifted in the graph in
\cref{fig:ex:rdf2}. Here, the object \texttt{run:l1} is the linked
object.

Semantic state access can be used to exhibit the state. For
example, the following query
% enables us to
extracts all objects containing kerogen (more precisely, all \SMOL
objects that are linked to an OWL object that contains kerogen).
\begin{sparqlframe}
    SELECT ?x { ?x [smol:links [domain:contains [a domain:Kerogen]]}
\end{sparqlframe}

% More importantly, we can also execute such
Queries can be executed from within the program to reflect on the
state.  We refer to such queries as \emph{semantic reflection},
because the domain ontology and the semantically lifted program state
are directly used in the program.  Consider the code in
\cref{fig:cooktrigger}, which queries for all \smol{Shale} objects
that are linked to a layer triggering the cooking process.  In our
work, we use semantic reflection to facilitate the
following:\looseness=-1
\begin{itemize}
\item \textbf{A separation of concerns} between the modeling of
  structure, such as layers, their properties and relations to each
  other, and the modeling of behavior, i.e., changes in these
  structures.
\item \textbf{A prevention of redundancy}: the properties of the
  layers must not be expressed in both the program and the
  ontology. Instead, the ontology is used directly.
\item \textbf{A semantic view:} The queries are expressed in the
  terminology of the domain, using standard semantic technologies
  accessible to domain experts.
\end{itemize}

\begin{figure}[t]
\begin{smolframe}
List<Shale> layers = member("<smol:links> some <domain:CookingTrigger>");
while(layers != null) do
  Shale layer = layers.content;
  layer.cook();
  layers = layers.next;
end
\end{smolframe}
\caption{\label{fig:cooktrigger}Executing the cooking process.}
\end{figure}

\section{SMOL: An Object-Oriented Language with Semantic Lifting}\label{sec:lang}
This section introduces semantic lifting by defining a
programming language and its runtime semantics, allowing us to
formalize the mapping from program state to knowledge graph and detail
the consequences of this mechanism for programming language design.
As mainstream object-oriented languages, such as Java, are
unnecessarily complex to present their complete and formal runtime
semantics here, we do so by introducing \SMOL (short for
\emph{Semantic Micro Object Language}), a small object-oriented
language with an ALGOL-inspired syntax, enhanced with semantic
lifting.\looseness=-1

We introduce
% the surface syntax of
\SMOL, emphasizing syntactic
support for semantic lifting, and formally define \SMOL in terms of
\emph{surface syntax} and \emph{runtime syntax}.  The surface syntax
describes the program as written by the programmer, while the runtime
syntax describes its internal representation during execution.  The
runtime semantics, i.e., the rules to execute a program, is defined as
transitions between states described in the runtime syntax.  To focus
on semantic lifting, we elide many standard aspects of \SMOL's
semantics; for completeness, the full language semantics is included
in \cref{app:language}.  We will extend \SMOL to investigate semantic
reflection (i.e., the ability to access the knowledge graph generated
by the semantic lifting at runtime from within a program)
in~\cref{sec:internal}. \looseness=-1

\subsection{Surface Syntax}\label{sec:surface}

Assume the standard sets of literal values (i.e., constants), such
as integers $\{1, 2, \ldots\}$, Booleans
$\{\xsmol{true}, \xsmol{false}\}$ and the unit and null singletons
$\{\xsmol{unit}\}$ and $\{\xsmol{null}\}$, respectively, given; we refer to
the names \smol{Int}, \smol{Boolean}, \smol{Unit}, \smol{Null} of
these sets as \emph{basic type names}.  For now, we consider basic
type names as purely syntactic constructs; we return to the type
system in \cref{sec:type}. In the sequel, let $\many{~\cdot~}$
denote comma-separated lists (i.e., zero or more repetitions), and
$[\cdot]$ denote optional constructs. \looseness=-1

\begin{definition}[Surface Syntax]\label{def:syntax}
  The syntax of \SMOL is given by the grammar in \cref{fig:syntax},
  where \smol{C}, \smol{f}, \smol{m}, \smol{v} range over
  \emph{class, type variable, field, method and variable names},
  respectively, which are strings. We also let $\mathtt{le}$ range over lists complying 
$\mathtt{predicateObjectList}$ production in Turtle syntax,\footnote{See~\texttt{https://www.w3.org/TR/turtle/\#grammar-production-predicateObjectList}; \newrev{the missing subject of the \texttt{predicateObjectList} will be filled in at runtime by an individual for the object.}}
  % Furthermore, we let
  \smol{b}
  % range
  over string literals, \smol{t} over basic type names, \smol{a} over
  literal values (including string literals), and $\mathit{op}$ over
  Boolean and arithmetic operators (such as $+$ and $\leq$).
\end{definition}

We use blue bold keywords to highlight syntax relevant for semantic
lifting, and black bold keywords for all other syntax highlighting.

\begin{figure}[t]
  \small
  % \noindent\resizebox{\textwidth}{!}{
    \begin{minipage}{\textwidth}
    \begin{align*}
    \mathsf{Prog} ::=\;& \many{\mathsf{Class}}~\xsmol{main}~\mathsf{Stmt}~\xsmol{end}&&\text{Programs}\\
    \mathsf{Class} ::=\;& \xsmol{class C}\big[\xsmol{extends C} \big]
    (\many{\mathsf{Field}})~[\mathsf{Linkage}]~\many{\mathsf{Met}}~\xsmol{end}&&\text{Classes}\\
    {\mathsf{Type} ::=\;}& \xsmol{t} \ssep \xsmol{C} \ssep \xsmol{List<C>} &&{\text{Types}}\\
    \mathsf{Field} ::=\;& \big[\xsmol{hidden} \ssep \xsmol{domain}\big]~{\mathsf{Type}}~\xsmol{f}&&\text{Fields}\\
    \mathsf{Linkage} ::=\;& \many{\xsmol{links(}\mathsf{Expr}\xsmol{) le;}}~\xsmol{links le;}&&\text{Domain linkage}\\
    \mathsf{Met} ::=\;& 
    {\mathsf{Type}}~\xsmol{m}(\many{{\mathsf{Type}}~\xsmol{v}})~{\mathsf{Stmt}}~\xsmol{end}&&\text{Methods}\\
    \mathsf{Stmt} ::=\;& 
    \mathsf{Loc}~\xsmol{=}~\mathsf{RHS} \xsmol{;} 
        \ssep\xsmol{if}~\mathsf{Expr}~\xsmol{then}~\mathsf{Stmt}~\xsmol{else}~\mathsf{Stmt}~\xsmol{end} 
    &&\text{Statements}\\
    &
    \ssep \mathsf{Expr}\xsmol{.m}(\many{\mathsf{Expr}})\xsmol{;} \ssep \xsmol{skip;} 
    \ssep \xsmol{while}~\mathsf{Expr}~\xsmol{do}~\mathsf{Stmt}~\xsmol{end} \\
    &\ssep\mathsf{Type}~\xsmol{v}~\xsmol{=}~\mathsf{RHS}\xsmol{;} \ssep \mathsf{Stmt~Stmt} \ssep \xsmol{return}~\mathsf{Expr} \xsmol{;} \\
    \mathsf{RHS} ::=\; & 
    \xsmol{new C}(\many{\mathsf{Expr}})~[\mathsf{Linkage}]
    \ssep \mathsf{Expr}.\xsmol{m}(\many{\mathsf{Expr}}) \ssep \mathsf{Expr} 
    &&\text{RHS expressions}\\
    \mathsf{Expr} ::=\;& \xsmol{this} \ssep \xsmol{null} \ssep \mathsf{Loc} \ssep \xsmol{a}
                     \ssep \mathsf{Expr}~\mathit{op}~\mathsf{Expr} \ssep \mathsf{Expr}~\xsmol{==}~\mathsf{Expr} \ssep \mathsf{Expr}~\xsmol{!=}~\mathsf{Expr} &&\text{Expressions}\\
      \mathsf{Loc} ::=\;& \mathsf{Expr}\xsmol{.f} \ssep \xsmol{v} 
    &&\text{Locations}
    \end{align*}
    \end{minipage}
% }
\caption{\label{fig:syntax}Surface syntax of \SMOL.}
\end{figure}

A program in \SMOL consists of a set of classes and a \smol{main}
block with a statement. A class declaration $\mathsf{Class}$ defines
fields and methods.  Classes can extend other classes (using single
inheritance).  For simplicity,
% Inheritance is kept simple:
if a class extends another, then all fields and methods of the
superclass are copied to the subclass. Inherited fields are placed
before newly declared fields.  Types are basic types, class names, or lists of class names. Thus, to avoid unnecessary complexity, we do not include generic types and restrict parametric types to lists of class names only.

Statements \smol{s} and expressions \smol{e} are standard, including a
\smol{null} reference and the self reference \smol{this}. Right-hand
sides \textsf{RHS} extend expressions with imperative constructs with
side effect. These include object creation and method calls. For
simplicity, these can only occur in assignments. Consequently, nested
object creation and method calls inside expressions need to be
encoded.
% are not supported.
Method calls can additionally occur as standalone statements (in which
case the return value from the method call is ignored).
% For simplicity,
Moreover, fields \smol{f} in \SMOL are publicly accessible, and field
access is always prefixed by the target object (e.g., \smol{this.f}).

\begin{figure}
    \centering
    \includegraphics[width=0.85\linewidth]{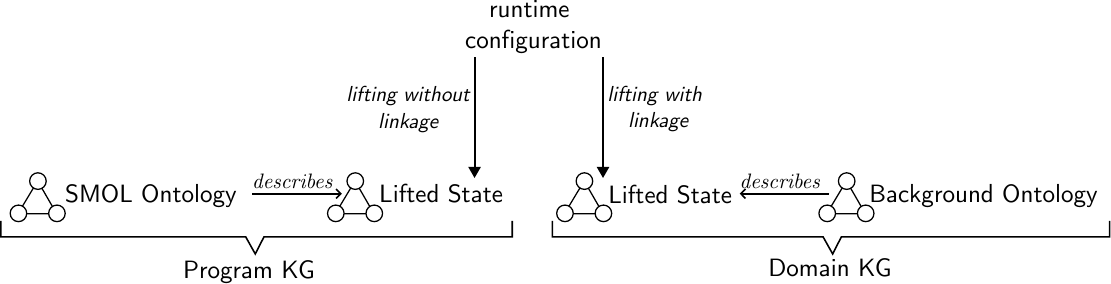}
    \caption{\newrev{High-level overview over the relation between lifting and different part of the knowledge graph.}}
    \label{fig:newfig}
\end{figure}

The constructs \smol{hidden}, \smol{domain} and \smol{links} are
specific to \SMOL. These constructs enable a certain control of the
semantic lifting.  We here introduce these constructs informally,
% describe them on a high level,
as their formal introduction requires the exact structure of the
semantic lifting (see \cref{sec:graph}).
% For now, it suffices to remark that
The lifted knowledge graph consists of two parts: the
\emph{program knowledge graph} that describes the state itself and the
\emph{domain knowledge graph} that describes context knowledge
provided by the user. \newrev{This is shown in \cref{fig:newfig}. The difference between the two parts is how the lifted configuration is described, either in terms of the \SMOL language (in the program knowledge graph) or in terms of the domain (in the domain knowledge graph).}\looseness=-1

Let us first introduce the optional field modifiers \smol{hidden} and
\smol{domain}.  The modifier \smol{hidden} excludes the field from
semantic lifting; i.e., the field will not have a counterpart in the
lifted knowledge graph.  The modifier \smol{domain} treats the field
not as part of the program knowledge graph, but as additional
information in the domain knowledge graph.  In addition, \SMOL
supports \emph{domain} $\mathsf{Linkage}$ as a programming
construct with the \smol{links} keyword, which connects the program
knowledge graph explicitly to the domain knowledge graph.
% Note that
Domain linkage  can also be used with object creation.

\begin{example}[A \SMOL Program]\label{ex:ddl1}
  We consider a program $\mathsf{Prog}_{\dll}$ modelling
  urban infrastructure, shown in \cref{fig:double1}.
  % , as the running example in this section. The program
  % $\mathsf{Prog}_{\dll}$ realizing it is given in
  % \cref{fig:double1}. As the very basis, $\mathsf{Prog}_{\dll}$
  The program defines classes \smol{Room}, \smol{Building} and
  \smol{Street} that include references to each other, as well as the
  size of a room and the accumulated size of a building. The
  \smol{main} statement block of $\mathsf{Prog}_{\dll}$ creates three
  rooms, which are in two buildings in a single street.

\begin{figure}[t]
\begin{smolframe}
class Room(Int size) end
class Building(List<Room> rooms, Int size, Street street) 
   Unit addRoom(Room room)
     this.rooms = Cons(room, this.rooms);
     this.size = this.size + room.size;
  end
end
class Street(List<Building> buildings, String name)
  Unit addBuilding(Building building)
     this.buildings = Cons(building, this.buildings);
     buildings.street = this;
  end
end
main
  Room r1 = new Room(10);
  Room r2 = new Room(20);
  Room r3 = new Room(30);
  Building b1 = new Building(null, 0, null); b1.addRoom(r1);
  Building b2 = new Building(null, 0, null); b2.addRoom(r2); b2.addRoom(r3);
  Street s1 = new Street(null, "Problemveien");
  s1.addBuilding(b1); s1.addBuilding(b2);
end
\end{smolframe}
\caption{\label{fig:double1}A \SMOL program $\mathsf{Prog}_{\dll}$ for
  urban infrastructure.}
\end{figure}

\end{example}

\subsection{Runtime Syntax and Semantics of SMOL without Reflection}
% So far, we have described the programmer's view on a \SMOL program;
% next we introduce its
% We now consider
We briefly introduce the \emph{runtime syntax} and \emph{semantics} of
\SMOL programs, the formalisms used to define program execution,
before semantic lifting is detailed in \cref{sec:graph} and semantic
reflection in \cref{sec:internal}.  Runtime syntax describes
\emph{runtime configurations}, i.e., terms representing the states of
a program at different steps of the program execution.  The runtime
semantics of \SMOL formalizes program execution by defining an
evaluation function on expressions and a transition system between
configurations.  This transition system itself is given in
\cref{app:language}, as the transitions without the concepts
of semantic reflection are standard.
% introduced in \cref{sec:internal} are completely standard.

Compared to the surface syntax given in \cref{sec:surface}, the
runtime configurations, which are specified by the runtime syntax,
% require some additional structure to represent all aspects of a
% program state during execution.  In particular, runtime
% configurations
describe the statements left to execute, the class table, the process
stack, the memory store of each object and the local memory store of
each process on the stack.

Lists \smol{List<C>} are a special construct in the syntax of \SMOL,
which enforces that lists cannot be nested and avoids full generics,
but allows lists to be treated as classes when it comes to typing and
runtime semantics: A list type \smol{List<C>} is treated as a class
without methods and two fields: \smol{C content} and \smol{List<C>
  next}.  In the sequel, we include lists whenever we refer to
classes.

We start with the formal definition of a \emph{class table}, which
represents static information about the fields and methods of the
classes defined in a program.

\begin{definition}[Class Table]\label{def:classtable}
  The \emph{class table} \classtable is a map from class names to sets
  of field declarations and method declarations.  The lists of
  fields and methods of the (instantiations of the) classes specified by
  \classtable can be accessed, for a class $\xsmol{C}$, via functions
  $\mathsf{fields}_\classtable (\xsmol{C})$ and
  $\mathsf{methods}_\classtable (\xsmol{C})$ respectively. Functions
  $\mathsf{vars}_\classtable (\xsmol{C}.\xsmol{m})$,
  $\mathsf{ret}_\classtable (\xsmol{C}.\xsmol{m})$ and
  $\mathsf{body}_\classtable (\xsmol{C}.\xsmol{m})$ are used to access
  the list of variables, return type and body of a method
  \smol{m} in $\xsmol{C}$, respectively.
\end{definition}

In addition to the static information about a program captured in its
class table, program states at runtime need to represent dynamically
created information, including the program's objects and process call
stack.  Let a \emph{domain element} (\emph{DE}) be either a literal
value (for a basic type) or an \emph{object reference} (for a class).
% which is a string.
The formal representation of a program state is a
\emph{runtime configuration}, defined as follows.

\begin{definition}[Runtime Configurations]\label{def:conf}
  A \emph{local store} $\sigma$ is a map from variables to DEs and an
  \emph{object store} $\rho$ is a map from fields to DEs.  Let
  $\mathsf{CT}$ be a class table and $\mathtt{X}$ range over object
  identifiers (the remaining terms
  are defined in \cref{def:syntax}).  \emph{Configurations} $\conf$,
  \emph{objects} $\mathsf{obs}$ and \emph{processes} $\mathsf{prs}$
  are defined by the following grammar:
$$\begin{array}{l@{\,::=\,}l@{\qquad}l@{\,::=\,}l@{\qquad}l@{\,::=\,}l}
    \conf & \classtable~\mathsf{obs}~\mathsf{prs} 
    &\mathsf{rs} & \mathsf{Stmt} \ssep \mathsf{Loc} \leftarrow
                   \xsmol{stack}\xsmol{;}~\mathsf{Stmt}
           & \mathsf{Cl} & \xsmol{C} \ssep \xsmol{List<C>} \\
    \mathsf{obs} & \many{(\mathsf{Cl},\rho)_\mathtt{X}}
          & \mathsf{prs} & \many{(\xsmol{m},\mathtt{X},\mathsf{rs},\sigma)}             
\end{array}$$
\end{definition}

Besides the class table \classtable, a runtime configuration $\conf$
contains objects ${\mathsf{obs}}$ and processes ${\mathsf{prs}}$.  An
object $(\mathsf{Cl},\rho)_\mathtt{X} $ has a unique identifier (i.e., name) $\mathtt{X}$
and contains its class \smol{C} (or a list \smol{List<C>}) and the
object's store $\rho$.  A process
$(\xsmol{m},\mathtt{X},\mathsf{rs},\sigma)$ contains the name $\mathtt{m}$ of the method it is executing, the
identifier $\mathtt{X}$ of the object in which it executes, a runtime
statement $\mathsf{rs}$ which remains to be executed and a local store
$\sigma$.
% In contrast to objects,
The list of processes in a configuration
may be seen as a stack corresponding to nested method calls.  To
capture the transfer of return values between method calls at runtime,
we use \emph{runtime statements} $\mathsf{rs}$, which extend the
statements $\mathsf{Stmt}$ with an additional statement
$\mathsf{Loc} \leftarrow \xsmol{stack}$ that identifies the location
$\mathsf{Loc}$ that is waiting for a return value from the next
process on the stack.  Each process on the stack, except for the top
process, starts with this runtime statement.

The connection between surface and runtime syntax is established when
execution starts: the program (in surface syntax) is translated into
an \emph{initial} runtime configuration, defined as
follows.\looseness=-1

\begin{definition}[Initial Configuration]\label{def:semantics}
  Let $\mathtt{E}$ be an object identifier. The initial configuration
  of a program $\mathsf{Prog}$ is
  $\classtable_{\mathsf{Prog}}~(\xsmol{Entry},\emptyset)_\mathtt{E}~(\xsmol{entry},\mathtt{E},\mathsf{Stmt},\emptyset)$,
  where $\classtable_{\mathsf{Prog}}$ is the class table for
  % from class names to their fields and methods as declared in
  $\mathsf{Prog}$, extended with an additional class \smol{Entry} that
  has a single, parameter-free method \smol{entry} with the statement $\mathsf{Stmt}$ of the main
  block as its body.\footnote{We assume, without loss of generality, that no program explicitly declares a class with name \smol{Entry}.}
\end{definition}

\noindent
In initial configurations, the empty sets denote the initially empty
stores.

\begin{example}[Initial Configuration]\label{ex:ddl2}
  \Cref{fig:ex:ddl2} shows the class table $\classtable_\dll$ for the
  program $\mathsf{Prog}_{\dll}$ from \cref{ex:ddl1}, where
  $\mathsf{Stmt}_\xsmol{m}$ is the method body of \smol{m} and
  $\mathsf{Stmt}_\dll$ the statement of the main block.  The initial
  configuration of $\mathsf{Prog}_{\dll}$ is then
  $\classtable_\dll~(\xsmol{Entry},\emptyset)_\mathtt{E}~(\xsmol{entry},\mathtt{E},\mathsf{Stmt}_\dll,\emptyset)$.
\end{example}

\begin{figure}[t]
\begin{align*}
\mathsf{fields}(\classtable_\dll,\xsmol{Room}) &= \{\xsmol{Int size}\}\\
\mathsf{fields}(\classtable_\dll,\xsmol{Building}) &= \{\xsmol{List<Room> rooms}, \xsmol{Int size},\xsmol{Street street}\} \\
\mathsf{fields}(\classtable_\dll,\xsmol{Street}) &= \{\xsmol{List<Building> buildings}, \xsmol{String name}\} \\
\mathsf{methods}(\classtable_\dll,\xsmol{Room}) &= \emptyset\\
\mathsf{methods}(\classtable_\dll,\xsmol{Building}) &= \{\xsmol{Unit addRoom(Room room)}~\mathsf{Stmt}_\xsmol{addRoom}~\xsmol{end}\}\\
\mathsf{methods}(\classtable_\dll,\xsmol{Street}) &= \\
&\hspace{-8mm}\{\xsmol{Unit addBuilding(Building building)}~\mathsf{Stmt}_\xsmol{addBuilding}~\xsmol{end}\}\\
\mathsf{vars}(\classtable_\dll,\xsmol{Building},\xsmol{addRoom}) &= \{\xsmol{Room room}\}\\
\mathsf{vars}(\classtable_\dll,\xsmol{Street},\xsmol{addBuilding}) &= \{\xsmol{Building building}\}\\
\mathsf{vars}(\classtable_\dll,\xsmol{Entry},\xsmol{entry}) &= \emptyset\\
\mathsf{body}(\classtable_\dll,\xsmol{Building},\xsmol{addRoom}) &= \mathsf{Stmt}_\xsmol{addRoom}\\
\mathsf{body}(\classtable_\dll,\xsmol{Street},\xsmol{addBuilding}) &= \mathsf{Stmt}_\xsmol{addBuidling}\\
\mathsf{body}(\classtable_\dll,\xsmol{Entry},\xsmol{entry}) &= \mathsf{Stmt}_\dll
\end{align*}
\caption{\label{fig:ex:ddl2}Class table for program
  $\mathsf{Prog}_{\dll}$ from \cref{ex:ddl1}.}
\end{figure}

At every point during execution, the state of a program can be
represented by means of runtime syntax. We return to the rules that
capture program execution in \cref{sec:internal}, when the full
language including semantic reflection has been introduced.

\section{Graph-Based State Semantics}\label{sec:graph}
Let us now consider the \SMOL ontology, which describes the OWL
classes and properties needed to describe the runtime configurations
of executing \SMOL programs, and then semantic lifting, a direct
mapping that translates such runtime configurations into a set of
triples.  Semantic lifting allows a runtime configuration to be
interpreted as a knowledge graph by serializing it in RDF, using the
vocabulary introduced below, and adding the triples needed for domain
linking.

\subsection{An Ontology for SMOL}\label{sec:smol-ontology}
The \SMOL-ontology\footnote{Note that \smolkb is a knowledge graph --- the term
  ontology here expresses that it contains general knowledge, which is
  applicable to a whole range of programs and configurations.}
\smolkb consists of a \emph{language layer} that describes elements
present in all programs, such as classes, fields and methods, and a
\emph{runtime layer} that describes the objects of a specific runtime
configuration. Statements, expressions and processes are not
lifted.\looseness=-1

The IRIs of all entities in our ontology share a common prefix, which
is added to the IRI by means of a function: $\smolpre{\,\cdot\,}$. For
readability, we use the prefix \texttt{smol:} in examples, or omit the
prefix altogether if it is clear from the context that we are
concerned with the language layer.

During semantic lifting, two additional prefixes are used to
distinguish knowledge about the program and about a specific runtime
state. Given a program, the function $\progpre{\,\cdot\,}$ (example prefix \texttt{prog:}) generates a
fresh IRI based on the current program --- two programs that share
some code can still be distinguished this way. The function
$\runpre{\,\cdot\,}$ (example prefix \texttt{run:}) generates fresh IRIs based on the current state.
% a concrete execution.
Two states during a run of the same program are thus lifted
into separate entities, connected by the entities of the common lifted
program.

% To express that a set of concepts is pairwise disjoint, we define
% \begin{align*}
% \mathtt{AllDisjoint}(A_1, \dots A_n) = \bigwedge_{\substack{1 \leq i,j \leq n\\A_i \neq A_j}}A_i \sqcap A_j = \bot
% \end{align*} %Use AllDisjointClasses

 \begin{definition}[\SMOL Ontology]\label{def:smolkb}
   % The \SMOL-ontology
   Let \smolkb be the union of the axioms in
   \cref{fig:app:axioms:oo,fig:app:axioms:run}.
 \end{definition}

 The \emph{language layer} consists of classes
 ($\smolpre{\mathtt{Class}}$), methods ($\smolpre{\mathtt{Method}}$)
 and fields ($\smolpre{\mathtt{Field}}$). Each class has a string as
 its name ($\smolpre{\mathtt{hasName}}$), and fields and methods are
 connected to the class in which they are declared. Fields and methods
 can be connected to more than one class, due to inheritance.  All
 these concepts are disjoint. Finally, we define the classes
 $\smolpre{\mathtt{Any}}$, $\smolpre{\mathtt{Unit}}$ and
 $\smolpre{\mathtt{List}}$.  \Cref{fig:app:axioms:oo} gives the axioms
 formally.  We use an object property $\smolpre{\mathtt{subClass}}$ to
 express inheritance between \SMOL classes, thus avoiding interactions
 between inheritance in OWL and object-oriented programming.

\begin{figure}[p]
\begin{owlframe}
Class: $\smolpre{\mathtt{Class}}$
Class: $\smolpre{\mathtt{Method}}$
Class: $\smolpre{\mathtt{Field}}$ 

DataProperty: $\smolpre{\mathtt{hasName}}$ 
  Domain: $\smolpre{\mathtt{Class}}$ and $\smolpre{\mathtt{Method}}$ and $\smolpre{\mathtt{Field}}$ Range: xsd:String
  Characteristics: Functional
  
ObjectProperty: $\smolpre{\mathtt{subClass}}$ Domain: $\smolpre{\mathtt{Class}}$ Range: $\smolpre{\mathtt{Class}}$ 
  Characteristics: Transitive

ObjectProperty: $\smolpre{\mathtt{hasMethod}}$ Domain: $\smolpre{\mathtt{Class}}$ Range: $\smolpre{\mathtt{Method}}$

ObjectProperty: $\smolpre{\mathtt{hasField}}$ Domain: $\smolpre{\mathtt{Class}}$ Range: $\smolpre{\mathtt{Field}}$ 

Individual: $\smolpre{\mathtt{Any}}$ Types: $\smolpre{\mathtt{Class}}$ 
Individual: $\smolpre{\mathtt{List}}$ Types: $\smolpre{\mathtt{Class}}$ Facts: $\smolpre{\mathtt{subClass}}$ $\smolpre{\mathtt{Any}}$, $\smolpre{\mathtt{hasName}}$ "List"
Individual: $\smolpre{\mathtt{Unit}}$ Types: $\smolpre{\mathtt{Class}}$ Facts: $\smolpre{\mathtt{subClass}}$ $\smolpre{\mathtt{Any}}$
  
AllDisjointClasses($\smolpre{\mathtt{Class}}$, $\smolpre{\mathtt{Method}}$, $\smolpre{\mathtt{Field}}$)
\end{owlframe}
\caption{\label{fig:app:axioms:oo}Axioms for the language layer of \smolkb.}
%  \end{figure}
% \begin{figure}[htbp]
\begin{owlframe}
Class: $\smolpre{\mathtt{Object}}$
Class: $\smolpre{\mathtt{MemoryEntry}}$
  
ObjectProperty: $\smolpre{\mathtt{implements}}$ 
  Domain: $\smolpre{\mathtt{Object}}$ Range: $\smolpre{\mathtt{Class}}$ Characteristics: Functional
  
ObjectProperty: $\smolpre{\mathtt{hasEntry}}$ 
  Domain: $\smolpre{\mathtt{Object}}$ Range: $\smolpre{\mathtt{MemoryEntry}}$  Characteristics: InverseFunctional
  
ObjectProperty: $\smolpre{\mathtt{hasPointer}}$ 
  Domain: $\smolpre{\mathtt{MemoryEntry}}$ Range: $\smolpre{\mathtt{Object}}$ Characteristics: Functional
  
DataProperty: $\smolpre{\mathtt{hasValue}}$ 
  Domain: $\smolpre{\mathtt{MemoryEntry}}$ Characteristics: Functional
  
ObjectProperty: $\smolpre{\mathtt{entryOf}}$ 
  Domain: $\smolpre{\mathtt{MemoryEntry}}$ Range: $\smolpre{\mathtt{Field}}$  Characteristics: Functional

ObjectProperty: $\smolpre{\mathtt{links}}$ 
  Domain: $\smolpre{\mathtt{Object}}$ Characteristics: Functional, InverseFunctional

Individual: $\smolpre{\mathtt{null}}$ Types: $\smolpre{\mathtt{Object}}$ Facts: $\smolpre{\mathtt{implements}}$ $\smolpre{\mathtt{Any}}$
\end{owlframe}
\caption{\label{fig:app:axioms:run}Axioms for the runtime layer of \smolkb.}
 \end{figure}

 The \emph{runtime layer} consists of objects
 ($\smolpre{\mathtt{Object}}$).  An important individual introduced
 here is $\smolpre{\mathtt{null}}$, which implements the type
 $\smolpre{\mathtt{Any}}$. Membership of \SMOL objects to \SMOL
 classes is expressed through $\smolpre{\mathtt{implements}}$.
 \Cref{fig:app:axioms:run} gives the axioms formally and introduces
 the $\smolpre{\mathtt{links}}$ relation used for domain linkage. Note
 that we define its domain, but not its range, which depends on a
 specific application.  The class $\smolpre{\mathtt{MemoryEntry}}$ and
 the properties $\smolpre{\mathtt{hasEntry}}$,
 $\smolpre{\mathtt{hasValue}}$, $\smolpre{\mathtt{hasPointer}}$, and
 $\smolpre{\mathtt{entryOf}}$ are used to model the memory of an
 object, where $\smolpre{\mathtt{hasPointer}}$ is used for fields of
 object type and $\smolpre{\mathtt{hasValue}}$ for fields of a basic
 data type.\looseness=-1

 \begin{example}[Semantically Lifted Memory]\label{ex:lifted-memory}
   Consider two objects \smol{o1} and \smol{o2} of a class \smol{C}, where a
   field \smol{f} of \smol{o1} points to \smol{o2}.  Semantic lifting will
   generate a graph where the prefixes mirror the origin of the
   different elements: the relations \smol{hasEntry}, \smol{entryOf}
   and \smol{hasPointer} are from the ontology (prefixed by
   $\smolpre{\cdot}$),
   % , namely the runtime layer, while
   the field \smol{f} is part of the program (prefixed by
   \progpre{\cdot}), while the objects \smol{o1} and \smol{o2} and
   % entry  are part of the runtime configuration objects and
   the memory entry \smol{e1} are from the runtime configuration
   (prefixed by \runpre{\cdot}).
\begin{rdfframe}[]
run:o1 smol:hasEntry run:e1.
run:e1 smol:entryOf prog:f.
run:e1 smol:hasPointer run:o2.
\end{rdfframe}

An alternative design would here be to use the  \emph{punning} feature of OWL 2~\cite{DBLP:journals/ws/GrauHMPPS08,OWL2NewFeatures} that allows using the same URI for an individual, a property, and a class.  These will be separate entities semantically, only sharing an identifier, and which one is meant can always be deduced from the context.  We can thus let
\texttt{prog:f} be both an OWL object and an OWL property. This
approach, which we also used in \cref{sec:overview}, allows the
following, more succinct lifting.

\begin{rdfframe}
    run:$o_1$ prog:f run:$o_2$.
\end{rdfframe}
\end{example}

There are no consequences for reasoning since the individual and the property
will be kept apart.  But this technique allows for more
intuitive queries without the need for a specific query interface for,
e.g., debugging. For these reasons, \SMOL supports both kinds of
semantic lifting;\footnote{The implementation has an option to switch
  between the two kinds of lifting.} we will continue to use punning
in examples.

\subsection{Domain Linkage}

Before detailing the technical aspects of semantic lifting itself, we
explain another novel aspect of \SMOL: the \smol{links} clause. The
purpose of the \smol{links} clause is to connect the program knowledge
graph to the domain knowledge graph, thereby associating domain
knowledge directly to the runtime state of \SMOL programs.  The
\smol{links} clause works similarly to \smol{case}
% or \smol{switch}
statements in imperative languages: it defines a sequence of guarded
expressions, where each guard is a Boolean expression. Additionally,
it contains an unguarded expression, which we represent by the guard
\smol{true}.  The semantics of the \smol{links} clause is that, during
lifting, link guards are evaluated in the listed order, and the link
expression of the first guard that evaluates to \smol{true} is used to
generate an additional axiom in the knowledge graph.\looseness=-1

To this aim, we introduce expressions with holes and substitution of terms for holes in these expressions. \newrev{Here, a hole is an expression that needs to be filled with a term before the expression can be evaluated (the terminology stems from Felleisen and Hieb's work on context-reduction semantics \cite{felleisen92tcs}).} Let $\bullet$ denote a hole in an expression $\mathsf{Expr}$ and $\mathsf{Expr}[X]$ the corresponding substitution of the hole by a term $X$. Thus, for example, $\bullet~\xsmol{!=}~5$ is an expression with a hole and the substitution $(\bullet~\xsmol{!=}~5)[2]$ reduces to the expression $2~\xsmol{!=}~5$.

\begin{definition}[Domain Linkage]
  Let $X$ be an object identifier, $\mathtt{e}$ a Boolean expression and
  $\conf$ a runtime configuration.
  \begin{itemize}
  \item A \emph{link expression} $\mathtt{le}$ is an axiom with a hole
    for its subject.
  \item Let $\mathtt{le}[X]$ denote the axiom obtained by filling the
    hole in $\mathtt{le}$ by $\runpre{X}$.
  \item A \emph{guarded} link expression is a pair
    $(\mathtt{e},\mathtt{le})$.
  \item A \emph{domain linkage} $\mathbb{L}$ is a sequence of guarded
    link expressions.
  \end{itemize}
  We denote by $\mathbb{L}[X,\conf]$ the axiom $\mathtt{le}[X]$
  obtained by filling the hole in the first guarded link expression
  $(\mathtt{e},\mathtt{le})$ in $\mathbb{L}$ such that $\mathtt{e}$
  evaluates to \smol{true} in the runtime configuration $\conf$.
\end{definition}

For a given program, all link expressions in rule $\mathsf{Linkage}$
in the grammar of \cref{def:syntax} will form a domain linkage, where
the last case $\xsmol{links}$ $\mathtt{le}$ is interpreted as the link
expression $(\mathtt{true},\mathtt{le})$.

\begin{example}\label{def:modelbridge}
  Consider a production by the rule $\mathsf{Linkage}$ in the grammar of
  \cref{def:syntax} of the form

  \medskip
  
\smol{links(e}$_1$\smol{) le}$_1$\smol{; ... links(e}$_{n-1}$\smol{) le}$_{n-1}$\smol{; links le}$_n$\smol{;}

\medskip

\noindent
This production gives rise to the domain linkage
\[ \big((\mathtt{e}_1, \mathtt{le}_1), \ldots,(\mathtt{e}_{n-1},
  \mathtt{le}_{n-1}),(\xsmol{true}, \mathtt{le}_n)\big)\]
\end{example}

Domain linkages can be associated with \SMOL classes as well as with
individual \SMOL objects (by annotating the \smol{new} constructor).
Given a class \texttt{C}, we denote by $\mathsf{links}(\mathtt{C})$
its associated domain linkage.  Similarly, given an object with
identifier \texttt{X}, we represent by $\mathsf{links}(\mathtt{X})$
its associated domain linkage, which is by default that of its class.
However, if an object has its own domain linkage, this linkage
overrides the domain linkage of its class.

Since \SMOL uses predicate object lists in Turtle syntax for link
expressions without a subject, the holes are left implicit and the
operation $\xsmol{le}[\mathtt{iri}]$ is realized by simply
concatenating $\mathtt{iri}$ as a prefix to the link expression
$\xsmol{le}$.
% Turtle syntax is used to capture link expressions without a subject,
% leaving the hole implicit.

\begin{example}[Domain Linkage]
  Consider the following variant of the \smol{Building} from
  \cref{ex:ddl1}, that links to the domain based on the accumulated
  size of all its rooms. 

  \begin{smolframe}
class Building(List<Room> rooms, Int size, Street street) 
  links (this.size >= 100) "a domain:BigHouse.";
  links "a domain:SmallHouse."; 
   Unit addRoom(Room room) ... end
end
\end{smolframe}

\noindent
The corresponding domain linkage for instances of class
\smol{Building} is defined by
\[\big(
  (\xsmol{this.size >= 100},\xsmol{"a domain:BigHouse."}),
  (\xsmol{true},\xsmol{"a domain:SmallHouse."})\big)\] Given an IRI
$\mathtt{domain:obj1}$ (which is not $\xsmol{run:obj1}$, see
\cref{sec:lifting}) and a runtime configuration $\conf$ in which
$\xsmol{obj1.size} = 20$,
$$\begin{array}{l@{\;}l}
    \mathbb{L}[\mathtt{domain:obj1},\conf] &=
    \bullet~\texttt{a~domain\!:\!SmallHouse.}[\texttt{domain\!:\!obj1}]\\
    & =
      \texttt{domain\!:\!obj1~a~domain\!:\!SmallHouse.}
  \end{array}$$
  since the first
  guard evaluates to \smol{false}, and the (implicit) second guard evaluates to
  \smol{true} in
% the runtime configuration
$\conf$.
\end{example}
% While we discuss the lifting formally in the next section,

We denote by $\mathbb{L}_\mathtt{X}$ the domain linkage for an object
$\mathtt{X}$.
\Cref{fig:ox} illustrates different semantic liftings of an object,
depending on its state, domain linkage and \smol{domain}
annotations.\footnote{The notation $\%f$ is analogous to non-answer
  variables in queries and replaced by the literal stored in the field
  at the moment of lifting. For simplicity, we omit this notation in
  our formalization (but it is implemented in the \SMOL interpreter).}

\begin{figure}[p]
\begin{minipage}{0.42\textwidth}
\begin{smolframe}[numbers=none]
// this.f == 0
class C(Int f) 


end
\end{smolframe}
\end{minipage}\hfill
\begin{minipage}{0.56\textwidth}
\begin{tikzpicture}\tikzstyle{every node}=[font=\small\tt]
\node [circle, inner sep=4pt, fill=blue, label=above:{run:obj}] (obj) {};
\node [circle, inner sep=4pt, fill=blue,label=below:{prog:C}] (c) [below=10mm of obj] {};
\node [circle, draw=none] (valf) [left=18mm of obj] {0};
\node [circle, inner sep=4pt, fill=blue, label=above:{domain:obj}] (dobj) [right=25mm of obj] {};
\path[->,shorten >=2pt,shorten <=2pt, very thick](obj) edge node[left]{a}(c);
\path[->,shorten >=2pt,shorten <=2pt, very thick](obj) edge node[below]{smol:links}(dobj);
\path[->,shorten >=2pt,shorten <=2pt, very thick](obj) edge node[below]{prog:f}(valf);
\end{tikzpicture}
\end{minipage}

\bigskip 
\bigskip 

\begin{minipage}{0.42\textwidth}
\begin{smolframe}[numbers=none]
// this.f == 0, this.g == 1
class C(Int f, domain Int g) 
  links "a domain:D"

end
\end{smolframe}
\end{minipage}\hfill
\begin{minipage}{0.56\textwidth}
\begin{tikzpicture}\tikzstyle{every node}=[font=\small\tt]
\node [circle, inner sep=4pt, fill=blue, label=above:{run:obj}] (obj) {};
\node [circle, inner sep=4pt, fill=blue,label=below:{prog:C}] (c) [below=10mm of obj] {};
\node [circle, draw=none] (valf) [left=18mm of obj] {0};
\node [circle, inner sep=4pt, fill=blue, label=above:{domain:obj}] (dobj) [right=25mm of obj] {};
\node [circle, draw=none] (valg) [right=20mm of dobj] {1};
\node [circle, inner sep=4pt, fill=blue,label=below:{domain:D}] (d) [below=10mm of dobj] {};
\path[->,shorten >=2pt,shorten <=2pt, very thick](obj) edge node[left]{a}(c);
\path[->,shorten >=2pt,shorten <=2pt, very thick](obj) edge node[below]{smol:links}(dobj);
\path[->,shorten >=2pt,shorten <=2pt, very thick](obj) edge node[below]{prog:f}(valf);
\path[->,shorten >=2pt,shorten <=2pt, very thick](dobj) edge node[below]{domain:g}(valg);
\path[->,shorten >=2pt,shorten <=2pt, very thick](dobj) edge node[left]{a}(d);
\end{tikzpicture}
\end{minipage}

\bigskip
\bigskip 

\begin{minipage}{0.42\textwidth}
\begin{smolframe}[numbers=none]
// this.f == 0,
class C(hidden Int f) 
  links "domain:g %f"

end
\end{smolframe}
\end{minipage}\hfill
\begin{minipage}{0.56\textwidth}
\begin{tikzpicture}\tikzstyle{every node}=[font=\small\tt]
\node [circle, inner sep=4pt, fill=blue, label=above:{run:ob}] (obj) {};
\node [circle, inner sep=4pt, fill=blue,label=below:{prog:C}] (c) [below=10mm of obj] {};
\node [circle, draw=none] (valf) [left=18mm of obj] {\phantom{0}};
\node [circle, inner sep=4pt, fill=blue, label=above:{domain:obj}] (dobj) [right=25mm of obj] {};
\node [circle, draw=none] (valg) [right=20mm of dobj] {0};
\node [circle, inner sep=4pt, fill=blue,label=below:{domain:D}] (d) [below=10mm of dobj] {};
\path[->,shorten >=2pt,shorten <=2pt, very thick](obj) edge node[left]{a}(c);
\path[->,shorten >=2pt,shorten <=2pt, very thick](obj) edge node[below]{smol:links}(dobj);
\path[->,shorten >=2pt,shorten <=2pt, very thick](dobj) edge node[below]{domain:g}(valg);
\path[->,shorten >=2pt,shorten <=2pt, very thick](dobj) edge node[left]{a}(d);
\end{tikzpicture}
\end{minipage}

\bigskip
\bigskip 

\begin{minipage}{0.42\textwidth}
\begin{smolframe}[numbers=none]
// this.f == 1,
class C(hidden Int f) 
  links(this.f > 0) "domain:g %f"
  links "a domain:D"
end
\end{smolframe}
\end{minipage}\hfill
\begin{minipage}{0.56\textwidth}
\begin{tikzpicture}\tikzstyle{every node}=[font=\small\tt]
\node [circle, inner sep=4pt, fill=blue, label=above:{run:obj}] (obj) {};
\node [circle, inner sep=4pt, fill=blue,label=below:{prog:C}] (c) [below=10mm of obj] {};
\node [circle, draw=none] (valf) [left=18mm of obj] {\phantom{0}};
\node [circle, inner sep=4pt, fill=blue, label=above:{domain:obj}] (dobj) [right=25mm of obj] {};
\node [circle, draw=none] (valg) [right=18mm of dobj] {1};
\path[->,shorten >=2pt,shorten <=2pt, very thick](obj) edge node[left]{a}(c);
\path[->,shorten >=2pt,shorten <=2pt, very thick](obj) edge node[below]{smol:links}(dobj);
\path[->,shorten >=2pt,shorten <=2pt, very thick](dobj) edge node[below]{domain:g}(valg);
\end{tikzpicture}
\end{minipage}

\bigskip
\bigskip 

\begin{minipage}{0.42\textwidth}
\begin{smolframe}[numbers=none]
// this.f == 0,
class C(hidden Int f) 
 links(this.f > 0) "domain:g %f"
 links "a domain:D"
end
\end{smolframe}
\end{minipage}\hfill
\begin{minipage}{0.56\textwidth}
\begin{tikzpicture}\tikzstyle{every node}=[font=\small\tt]
\node [circle, inner sep=4pt, fill=blue, label=above:{run:obj}] (obj) {};
\node [circle, inner sep=4pt, fill=blue,label=below:{prog:C}] (c) [below=10mm of obj] {};
\node [circle, draw=none] (valf) [left=18mm of obj] {\phantom{0}};
\node [circle, inner sep=4pt, fill=blue, label=above:{domain:obj}] (dobj) [right=25mm of obj] {};
\node [circle, inner sep=4pt, fill=blue,label=below:{domain:D}] (d) [below=10mm of dobj] {};
\path[->,shorten >=2pt,shorten <=2pt, very thick](obj) edge node[left]{a}(c);
\path[->,shorten >=2pt,shorten <=2pt, very thick](obj) edge node[below]{smol:links}(dobj);
\path[->,shorten >=2pt,shorten <=2pt,very thick](dobj) edge node[left]{a}(d);
\end{tikzpicture}
\end{minipage}
\bigskip
\caption{Dynamic variations of semantic lifting, depending on domain linkage
  and annotations. The \texttt{prog:f} and \texttt{domain:g} edges are
  short notation for the entities. \label{fig:ox}}
\end{figure}

%%% Local Variables: 
%%% mode: latex
%%% TeX-master: "../main"
%%% End: 

\subsection{Semantic Lifting}\label{sec:lifting}
We define a direct mapping to lift runtime configurations into
knowledge graphs, extending the \SMOL ontology $\smolkb$ of
\Cref{def:smolkb}.
The availability of domain knowledge then enables the runtime state of
the program to be accessed externally (i.e., via the knowledge graph),
in terms of the vocabulary and axioms of the domain,
% program with respect to domain knowledge,
formalized as an ontology.
% The programmer can thus \emph{directly} access vocabulary and axioms
% of the domain and perform computations with respect to them.
%
In order to connect the resulting program knowledge graph to a domain
knowledge graph, the domain knowledge graph needs to be a conservative
extension~\cite{DBLP:conf/ijcai/LutzWW07} of \smolkb, to ensure that
the domain knowledge cannot introduce inconsistencies in the lifted
runtime configurations (assuming that the domain knowledge graph is
consistent in the first place):

\begin{definition}[Domain Knowledge Graph]\label{def:userkb}
  Domain knowledge is given as a knowledge graph \userkb, and a
  function $\userpre{\,\cdot\,}$ that adds a prefix to IRIs, such that
  \userkb is a conservative extension of \smolkb.
\end{definition}

% The condition ensures that formalized domain knowledge and the lifted
% runtime configuration of the program will be able to interact in a
% meaningful way, without causing inconsistency (if \userkb is
% consistent in the first place).

% \todo{I think we need to explain the encoding of basic types and their values}

The direct mapping generates the remaining part of the knowledge
graph, namely the graph lifted from the current runtime
configuration. Recall from \cref{ex:lifted-memory} how the different
prefixes mirror the origin of the different lifted elements. The two
layers have mutually exclusive prefixes, added by functions
$\progpre{\,\cdot\,}$ and $\runpre{\,\cdot\,}$.
% In the
% example, we denote these prefixes with \texttt{prog:} and
% \texttt{run:}.

\begin{definition}[Direct Mapping]
  Given a runtime configuration
  $\mathtt{conf}=\classtable~\mathsf{ob}_1\dots\mathsf{ob}_n~\mathsf{prs}$,
  the \emph{direct mapping} $\mu$ is a function from runtime configurations
  to knowledge graphs defined as follows:
\[
\mu(\mathtt{conf}) = 
\bigcup_{\xsmol{C} \in \mathbf{dom}(\classtable)} \mu(\mathtt{C}) \cup \bigcup_{1 \leq \mathtt{X}\leq n}\big(\mu(\mathsf{ob}_\mathtt{X}) \cup \mathbb{L}_\mathtt{X}[\runpre{\mathtt{X}},\mathtt{conf}] \big)  \cup \mathit{close}\ .
\]
% where $\mathbb{L}_\mathtt{X}$ denotes the domain linkage for object
% $\mathtt{X}$.
The mapping $\mu(\mathtt{C})$ of a class $\mathtt{C}$ is defined in
\cref{fig:mapping:newct} and the mapping $\mu(\mathsf{ob}_\mathtt{X})$ of an object with identifier $\mathtt{X}$ in
\cref{fig:mapping:newobj}.  The axiom set $\mathit{close}$ is defined
as follows. Let $\mathtt{C}_1,\dots$ be all classes in \classtable,
$\mathtt{m}_1,\dots$ all methods in \classtable, $\mathtt{f}_1,\dots$
all fields, and $\mathtt{X}_1,\dots$ all object identifiers, then the following axioms are part of the ontology.
\begin{owlframe}
$\smolpre{\mathtt{Class}}\;\;$ EquivalentTo: { $\progpre{\mathtt{C}_1}$, ... }
$\smolpre{\mathtt{Method}}$ EquivalentTo: { $\progpre{\mathtt{m}_1}$, ... }
$\smolpre{\mathtt{Field}}\;\;$ EquivalentTo: { $\progpre{\mathtt{f}_1}$, ... }
$\smolpre{\mathtt{Object}}$ EquivalentTo: { $\runpre{\mathtt{X}_1}$, ... }
\end{owlframe}
\end{definition}

The axioms added by $\mathit{close}$ are used to explicitly state all
members of the classes in the \SMOL ontology.  Intuitively, these
axioms ensure that despite an open world assumption, one cannot infer
the existence of objects that must exist (according to the domain
knowledge graph),
% user-provided ontology or query),
unless they also exist in the given runtime configuration.

The lifting of classes in \cref{fig:mapping:newct} follows the
structure of the class table.  Inherited methods and fields are
considered different between super- and subclass, as they are
redeclared in the subclass. The lifting of objects in
\cref{fig:mapping:newobj} differentiates between fields holding values
of basic data types and fields pointing to other objects, because of
the distinction between data and object property in OWL. We assume
that for every basic data type $\mathtt{T}$ there is an xsd equivalent
that can be retrieved with $\mathsf{xsd}(\mathtt{T})$, and analogously
for literals.

We illustrate how the runtime configuration of a program can be
accessed in terms of a formalized domain vocabulary in the following
example.

\begin{example}[Querying Runtime States with Domain
  Knowledge]\label{ex:building2}
  Recall the class \smol{Building} from \cref{ex:ddl1}:
  \begin{smolframe}
class Building(List<Room> rooms, domain Int size, Street street) 
   Unit addRoom(Room room) ... end
end
\end{smolframe}
Now assume that a \emph{villa} is a building with a surface of more
than 300 square meters. This assumption can be expressed in the domain
knowledge graph as follows:
\begin{owlframe}
    domain:Villa EquivalentTo: domain:size some xsd:integer[<= 300]
\end{owlframe}
Although villas are not defined in the \SMOL program, objects in the
runtime configuration of the program that qualify as villas can
nevertheless be retrieved from the combined knowledge graph by the
following query:
% The query to do so is the following.
% Note that the query does not retrieve the villas, but the \SMOL
% objects that are linked to them.
\begin{sparqlframe}
    SELECT ?obj {?obj smol:links [a domain:Villa] }
\end{sparqlframe}

\noindent
The query returns the \SMOL objects that are linked to villas.
\end{example}

Recall from \cref{sec:smol-ontology} that fields may also be lifted as
properties (so-called punning). The additional axioms are given in
\cref{fig:mapping:punning}, again differentiating between data and
object properties.

\begin{figure}[t]
\begin{owlframe}
Individual: $\progpre{\mathtt{C}}$ 
  Facts: a $\smolpre{\mathtt{Class}}$, $\smolpre{\mathtt{hasName}}$ "C",
  [$\smolpre{\mathtt{subClass}}$ $\progpre{\mathtt{D}}$],     // if $\mathtt{C}$ extends $\mathtt{D}$
  [$\smolpre{\mathtt{subClass}}$ $\smolpre{\mathtt{Any}}$],    // otherwise
  $\smolpre{\mathtt{hasMethod}}$ $\progpre{\mathtt{m}_1}$,  $\dots$, $\smolpre{\mathtt{hasMethod}}$ $\progpre{\mathtt{m}_n}$,   
  $\smolpre{\mathtt{hasField}}$ $\progpre{\mathtt{f}_1}$,  $\dots$, $\smolpre{\mathtt{hasField}}$ $\progpre{\mathtt{f}_k}$   

Individual: $\progpre{\mathtt{m}_1}$ Facts: a $\smolpre{\mathtt{Method}}$, $\smolpre{\mathtt{hasName}}$ "$\mathtt{m}_1$"
$\dots$
Individual: $\progpre{\mathtt{f}_1}$ Facts: a $\smolpre{\mathtt{Field}}$, $\smolpre{\mathtt{hasName}}$ "$\mathtt{f}_1$"
$\dots$
\end{owlframe}
\caption{The lifting of a class $\mathtt{C}$ with methods
  $\mathtt{m}_1$,\dots,$\mathtt{m}_n$ and fields
  $\mathtt{f}_1$,\dots,$\mathtt{f}_k$.\label{fig:mapping:newct}}
\end{figure}

\begin{figure}[tbh]
\begin{owlframe}
Individual: $\runpre{\mathtt{X}}$
    Facts: a $\smolpre{\mathtt{Object}}$, $\smolpre{\mathtt{implements}}$ $\progpre{\mathtt{C}}$,
    $\smolpre{\mathtt{links}}$ $\userpre{\mathtt{X}}$,
    $\smolpre{\mathtt{hasEntry}}$ $\progpre{e_{\mathtt{f}_i}}$, 
    $\dots$  // for every class that is not annotated domain or hidden

Individual: $\userpre{\mathtt{X}}$
    $\smolpre{\mathtt{hasEntry}}$ $\progpre{e_{\mathtt{f}_i}}$, 
    $\dots$  // for every class that is annotated domain but not hidden

Individual: $\progpre{e_{\mathtt{f}_1}}$
    Facts: a $\smolpre{\mathtt{MemoryEntry}}$, 
    [$\smolpre{\mathtt{hasPointer}}$ $\progpre{\rho(\mathtt{f}_1)}$],     // if $\rho(\mathtt{f}_1)$ is an object identifier
    [$\smolpre{\mathtt{hasValue}}$ $\mathsf{xsd}( \rho(\mathtt{f}_1))$],      // otherwise
    $\smolpre{\mathtt{entryOf}}$ $\progpre{\mathtt{f}_1}$, $\dots$
\end{owlframe}
\caption{The lifting of an object $(\mathtt{C},\rho)_\mathtt{X}$,
  where class $\mathtt{C}$ has fields
  $\mathtt{f}_1$,\dots,$\mathtt{f}_k$.\label{fig:mapping:newobj}}
\end{figure}

\begin{figure}[tbh]
\begin{owlframe}
ObjectProperty: $\progpre{\mathtt{f}_i}$ Domain: $\progpre{\mathtt{C}}$ Range: $\progpre{\mathtt{D}}$ // if the field has type $\mathtt{D}$
DataProperty: $\progpre{\mathtt{f}_i}$ Domain: $\progpre{\mathtt{C}}$ Range: $\mathsf{xsd}(\mathtt{T})$  // if the  field has data type $\mathtt{T}$ 
\end{owlframe}
\begin{owlframe}
Individual: $\runpre{\mathtt{X}}$
    Facts: a $\smolpre{\mathtt{Object}}$, $\smolpre{\mathtt{implements}}$ $\progpre{\mathtt{C}}$,
    [$\smolpre{\mathtt{f}_1}$ $\progpre{\rho(\mathtt{f}_1)}$],          // if $\rho(\mathtt{f}_1)$ is an object identifier
    [$\smolpre{\mathtt{f}_1}$ $\mathsf{xsd}(\rho(\mathtt{f}_i))$],         // otherwise
    $\dots$
\end{owlframe}
    \caption{Alternative liftings of objects and classes if fields are modeled as both object properties and individuals using punning.}
    \label{fig:mapping:punning}
\end{figure}

%%% Local Variables: 
%%% mode: latex
%%% TeX-master: "../main"
%%% End: 

\section{Semantic Reflection}\label{sec:internal}

In this section, we explain semantic reflection by showing how a
running \SMOL program can interact directly with the knowledge graph
obtained by semantic lifting from its own runtime configuration.  We
have seen in \cref{sec:graph} how semantic lifting allows the
representation of a program state in the knowledge graph to be
controlled, using additional structures in the programming language to
connect the program knowledge graph to a domain knowledge graph.
Semantic lifting enables \emph{external} queries to investigate a
program state through a semantic, domain specific lens from the
outside, which can be used for debugging or to access computation
results after a program execution. In contrast, semantic reflection
enables the program itself to directly interact with the semantically
lifted runtime state and the domain knowledge during execution.

Semantic reflection is a powerful technique that enables semantic
state access from \emph{within} the program, which gives programs the
ability to explore their own runtime state through a domain-specific
lens, and to use this exploration to influence program behavior.
Technically, we combine the semantic lifting of configurations during
execution with language support to perform operations on the knowledge
graph.
% In particular, we use SPARQL queries and OWL classes to retrieve
% lists of individuals, and SHACL shapes for validation checks.

% Another technique we introduce in this section is \emph{computational}
% semantic state access (CSSA): We generate additional axioms for each
% object, which are computed by \emph{executing} marked methods.  By
% using methods to compute these axioms, we avoid to duplicate
% information solely for the sake of lifting and give the programmer a
% natural way to express extension of the semantic lifting.

\subsection{Language Support for Semantic Reflection}\label{ssec:reflect}
We consider language extensions that operate on knowledge graphs.
These extensions only extend the grammar of \SMOL (see
\cref{fig:syntax}) with additional RHS expressions.

To allow dynamic, but type-safe queries, we consider expressions
\smol{access} to ask for objects that satisfy a SPARQL query,
\smol{member} to ask for objects that are members of an OWL concept,
and \smol{validate} to check if the knowledge graph satisfies a
particular SHACL shape.  In these queries, we use a slightly extended
version of SPARQL by allowing, at every point in the grammar of SPARQL
where a variable may occur in a graph
pattern,\footnote{See~\url{https://www.w3.org/TR/sparql11-query/\#GraphPattern}}
% this extension supports
the use of a \emph{parameter variable} $\%i$. These parameter
variables are replaced by IRIs before the query is executed.  This is
analogous to SQL prepared statements in, e.g., Java
libraries.\footnote{See~\url{https://docs.oracle.com/javase/8/docs/api/java/sql/PreparedStatement.html}}
In particular, we require that graph patterns $\mathtt{P}$ in SPARQL SELECT
queries are such that (1) the set of parameter variables form an
interval $[\%1,\dots,\%n]$ for some $n\in\mathbb{N}$ and (2) there is
a query substitution mechanism, denoted $\mathtt{P}(v_1,\dots,v_n)$,
that syntactically replaces these variables by $n$ values
$v_1,\dots,v_n$.

\begin{definition}[Extended Surface Syntax]\label{def:esyntax}
The grammar in \cref{def:syntax} is extended as follows:

\noindent
\begin{align*}
\mathsf{RHS} ::=\;&  \ldots \ssep \xsmol{access}(\mathtt{sparql},\many{\xsmol{Expr}}) \ssep \xsmol{member}(\mathtt{owl}) \ssep \xsmol{validate}(\xsmol{shacl})     &&\text{RHS Expressions} 
\end{align*}
    
\noindent
where $\mathtt{sparql}$ ranges over the extended SPARQL SELECT queries with one
answer variable, $\mathtt{owl}$ over OWL concepts and $\mathtt{shacl}$
over SHACL shapes.
\end{definition}

The \smol{access} expression returns a list of objects, resulting from
the extended SPARQL query given as the first parameter.  These objects
\emph{must} exist prior to the execution of this expression.  The
other parameters to this expression are query parameters; the query
substitution mechanism reduces them to a standard SPARQL query.  The
\smol{member} expression returns the list of objects which are members
of the OWL concept in its parameter.  The \smol{validate} expression
applies the SHACL shape in its parameter and returns a Boolean,
depending on whether the knowledge graph of the semantic lifting
satisfies this shape or not.

Before we formalize semantic reflection, we illustrate its use by an
example to show how domain knowledge about the runtime configuration
of a program can be accessed directly in the program.

\begin{figure}[t]
\begin{smolframe}
class Inspector()
 Unit inspect(String streetName)
  List<Building> over = 
     access("SELECT ?x {?x smol:links [a domain:Villa]; 
                           prog:street [prog:name %1]. }", 
            this.streetName);
  for b in over do 
    this.inspectBuilding(b);
  end
 end
end
\end{smolframe}
\caption{\label{fig:mediate}Using domain knowledge to influence the
  execution in \SMOL.}
\end{figure}

\begin{example}[Programmer Access to the Domain Knowledge Graph]
  Assume that we need to perform an inspection of all villas in a
  given street, continuing from \cref{ex:ddl1,ex:building2}.  The code
  in \cref{fig:mediate} illustrates a possible implementation in \SMOL
  using semantic reflection.  It is left to the domain knowledge to
  define the meaning of \texttt{domain:Villa}, which can consequently
  be changed according to different scenarios outside of the \SMOL
  program.
  % Exchanging the background knowledge can be used to change
  % the specification of when a building is considered a villa.
  In the query, the variable $\%1$ is replaced by the literal passed
  as the second argument to the method.
\end{example}

% Semantic lifting makes both the class table and the runtime stack
% structure are available in the knowledge graph.  We can reason about
% these structures at runtime \emph{in terms of the formal (runtime)
% semantics and the domain}.  This does
The example shows how semantic lifting not only exposes the structure
of the implementing runtime environment but adds domain knowledge,
which one can access and use in the programs themselves by means of
semantic reflection.

We now discuss how semantic reflection can be realized operationally
by formalizing its behavior.  To this aim, we define a semantics for
the execution of \SMOL programs that captures both semantic lifting
and semantic reflection. We here only consider the essential aspects
of these operations; the full structural operational
semantics~\cite{Plotkin} is included in \cref{app:language}.

Let us consider a transition relation
$\mathsf{conf}_1 \rightarrow_\mathsf{er}^\userkb \mathsf{conf}_2$
defined by a set of transition rules, where $\mathsf{conf}_1$ and
$\mathsf{conf}_2$ are runtime configurations of \SMOL, $\mathsf{er}$
is a SPARQL entailment
regime\footnote{See~\url{https://www.w3.org/TR/sparql11-entailment/}}
and \userkb is some domain knowledge according to \cref{def:userkb}.
Let $\mathsf{conf}_1 \rightsquigarrow^\userkb_\mathsf{er} \mathsf{conf}_2$
 denote the
\emph{reachability} in the operational semantics, i.e., the transitive
closure of the transition relation, and by
$\mathsf{conf}_1 \Downarrow^{\userkb}_\mathsf{er} \mathsf{conf}_n$ the
\emph{maximal reflexive-transitive closure} of this relation.
% Finally, we say that a configuration $\mathsf{conf}$ is
% \emph{successfully terminated} if has the form
% $\classtable~\mathsf{obs}~\emptyset$, i.e., the runtime configuration
% has no processes.

We denote by $\mathsf{listify}(\mathit{des})$ an auxiliary function
that takes a set $\mathit{des}$ of domain elements and returns a \SMOL
list $\mathsf{obs}_\mathtt{Y}$ containing an object for each of the
domain elements,
% set of objects that wraps the input set into objects of \smol{List}
% class,
where the subscript $\mathtt{Y}$ denotes the object identifier of the
head of the list. The objects in the list are fresh in the usual sense
of object creation: they have new and unique identifiers. The function
$\mathsf{listify}$ fails (i.e., it is undefined) if the input list
mixes different literal types, or mixes literals with object
identifiers.

We first explain the behavior of \smol{validate}. In this case, the
next statement to be executed contains a \smol{validate} expression
with some shape \smol{shacl}. After the transition, the
\smol{validate} expression is replaced by a (side-effect-free)
assignment to the same location, but with the query-result
$\mathsf{res}$ as its RHS. This Boolean literal results from
evaluating the conformity of the lifted configuration together with
the SMOL ontology and the domain knowledge graph. Otherwise, the
objects and processes of the configuration are not changed.

\begin{definition}[Semantics of \textbf{validate}]\label{def:validate}
  Let $\userkb$ be a knowledge graph, $\mathsf{er}$ an entailment
  regime and $\mathsf{conf}$ a configuration of the form
    \[\mathsf{conf} = \mathsf{CT}~\mathsf{obs}~\mathsf{prs}, (\xsmol{m},\,\mathtt{X},\,\mathsf{Loc}~\xsmol{= validate(shacl);}~\mathsf{Stmt},\,\sigma)\]
    where the next statement to execute in the top process contains a
    \smol{validate} expression.  Let $\mathsf{res}$ be the result of
    checking the lifted configuration against the SHACL shape(s)
    \smol{shacl}:
    \[\mathsf{res} = \mathsf{Sha}\big(\smolkb \cup \userkb \cup
      \mu(\mathsf{conf}), \xsmol{shacl}\big)\ .\]
    Recall that $\mathsf{res}$ is a Boolean value in \SMOL.  The
    transition from $\conf$ is defined as
    \[\mathsf{conf} \rightarrow_\mathsf{er}^\userkb
      \mathsf{CT}~\mathsf{obs}~\mathsf{prs},
      (\xsmol{m},\,\mathtt{X},\,\mathsf{Loc}~\xsmol{=}~\mathsf{res}\xsmol{;}~\mathsf{Stmt},\,\sigma)\ .\]
\end{definition}

The behavior of \smol{member} is similar to \smol{validate} in the
sense that execution returns an object identifier $\mathtt{Y}$, which
is the head of a list of \SMOL objects.
% Rule \rulename{member} is mostly analogous, but matches on
% \smol{member} and assign an object identifier $\mathtt{Y}$.  This
% identifier is the head of the list of \SMOL objects, generated from
% the list of object identifiers retrieved by the membership query
% \smol{owl}.  The objects needed to create the list are also added to
% the configuration ($\mathsf{obs}_\mathtt{Y}$).

\begin{definition}[Semantics of \textbf{member}]\label{def:member}
  Let $\userkb$ be a knowledge graph, $\mathsf{er}$ an entailment
  regime and $\mathsf{conf}$ a configuration of the form
    \[\mathsf{conf} = \mathsf{CT}~\mathsf{obs}~\mathsf{prs}, (\xsmol{m},\,\mathtt{X},\,\mathsf{Loc}~\xsmol{= member(owl);}~\mathsf{Stmt},\,\sigma)\]
    where the next statement to execute in the top process contains a
    \smol{member} expression.  Let $\mathsf{res}$ be the result of
    performing the membership query \smol{owl} on the lifted
    configuration:
    \[\mathsf{res} = \mathsf{Mem}\big(\smolkb \cup \userkb \cup
      \mu(\mathsf{conf}), \xsmol{owl}\big)\ ,\] which is a set of
    IRIs.  Let $\mathsf{obs}_\mathtt{Y} =
      \mathsf{listify}\big(\mathsf{res}\big)$ be the representation of
      this set as a \SMOL list.
    %   be as follows:
    %   \[\mathtt{Y},\mathsf{obs}_Y = \mathsf{listify}\big(\mathsf{res}\big)\ .\]
      If $\mathsf{obs}_\mathtt{Y}$ is defined, then the transition
      from $\conf$ is defined as
    \[\mathsf{conf} \rightarrow_\mathsf{er}^\userkb
      \mathsf{CT}~\mathsf{obs}~\mathsf{obs}_\mathtt{Y}~\mathsf{prs},
      (\xsmol{m},\,\mathtt{X},\,\mathsf{Loc}~\xsmol{= Y},\,\sigma)\ .\]
\noindent
If $\mathsf{obs}_\mathtt{Y}$ is not defined (see above), then
the behavior of \smol{member} is also not defined.
\end{definition}

% The rule \rulename{access} functions as follows.
We finally explain the behavior of \smol{access}.  In this case, the
next statement to be executed contains an \smol{access} expression with some
query \smol{sparql} and expressions
$\mathsf{Expr}_1,\dots,\mathsf{Expr}_n$.  The expressions
$\mathsf{Expr}_1,\dots,\mathsf{Expr}_n$ are evaluated in the current
state and their results substituted for the parameter variables in the
query. The resulting query is evaluated using the semantically lifted
configuration, producing a set of domain elements $\mathit{des}$ from
which a \SMOL list with objects $\mathsf{obs}_\mathtt{Y}$ and head
$\mathtt{Y}$ is constructed.  These objects are then added to the
configuration and the statement reduced to an assignment of
$\mathtt{Y}$ into the target location.

\begin{definition}[Semantics of \textbf{access}]\label{def:access}
  Let $\userkb$ be a knowledge graph, $\mathsf{er}$ an entailment
  regime and $\mathsf{conf}$ a configuration of the form,
    \[\mathsf{conf} = \mathsf{CT}~\mathsf{obs}~\mathsf{prs},
      (\xsmol{m},\,\mathtt{X},\,\mathsf{Loc}~\xsmol{=
        access(sparql},\mathsf{Expr}_1,\dots,\mathsf{Expr}_n\xsmol{);}~\mathsf{Stmt},\,\sigma)\  ,
\]
where the next statement to execute in the top process contains an
\smol{access} expression.  Let
$\sem{\mathsf{Expr}}_\mathtt{X}^{\sigma,\mathsf{obs}}$ denote the
result of evaluating an expression $\mathsf{Expr}$ and $\mathsf{res}$
the result of performing the SPARQL query \smol{sparql} on the
semantically lifted configuration, with all parameter variables
replaced by the literals resulting from the corresponding expressions:
\[\mathsf{res} =\mathsf{Ans}_\mathsf{er}\big(\smolkb \cup \userkb \cup \mathcal{K}_\mathsf{conf},   \xsmol{sparql}[\sem{\mathsf{Expr}_1}_\mathtt{X}^{\sigma,\mathsf{obs}}\dots
  \sem{\mathsf{Expr}_n}_\mathtt{X}^{\sigma,\mathsf{obs}}]\big)\ ,
\]
which is a set of IRIs.\footnote{\newrev{We remind that we only allow SPARQL queries with a single answer variable here, see~\cref{def:esyntax}.}}  Let
$\mathsf{obs}_\mathtt{Y} = \mathsf{listify}\big(\mathsf{res}\big) $
be the representation of this set as a \SMOL list.  If
$\mathsf{obs}_\mathtt{Y}$ is defined, then the transition from
$\conf$ is defined as
    \[
    \mathsf{conf} \rightarrow_\mathsf{er}^\userkb
    \mathsf{CT}~\mathsf{obs}~\mathsf{obs}_\mathtt{Y}~\mathsf{prs},
    (\xsmol{m},\,\mathtt{X},\,\mathsf{Loc}~\xsmol{= Y},\,\sigma)\ .
\]
If $\mathsf{obs}_\mathtt{Y}$ is not defined (see above), then the
semantics of \smol{access} is not defined.
\end{definition}

%%% Local Variables: 
%%% mode: latex
%%% TeX-master: "../main"
%%% End: 

\subsection{Eliminating Runtime Failures for Semantically Reflected Programs}\label{sec:type}
% Having established semantic lifting and semantic reflection, we turn our attention towards the tools necessary to support the programmer in using them.  

The clash of two different class models\footnote{Remark that the
  \emph{impedance mismatch} (or semantic gap) between object-oriented
  and ontology/database class models is a general
  phenomenon~\cite{DBLP:journals/ijseke/BasetS18}, and not specific to
  semantic reflection.} and the interaction between the programming
and semantic layers may be challenging for the programmer. We here
consider static techniques to ensure that interaction between these
layers happens correctly. At the level of syntax, we can enforce some
constraints on statements with \smol{access}, \smol{member} and
\smol{validate} expressions to avoid programming errors; for example,
the language extensions for semantic reflection should contain
syntactically correct SPARQL queries, OWL concepts and SHACL shapes.  A
particular concern is that the answers to queries over a knowledge
graph are generally (untyped) multisets of IRIs, whereas \SMOL
programs are otherwise typed.  In this section, we consider the
following failures that are specific to semantically reflected
programs.
% the following three kinds of failure may still occur:
\begin{itemize}
\item \textbf{Representation Failure:} When executing an \smol{access}
  expression, the query may return a set of IRIs that cannot be
  represented as values at runtime.  For example, the query
\[ \mathtt{SELECT~?x}~\{\mathtt{prog:obj1~?x~1}\}\]
returns a set of predicates, which cannot be translated to \SMOL objects.

\item \textbf{Location Failure:} While representation failures
  manifest when the semantic reflection is performed, a
  failure to respect the type of the target location may lead to later
  runtime errors.
  % Consider a query that returns a set of IRIs (or literals) that can
  % be represented, but is not uniform or does not respect the
  % expected type of the location in which the results are loaded
  % into.
  For example, a program could execute a query that returns string
  literals and then perform numerical operations on the elements of
  the result list.  This will cause a delayed error, once the first
  string in the list is accessed and used for an operation expecting
  an integer.

\begin{smolframe}
class C (String str) end
...
C c = new C("a");
List<Int> res = access("SELECT ?x {?o prog:str ?x}");
print(res.content + 1); //runtime error
\end{smolframe}

Assuming that the rest of the program is correct,
% well-typed,
the problem is that the query loads the results into a location of
type \smol{List<Int>}. \newrev{If we assume that types are not represented at runtime, then this error occurs in line 5 when the data is processed, as the addition operator fails. Storing the results in line 4 will succeed.}

\item \textbf{Inconsistency:} If a semantically lifted state results
  in an inconsistent knowledge base, then query answering is not
  defined. As we lift the type of fields, the following program
  results in a query access over an inconsistent knowledge base.  The
  knowledge graph contains the axiom
  $\progpre{\mathtt{D}} \sqcap \progpre{\mathtt{C}} \sqsubseteq \bot$,
  which stems from the class hierarchy.  From the class table, the
  following axiom for the field \smol{D.c} is generated:
  $\top \sqsubseteq
  \forall\progpre{\xsmol{D\_c}}.\progpre{\mathtt{C}}$.  and the sole
  created object is lifted as an individual $i$ with
  $\progpre{\mathtt{D}}(i)$.  These three axioms form an inconsistent
  knowledge graph.

\begin{smolframe}
class C() end
class D(C c) end

main
    D d = new D(null);
    d.c = d; 
    List<C> l = access(...);
end
\end{smolframe}

This is a different failure than location failure: while location
failure leads to an error in the runtime semantics of the program,
inconsistency leads to an error in the query answering. For example,
in the above, the location failure does not lead to a runtime error
because the field is never \emph{read}.
\end{itemize}

\subsubsection{A Type System for Semantic Reflection}\label{sec:typesystem}
A static type system \emph{``makes sure a program does not go
  wrong''}~\cite{DBLP:books/daglib/0005958}, for some notion of
\emph{``going wrong''}.  We present a type system for \SMOL that
eliminates errors related to representation failure, location failure
and inconsistency.  Specifically, the type system for \SMOL ensures
that semantic queries to the semantically lifted program return a list
of IRIs and literals that can be represented by a value of the type of
the target location (or a sub-type thereof) in the program. This is
sufficient to guarantee consistency of the knowledge graph.  The
presented type system focuses on the program knowledge
graph. Consequently, it does not cover domain linkage, which would
require a deeper analysis depending on guard expressions
% , akin to dependent types~\cite{Pierce2}; this goes beyond the basic
% principles of semantic reflection addressed here,
--- such an analysis has been left for future work.

We exploit the fact that each type in \SMOL has a \emph{direct} correspondence
to a class in the knowledge graph, and tackle the three different kinds
of semantic reflection as follows:
\begin{itemize}
\item \textbf{SHACL:} The \smol{validate} expression should always
  return a Boolean if the expression's SHACL shape is syntactically
  well-formed. \newrev{This can encode other reasoning or validation tasks as well~\cite{DBLP:conf/lpnmr/BogaertsJB22,DBLP:journals/lmcs/BogaertsJB24}.}
\item \textbf{OWL:} Given a statement \smol{l = member(C);}, where
  \smol{l} has type \smol{List<D>}, type checking is performed by
  concept subsumption: the parameter concept \smol{C} must be a
  subsumed by $\mathtt{prog:D}$. This can be checked directly using a
  reasoner.
\item \textbf{SPARQL:} The most involved form of semantic reflection
  is querying with an \smol{access} expression.  Given a statement
  \smol{l = access(Q);}, where \smol{l} has type \smol{List<D>}, type
  checking amounts to query containment under an entailment regime:
  Obviously, loading all elements of $\mathtt{prog:D}$ would be safe
  (i.e., representable in the runtime and respecting the type of the
  target location).  This is equivalent to the query
  $Q_\mathtt{D} = \mathtt{SELECT~?x~\{?x~a~prog:D\}}$.  Consequently,
  we check whether the query returns a subset of $Q_\mathtt{D}$, using
  the entailment regime for reasoning. \newrev{We will introduce this kind of type checking in the following section.}
\end{itemize}

These checks will be performed at compile time, i.e., without a
specific state --- it suffices to show that every reachable
configuration is consistent with the \SMOL-ontology $\smolkb$.  If
domain knowledge is used, it must be in the form of a conservative
extension of this ontology~\cite{DBLP:conf/ijcai/LutzWW07}.

We here focus on the handling of \smol{access}.
% , which is handled by the rule \rulename{T-access}.
We do not detail the general setup and soundness proofs here; besides
the cases for the semantic reflection rules, these are standard.  For
the full formal treatment of the type system, see \cref{app:type}.  We
first introduce the typing environment that we use to keep track of
field, variable and parameter types.  Together with the class table,
the typing environment provides context for typing judgments.

\begin{definition}[Typing Environment]
  A \emph{typing environment} $\Gamma$ is a partial function, mapping
  \emph{locations} (fields, variables, method parameters) to types.
  The empty typing environment is denoted $\emptyset$.\looseness=-1

  Given a program with a statement $\mathsf{Stmt}$, we use
  $\Gamma_\mathsf{Stmt}$ to denote the typing environment that maps (a) all
  fields of the class containing $\mathsf{Stmt}$ to their declared
  types, (b) all method parameters of the method containing
  $\mathsf{Stmt}$ to their declared types, (c) all variables declared
  before $\mathsf{Stmt}$ in this method to their declared types, and
  (d) is undefined for all other locations.
\end{definition}

Let $\Gamma, \classtable \vdash^\userkb_\mathsf{er} \mathsf{Stmt}$
denote that a statement $\mathsf{Stmt}$ is well-typed in the context
of an environment $\Gamma$, class table $\classtable$, domain
knowledge graph $\userkb$ and entailment regime
$\mathsf{er}$. Similarly, let
$\Gamma, \classtable \vdash^\userkb_\mathsf{er} \mathsf{Expr} :
\mathsf{Type}$ denote that an expression $\mathsf{Expr}$ has type
$\mathsf{Type}$ in the given context.\looseness=-1

\begin{definition}[Typing Reflection]
  Given a typing environment $\Gamma$, a class table \classtable, a
  domain knowledge graph $\userkb$ and an entailment regime
  $\mathsf{er}$, the typing judgment
  \[\Gamma, \classtable \vdash^\userkb_\mathsf{er} \xsmol{List<c> v =
      access("SELECT
      ?obj}~\{\xsmol{P}\}\xsmol{"},\mathsf{Expr}_1\xsmol{,}\dots\xsmol{,}\mathsf{Expr}_n\xsmol{)}\]
  holds if the following conditions can be satisfied:
\begin{itemize}
\item The query parameters can be assigned types:
  $\Gamma, \classtable \vdash \mathsf{Expr}_i : \mathsf{Type}_i$ for
  $0<i \leq n$
\item Query containment holds for the given types of the query
  parameter variables:
  \[\begin{array}{l}
      \mathtt{SELECT~?obj~\{P.~?v_{\textrm{1}}~a~\mu(\mathsf{Type}_{\textrm{1}}).~\dots.~?v_{\textrm{n}}~a~\mu(\mathsf{Type}_{\textrm{n}})\}}\\
      \qquad\subseteq_{\mathsf{er}}^{\smolkb \cup \userkb}
      \mathtt{SELECT~?obj}~\{\mathtt{?obj~a}~\progpre{\xsmol{c}}\}\ .
    \end{array}\]
\end{itemize}
\end{definition}

% It formalizes the above intuition that the query can only retrieve
% objects of the class of the target variable. The added triples
% $?v_1~a~\mu(\mathsf{Type}_1)$ add information about the parameter
% variables: For the analysis, they are regarded as answer variables
% for which we can approximate the binding. The bound values will be
% members of the type of $\mathsf{Expr}_i$, which is exactly the
% information expressed by the added triples.

Here, the first condition assigns types to all expressions that are
used as parameters to the query.
% In its first condition, it type-checks first all expressions that are
% used as parameters.
The derived types are then used to approximate the schema variable
associated with this parameter in the second premise.  To this aim, an
\emph{approximating triple} of the form $?\mathtt{v}_i~\mathtt{a}~\mu(\mathsf{Type}_i)$
is generated for each parameter variable $v_i$ typed with $\mathsf{Type}_i$.
The approximating triple expresses that the value used to instantiate
the query must be of the given type.  The second condition adds the
approximating triples and checks (under the chosen entailment regime
and background knowledge) whether the resulting query is contained in
the query that retrieves all values of the type of the targeted
location.  In this case, every possible result of the query is a
member of the type of the targeted location; i.e., the query only
retrieves values of the correct type.  We write
$\vdash^{\userkb}_\mathsf{er}\mathsf{Prgm}$ to express
that all statements within a program $\mathsf{Prgm} $ are well-typed.

% \begin{figure}
% \TTTINFER{T-access}{
% \forall i \leq n .~\Gamma, \classtable \vdash \mathsf{Expr}_i : \mathsf{Type}_i
% }{
%     \mathtt{SELECT~?obj~\{P.~?v_1~a~\mu(\mathsf{Type}_1).~\dots.~?v_n~a~\mu(\mathsf{Type}_n)\}}
% }{
%     \hspace{35mm}
%    \subseteq_{\mathsf{er}}^{\smolkb \cup \userkb} \mathtt{SELECT~?obj}\{\mathtt{?obj~a}~\progpre{\xsmol{c}}\}
% }{
% \Gamma, \classtable \vdash^\userkb_\mathsf{er}  \xsmol{List<c> v := access("SELECT ?obj}\{\xsmol{P}\}\xsmol{"},\mathsf{Expr}_1\xsmol{,}\dots\xsmol{,}\mathsf{Expr}_n\xsmol{)} 
% :\mathsf{Unit}
% }

% % \TTINFER{T-approx}{
% % \forall i \leq n .~\Gamma, \classtable \vdash \mathsf{Expr}_i : \mathsf{Type}_i
% % }{
% %     \mathtt{SELECT~?obj}~\{\mathtt{P}[\%i\mapsto\mathtt{v_i}]\mathtt{.~?v_1~a~\mu(\mathsf{Type}_1).~\dots.~?v_n~a~\mu(\mathsf{Type}_n)}\} \sqsubseteq^{\smolkb \cup \userkb} \mathsf{D}
% %     \qquad \mathsf{D} \sqsubseteq^\mathcal{K} \progpre{\xsmol{C}}
% % }{
% % \Gamma, \classtable \vdash^\userkb_\mathsf{er}  \xsmol{List<c>:= access("SELECT ?obj}\{\xsmol{P}\}\xsmol{"}
% % ,\mathsf{Expr}_1,\dots,\mathsf{Expr}_n\xsmol{)} \; : \; \xsmol{Unit}
% % }
% \caption{Typing rules for \smol{access}}
% \label{fig:type:main}
% \end{figure}

% It matches on some environment $\Gamma$, a classtable \classtable and some \smol{access} statement targeting a variable declaration.
% If all premises can be fulfilled, then the statement is considered well-typed.

The case for \smol{member} is similar, but operates on OWL classes
instead of SPARQL queries.  The case for \smol{validate} is
straightforward; since SHACL queries take no input and return only
true or false, the type of the expression must be a Boolean.
% \todo{Should we include these cases in an apapendix?}

\subsubsection{Optimizing Query Containment}
The type system outlined in \cref{sec:typesystem} is sound but not
complete: it provides a fine-grained, but only sufficient condition
for type safety of statements with \smol{access} expressions, while
necessity cannot be guaranteed since there are ABoxes that do not
correspond to configurations.

Moreover, applicability of the type system is limited in practice by
the fact that, as far as we are aware, there are no algorithms and
tools for checking query containment over
$\mathcal{SROIQ}(\mathbf{D})$ TBoxes under non-trivial entailment
regimes. To overcome this issue, we consider a stronger sufficient
condition, which is based on \emph{concept subsumption} rather than
query containment under entailment regimes. This approach is
advantageous since concept subsumption is a main reasoning task for
description logics, and there are practical systems (e.g.,
HermiT~\cite{GlimmHMSW14}) implementing efficient concept subsumption
for the description logic underying OWL2 DL (i.e., $\mathcal{SROIQ}(\mathbf{D})$)
and its fragments.

%Let us consider the case of unary conjunctive queries (CQ), i.e., queries with the patterns of the form $\mathtt{P}_1.~\dots.~\mathtt{P}_n$, where each $\mathtt{P}_i$ is a triple. 
A unary query ${\mathtt{Q}}$ is \emph{subsumed} by a concept $C$ with
respect to a knowledge graph (or TBox) $\mathcal K$, written
$\mathtt{Q} \sqsubseteq^\mathcal{K} C$, if
$s^\mathcal I \in C^\mathcal I$ for every certain answer $s$
to $\mathtt{Q}$ over $\mathcal K$ and each model $\mathcal I$ of
$\mathcal{K}$.
% For brevity, the theorem is stated only for assigned locations of
% class (or list) types; the case of datatypes is analogous.
In practice, we can syntactically construct a concept $D$ from the
query using a technique that guarantees the first subsumption
($\mathtt{Q} \sqsubseteq^\mathcal{K} D$ with respect to $\mathcal{K}$)
to hold, and then check the second subsumption
($D \sqsubseteq^\mathcal{K} C$) by a description logic reasoner. A
more specific (with respect to $\sqsubseteq^\mathcal{K}$) concept $D$
ensures a more fine-grained sufficient condition for type
safety. However, unless $\xsmol{Q}$ is equivalent (with respect to
$\sqsubseteq^\mathcal K$) to a concept, there is no most specific
concept $D$. Thus, there may be many techniques for constructing $D$
from the query. 

\newrev{To be more concrete, we can consider a variation of this technique, applicable when the query $\mathtt{Q}$ is a conjunctive query (i.e., when the body consists of only basic graph patterns not mentioning any IRIs with special semantics) and the 
entailment regime is OWL2 DL (i.e., $\mathcal{SROIQ}(\mathbf{D})$). In this case, the containment $\mathtt{Q} \sqsubseteq^\mathcal{K} C$ boils down to, essentially, ABox reasoning over the body of the query seen as the ABox; in particular, we can just check, using a standard OWL2 reasoner, whether $I_{?obj}$ belongs to $C$ over $\mathcal{K}$ extended with ABox part $\mathcal A_\mathtt{Q}$ that is obtained from the graph patterns of $\mathtt{Q}$ by replacing each variable $?x$ (including ${?obj}$) by a fresh IRI $I_{?x}$ (and translating them to an ABox in the standard way). Note that, formally, we do not have $\mathtt{Q} \sqsubseteq^\mathcal{K} D$ for the constructed ABox as $D$ any more (even if we can translate the ABox to a concept using the standard internalization method), but the approach is still sound because the IRIs are fresh.}

% A reasonable choice for constructing the concept $D$ from a query
% $\xsmol{Q}$, % which we make in our implementation, 
% is to take a \emph{repetition-free unraveling} of $\xsmol{Q}$; for
% datatype-free queries, this is concept equivalent, with respect to
% $\sqsubseteq^\emptyset$, to a maximal constant-free query $\xsmol{Q}'$
% that is tree-shaped on the one hand, and has a homomorphism to
% $\xsmol{Q}$ that does not identify atoms on the other (cf.~the
% \emph{rolling up} of Horrocks and
% Tessaris~\cite{DBLP:conf/aaai/HorrocksT00}).  Here, a query is
% \emph{tree-shaped} if the multigraph with the query's variables as
% nodes and its atoms $R(x, y)$ as edges $\{x, y\}$, forms a tree.  For
% example, for
% ${\xsmol{Q} \equiv \exists y, z.\, R(x, y) \land P(y, z) \land S(z,
%   x)}$, a possible unraveling is
% $\exists R. \exists P \sqcap \exists S^-$, with justifying query
% $\xsmol{Q}' \equiv \exists y, z, x'.\, R(x, y) \land P(y, z) \land
% S(z, x')$. This unraveling is not unique (e.g.,
% $\exists R \sqcap \exists S^-.\exists P^-$ is another possibility) but
% it always exists, and we can take any candidate to construct $C$. The
% generalization of this technique to queries with datatypes is
% straightforward. In fact,
% % it is straightforward to see that
% if
% $\xsmol{Q}$ is tree-shaped, which is common in practice, then this
% % such
% unraveling is always the most specific (for any TBox).

We can now state our type safety theorem that expresses two
properties:
\begin{enumerate}
\item Program execution does not get stuck when using reflection, in
  particular not due to failure to translate the results of a query
  into internal data structures. (We here ignore reasons for failure
  that correspond to exceptions, such as null pointer access and
  division by zero.)
\item Every configuration reachable from a well-typed program
lifts to a consistent knowledge graph.
\end{enumerate}
Thus, even without reflection, the type system can give guarantees to
programs that access a knowledge graph.

\begin{restatable}[Type Safety]{thm}{safety}\label{thm:safety}
%\begin{theorem}[Type Safety]\label{thm:safe}
  Let $\mathsf{Prog}$ be a program that is well-typed with respect to
  $\vdash^{\userkb}_\mathsf{er}$,
%  $\vdash^{\smolkb \cup \userkb}_\mathsf{er}$,
  where $\userkb$ is a conservative extension of $\smolkb \cup \mu(\classtable_\mathsf{Prog})$.
% \begin{enumerate}
% \item
 Every reachable configuration of $\mathsf{Prog}$ can be lifted
  to a consistent knowledge graph:
  \[\forall \mathsf{conf}.~\mathsf{init}_{\mathsf{Prog}}
    \rightsquigarrow^{\userkb}_\mathsf{er} \mathsf{conf}~
    ~\rightarrow~ ~\mu(\mathsf{conf}) \cup \smolkb \cup \userkb \text{
      is consistent.}\]
%     \item If $\mathsf{Prog}$ terminates, then it terminates successfully:
% \[\forall \mathsf{conf}.~\mathsf{init}_{\mathsf{Prog}} \Downarrow^{\userkb}_\mathsf{er} \mathsf{conf}~ ~\rightarrow~ \mathsf{conf} \text{ is successfully terminated}\]
% \end{enumerate}
%\end{theorem}
\end{restatable}

The proof is a standard inductive \emph{subject reduction}
proof~\cite{DBLP:books/daglib/0005958}, based on the structural
operational semantics of \SMOL (\cref{app:language}) and a type system
for runtime configurations (\cref{{app:type}}).

%%% Local Variables: 
%%% mode: latex
%%% TeX-master: "../main"
%%% End: 

\section{Discussion}\label{sec:eval}
In this section, we discuss how the implementation of \SMOL has been realized
(\cref{sec:implementation}), design choices for semantic lifting
(\cref{sec.extensions}) and applications of semantic reflection
(\cref{sec:epplications}).

\subsection{An Interpreter for  SMOL}
\label{sec:implementation}
\SMOL has been implemented as an interpreter in an open-source
project, available together with documentation and examples at
\href{https://smolang.org/}{\sf www.smolang.org}.\footnote{The version described in this work is available under \href{https://github.com/smolang/SemanticObjects/tree/prepare-1.0}{\sf https://github.com/smolang/SemanticObjects/tree/prepare-1.0}.}
% The source code of our \SMOL interpreter is available as an open
% source project, including examples showing how to use it.
The interpreter is mainly written in Kotlin, so it compiles to Java
class files and runs in the Java Virtual Machine (JVM) on most
platforms, including Windows, macOS, and Linux.  It uses ANTLR~\cite{parr2013definitive} to
parse program files, which ensures that the \SMOL grammar is followed
strictly.  For semantic data access, our implementation uses Apache Jena\footnote{\url{https://jena.apache.org/}},
OWL API~\cite{DBLP:journals/semweb/HorridgeB11}, and ONT-API\footnote{\url{https://github.com/owlcs/ont-api}}, and
HermiT~\cite{GlimmHMSW14} is used as the
default OWL reasoner.

Compared to the small language presented in this paper, the implemented
language has some additional features that are orthogonal to semantic
lifting, such as support for abstract classes, extended treatment of
generics, support for loading SHACL shapes from files instead of
string literals, further datatypes, a standard library, etc.  In
addition to the statements described in \cref{def:syntax},
the full language supports local variables, a statement \smol{destroy}
for manual memory management, an extension to integrate simulation
units~\cite{annsim} and some syntactic sugar for convenience (e.g.,
whole classes can be annotated with \smol{hidden}).
% For the modeling bridge, the string literal must be a predicate
% object list
% \footnote{Cf.~\url{https://www.w3.org/TR/turtle/\#grammar-production-predicateObjectList}}
% in turtle syntax.

In its simplest form, the interpreter takes a \SMOL program as input
and executes it by means of a process stack, a
static table, and a memory heap.
% , all of which are standard structures in most interpreters.
First, the interpreter clears the memory heap and scans the program to
generate the initial program stack and the static table.  Then it
considers each subroutine from the stack and performs the required
change to the memory heap until the stack is empty and the program
execution is done.  The project also includes a \emph{Read-Eval-Print
  Loop} (REPL): an interactive environment that can be used to control
and inspect the interpreter before, during, or after program
execution.  For example, the user can stop the program at any given
state, query the state using SPARQL, change which sources and
reasoners to use, check for consistency, or validate the data using
SHACL.

\begin{figure}[bt]
    \centering
    \resizebox{\textwidth}{!}{\begin{tikzpicture}

% Settings, colors and classes
\tikzset{rect/.style={shape=rectangle, draw, align=center}}
\definecolor{reasonerColor}{RGB}{233,154,154}
\definecolor{modelColor}{RGB}{165,195,242}
\definecolor{fileColor}{RGB}{183,214,170}
\definecolor{dataStructureColor}{RGB}{254,228,158}
\definecolor{ontologyColor}{RGB}{180,168,213}
\tikzset{reasoner/.style={shape=rectangle, draw, align=center, fill=reasonerColor}}
\tikzset{model/.style={shape=rectangle, draw, align=center, fill=modelColor}}
\tikzset{textC/.style={shape=rectangle, draw=none, align=center, fill=none}}

\node (start) at (0,0) {};
\node (headers) at (0,3.8cm) {};

% Sources
\node[rect, rounded corners, fill=dataStructureColor, text width=3cm, anchor=west, minimum height=1cm](s2) at (0,-0.75cm) {Static Table \\ \footnotesize{(data structure)}};
\node[rect, rounded corners, fill=dataStructureColor, below=0.5cm of s2, text width=3cm, minimum height=1cm] (s1){Heap \\ \footnotesize{(data structure)}};
\node[rect, rounded corners, fill=fileColor, text width=3cm, anchor=west, minimum height=1cm](s3) at (0,0.75cm) {Domain Ontology \\ \footnotesize{(file)}};
\node[rect, rounded corners, fill=fileColor, above=0.5cm of s3, text width=3cm, minimum height=1cm](s4) {SMOL Ontology \\ \footnotesize{(file)}};

% Models
% \node[model, right=1cm of s1, text width=3cm, minimum height=1cm](g1) {Heap Model \\ \footnotesize{(virtual)}};
% \node[model, right=1cm of s2, text width=3cm, minimum height=1cm](g2) {Static Table Model \\ \footnotesize{(virtual)}};
% \node[model, right=1cm of s3, text width=3cm, minimum height=1cm](g3) {Domain Ontology Model \\ \footnotesize{(materialized)}};
% \node[model, right=1cm of s4, text width=3cm, minimum height=1cm](g4) {SMOL Ontology Model \\ \footnotesize{(materialized)}};
\node[model, right=1cm of s1, text width=3cm, minimum height=1cm](g1) {Virtual Model};
\node[model, right=1cm of s2, text width=3cm, minimum height=1cm](g2) {Virtual Model};
\node[model, right=1cm of s3, text width=3cm, minimum height=1cm](g3) {Materialized Model};
\node[model, right=1cm of s4, text width=3cm, minimum height=1cm](g4) {Materialized Model};

% Final models and ontology
\node[model, right=15cm of start, anchor=east, text width=5.5cm, minimum height=1.4cm](m) {Integrated Model};
\node[model, anchor=north east, fill=ontologyColor, text width=3cm, minimum height=1.4cm](owlOntology) at (m.east|-s4.north) {OWL Ontology};
\node[model, anchor=south east, text width=3cm, minimum height=1.4cm](infModel) at (m.east|-s1.south) {Inference Model};

% Reasoners
\node[reasoner, rounded corners, left=0.4cm of owlOntology.west, anchor=east, text width=2.5cm, minimum height=0.7cm](owlReasoner) {OWL Reasoner};
\node[reasoner, rounded corners, left=0.4cm of infModel.west, anchor=east, text width=2.5cm, minimum height=0.7cm](jenaReasoner) {Jena Reasoner};

% Tasks
%\node[rect, draw=none, align=left, right=0.8cm of owlOntology, yshift=0.5cm, text width=3.2cm, minimum height=0.7cm](taskConsistency) {\small{Consistency Checking}};
%\node[rect, draw=none, align=left, right=0.8cm of owlOntology, yshift=0cm, text width=3.2cm, minimum height=0.7cm](taskType) {\small{Instance Retrieval}};
%\node[rect, draw=none, align=left, right=0.8cm of owlOntology, yshift=-0.5cm, text width=3.2cm, minimum height=0.7cm](taskClass) {\small{Type Checking}};
\node[rect, draw=none, align=left, right=0.8cm of owlOntology, yshift=0cm, minimum height=0.7cm](taskDLQuery) {\small{DL query}};
\node[rect, draw=none, align=left, right=0.8cm of infModel, yshift=1.0cm, minimum height=0.1cm](taskSPARQL) {\small{SPARQL} \\ \small{/SHACL}};

% Headers
% \node[textC, minimum height=1.4cm](v) at (s4.north|-headers) {\underline{\LARGE{Sources}}};
% \node[textC, minimum height=1.4cm](v) at (g4.north|-headers) {\underline{\LARGE{Models}}};
% \node[textC, minimum height=1.4cm](v) at (owlReasoner.north|-headers) {\underline{\LARGE{Reasoners}}};
% \node[textC, minimum height=1.4cm](v) at ([shift=({-0.3cm,0cm})]taskDLQuery.north|-headers) {\underline{\LARGE{Tasks}}};

% sources to models
\draw[-latex] (g1.west) -- (s1.east) node [pos=0.5,above] {\small{find}};
\draw[-latex] (g2.west) -- (s2.east) node [pos=0.5,above] {\small{find}};
\draw[-latex] (s3.east) -- (g3.west) node [pos=0.5,above] {\small{load}};
\draw[-latex] (s4.east) -- (g4.west) node [pos=0.5,above] {\small{load}};

% indivudal models to integrated model
\draw[] ([shift=({0cm,0.25cm})]g1.east) -- ([shift=({0cm,-0.3cm})]m.west);
\draw[] (g2.east) -- ([shift=({0cm,-0.1cm})]m.west);
\draw[] (g3.east) -- ([shift=({0cm,0.1cm})]m.west);
\draw[] ([shift=({0cm,-0.25cm})]g4.east) -- ([shift=({0cm,0.3cm})]m.west);

% reasoners to final models
\draw[] (jenaReasoner.east) -- (infModel.west|-jenaReasoner.east);
\draw[] (owlReasoner.east) -- (owlOntology.west|-owlReasoner.east);

% integrated model to inference models
\draw[-latex] (owlOntology.south|-m.north) -- (owlOntology.south) node [pos=0.5, left] {\small{ONT-API}};
\draw[] (infModel.north|-m.south) -- (infModel.north);

% Arrows from tasks to final models
\draw [-{Latex}] ([shift=({0.0cm,-0.06cm})]taskSPARQL.west) -- +(-0.2cm,0) |- node[pos=0.25] {} ([shift=({0.2cm,0cm})]infModel.east);
\draw [-{Latex}] ([shift=({0.0cm,0.06cm})]taskSPARQL.west) -- +(-0.2cm,0) |- node[pos=0.25] {} ([shift=({0.2cm,0cm})]m.east);
% \draw[-latex] (taskConsistency.west) -- ([shift=({0.2cm,0cm})]owlOntology.east|-taskConsistency.west);
% \draw[-latex] (taskType.west) -- ([shift=({0.2cm,0cm})]owlOntology.east|-taskType.west);
% \draw[-latex] (taskClass.west) -- ([shift=({0.2cm,0cm})]owlOntology.east|-taskClass.west);
\draw[-latex] (taskDLQuery.west) -- ([shift=({0.2cm,0cm})]owlOntology.east|-taskDLQuery.west);

\end{tikzpicture}}
    \caption{Diagram showing how data is accessed semantically.}\label{fig:semantic-data-access}
\end{figure}

\Cref{fig:semantic-data-access} gives an overview of how data from
different sources is accessed semantically in our system.  The
% four
available sources (left side) can be activated or deactivated
% all be switched on or off
independently. The active sources are combined into an
% one single
integrated model that can be queried directly.  If reasoning is
required, then querying can either be done via the available inference
model or via the OWL ontology interface.  The static table and the
heap, which are the two sources linked to the program state, are
accessed virtually, i.e., statements are not materialized, but
generated only when needed.  These virtual sources use a guard
mechanism to avoid traversing irrelevant parts of the internal data
structure corresponding to the program state.  The data access system
is built with Jena,
OWL API and
ONT-API. The key
parts of the data access system are detailed in the following.

\subsubsection{Sources}
Currently, the system supports four different data sources, but new
sources can be added in the future as needed.  Each of the four
sources gives access to a particular set of statements:
\begin{itemize}
\item the \SMOL \emph{ontology} is a \newrev{fixed} OWL file
  describing the domain model for runtime configurations \newrev{according to \cref{def:smolkb}};
\item the \emph{domain ontology} is an optional, static OWL file
  describing the model of the relevant domain (e.g., Geology in
  \cref{sec:overview});
\item the\emph{ static table} is an internal data structure containing
  static information about the current program, like the class
  hierarchy and each class' fields and methods; and
\item the \emph{heap} is an internal data structure containing all the
  objects constituting the current runtime configuration.
\end{itemize}
Both the \SMOL ontology and the domain ontology are static and
relatively small files, which are given as serializations of RDF.
Hence, it is unproblematic to materialize their statements during the
initialization of the system.  The two remaining sources, on the other
hand, are not provided as RDF statements, but as internal data
structures coupled with a mapping to a corresponding RDF
representation.  For each of these two sources, this RDF
representation is not materialized.  Instead, the statements are
accessed \emph{virtually}; i.e., the statements are only generated
when requested during query answering.  While the static table remains
static during runtime and is limited in size by the static parts of
the program, the heap is dynamic and could potentially become very
large.  This shows the importance of accessing the heap virtually:
materializing all RDF statements about the heap, which would need to
be done every time is it accessed by a query, is very demanding.

\subsubsection{Querying}
The simplest way of accessing data is by means of a SPARQL query or
by validating with SHACL directly on the integrated model.  Alternatively,
if a Jena reasoner is provided, it is possible to query with inference
via the provided inference model.  The third option is to send a
description logic query to the available OWL ontology, which must be
connected to a suitable OWL reasoner.  This third approach is used by
\SMOL's type checking mechanism.  While SPARQL can query collections
of RDF statements directly, DL queries instead require a set of OWL
axioms.  The translation from RDF to OWL in our system is done by
OWL-API: OWL axioms are created when \newrev{required},
while the leftover statements are just translated to general assertion axioms.  It should be clear that there is no limitation to who or what can access data, and when this can be done: queries can be posed externally by a user or internally by the program, and this can be done either before, during, or after the execution of the program.

\subsubsection{Virtualization}\label{sec:virtualization}
Queries posed to the integrated model are distributed to the models
that correspond to the different sources (see
\cref{fig:semantic-data-access}).  For the materialized models (here,
the \SMOL ontology and domain ontology), the query is simply evaluated
over the available statements.  For the virtual models, the system
uses the corresponding mapping to generate answers.  This mapping is
manifested as an implementation of a search method $\mathit{find}(t)$
for each source, which takes a search triple $t$ as input and returns
the set of statements matching this triple.  For simple queries with
just one triple, $\mathit{find}$ only needs to be called directly
once.  For more complex queries, the query planner must first split
the query into a set of multiple $\mathit{find}$ calls and then
combine the results from each such call into the final result set.  In
other words, the implementation of $\mathit{find}$ is only responsible
for traversing the relevant data structure and returning the
statements matching $t$, while the query planner is responsible for
transforming the query into $\mathit{find}$ calls and combining the
answers.

A naive way of implementing $\mathit{find}$ could be to always
traverse the whole internal data structure and collect statements
matching $t$.  Instead, our implementation carefully considers the
search triple $t$ and prevents the system from traversing the parts
that will only lead to irrelevant statements.  This is achieved by
guards \newrev{in the traversal by the interpreter}, which is a simple control mechanisms that
cannot be passed unless a given expression holds.  For example, if we
know that a given for-loop only generates statements of the form
\texttt{(?v :y ?w)} and \texttt{(?v :z ?w)}, where \texttt{?v} and
\texttt{?w} are variables, then there should be a guard in front of
the loop to check if $t$ matches either of the two forms. Placing
guards into the code in a way that improves the efficiency of
$\mathit{find}$ requires a good overview of the data structure and the
types of statements to which each part corresponds.  It is worth
noting that virtualization combined with reasoning does not work 
well in the current setup: many reasoners require initial
materialization of all statements, which conflicts with the idea of
virtual access.\looseness=-1

%%% Local Variables: 
%%% mode: latex
%%% TeX-master: "../main"
%%% End: 

\subsection{Design Choices for Semantic Lifting}\label{sec.extensions}
In this section, we discuss alternatives to the design of \SMOL's semantic
lifting approach that has been formalised in this paper.

\subsubsection{Abstraction}\label{sec:cssa}
To emphasize semantic lifting, the design of \SMOL supports
\emph{lifting by default} and the language offers hiding through
explicit annotations in its semantic lifting mechanism. Most often,
only a part of runtime configurations is actually queried in
practice. Therefore, we have opted for a virtual model with a
pull-based $\mathit{find}$-mechanism to lift parts of the heap upon
need (discussed in \cref{sec:virtualization}).  An alternative design,
which we believe is a reasonable trade-off in most applications,
would be to implement \emph{hiding by default}, and only expose
carefully selected pieces of information through the semantic lifting
mechanism. Obviously, \emph{hiding by default} can also profit from
virtualization as outlined above.

Semantic lifting can also be integrated with an abstraction function.
Taken together, hiding and abstraction would allow a more high-level
representation of objects in the knowledge graph. Taken to the
extreme, no actual fields would need to be lifted and an object could
be represented in the knowledge graph solely though an abstraction
function. Integrating abstraction in the semantic lifting process adds
an additional layer of computation to the semantic lifting
process. Pure methods can be used to operationally realize abstraction
for the semantic lifting process (pure methods are methods without
side-effects in object-oriented programming).

% Semantic lifting serializes the state\ebj{Could you check carefully
%   about how we use ``configuration'' vs ``state'' through the whole
%   paper?} of a program, thus all information that is not explicit must
% be derived from the knowledge graph.

It is possible to make information that is implicit in the state of an
object, explicit during the semantic lifting process by means of a
similar computational layer.
% In general, object-orientation can provide access to information
% that is not explicitly represented in the state of an object but
% rather computed from the state, i.e., information that is implicit
% but accessible.
To materialize such implicit information in a program, one typically
uses pure methods, as discussed above.  For the purpose of semantic
lifting, we could annotate such methods with a \xsmol{rule} modifier,
such that all annotated methods are executed on all objects from the
class of the method during the lifting process.

\begin{figure}[t]
\begin{smolframe}
class Building(List<Room> rooms, Int size, Street street) 
   Unit addRoom(Room room)
     this.rooms = Cons(room, this.rooms);
  end
  rule domain Int size()
    Int res = 0;
    for r in rooms do
      res = res + r.size;
    end
    return res;
  end
end
\end{smolframe}
\caption{\label{fig:rectangle}Using pure methods in semantic lifting.}
\end{figure}

\begin{example}[Materializing implicit information during semantic
  lifting]\label{ex:rectangle}
  Consider a version of class \xsmol{Building} from \cref{ex:ddl1}, in
  which instances of \xsmol{Building} do not explicitly maintain their
  size in a field. The code of this version is given in
  \cref{fig:rectangle}. Here, the size of a \xsmol{Building} instance
  needs to be computed using the \xsmol{size} method when needed, since
  it is not directly available.  By annotating this method with
  \xsmol{rule}, it is regarded as a field during semantic lifting;
  thus, the method is executed whenever the object is lifted and the
  result stored in a field \xsmol{size} in the lifted object.
\end{example}

\SMOL supported such computational semantic lifting in early versions
(e.g., \cite{DBLP:conf/esws/KamburjanKSJG21}), including a simple type
system to ensure that these methods do not alter the state during
lifting.  The implementation called the method using the current
function stack, executed it and stored the return value in the
knowledge graph.  This technique required arbitrary code to be
executed; it was removed due to its low performance --- while a
powerful modeling tool, adding this computational layer to the
semantic lifting process does not scale well for systems with many
objects.  Its role to derive information from the lifted state can
either be done by a reflective architecture (see \cref{sec:twins}) or
rule engine tools such as
SWRL.\footnote{\url{https://www.w3.org/submissions/SWRL-FOL/}}

\subsubsection{Integrating Black-Box Components}
For black-box components, the runtime state of a component will not
generally be available for semantic lifting. In this case, we suggest
to represent the state of the black-box component in the knowledge
graph in terms of a semantically lifted interface, which provides a
component descriptor and the input and output values that are
exchanged between the program and the black-box component. This can be
realized through a class definition that represents the component
descriptor and a proxy object that exchanges input and output values
between program and component.\looseness=-1

Let us concretize this approach to the semantic lifting of black-box
components by considering how functional mock-up units (FMUs) can be
integrated into \SMOL~\cite{annsim}, thereby allowing numerical
simulators to be embedded in \SMOL programs.  An FMU is a black-box
component for a numerical simulator, as defined by the functional
mock-up interface (FMI)~\cite{fmi}, allowing simulation units and
models to to exchanged for co-simulation~\cite{cosim}.  An FMU is
defined by a set of input and output ports. To perform the simulation,
it provides a set of procedures to advance the simulation (and thus,
update the output variables) for a certain amount of simulation
time. The information about input and output ports, such as type and
name, as well as other information, such as the tool used to generate
the FMU or the external references to guidelines, are stored in the
\emph{FMU model information}.  In \SMOL, each FMU is handled as a
special object, generated from its model description. The fields are
names and typed after the input and output ports.

\begin{example}[A Co-Simulation Scenario in SMOL and its Semantic Lifting]
  \Cref{fig:lv} is a simulation of a prey-predator system. It loads
  two FMUs, one each for prey and predators, which are stored in
  \xsmol{Prey.fmu} and \xsmol{Predator.fmu}.  The type \smol{FMO[in
    Double y, out Double x]} defines a functional mock-up object
  (FMO), which acts as a wrapper for the FMU and has two fields of
  type \smol{Double}, one of which can only be written (\xsmol{y}) and
  one only read ((\xsmol{x}). This information is checked against the
  model information in the FMU file.  The loop copies values between
  the simulators and calls the special \smol{tick} method that
  advances simulation time by the provided parameter (here, one time
  unit).

  The lifting of the configuration includes the lifting of the FMOs.
  Each such lifting contains the variables of the FMO (analogous to
  the lifting of fields of normal objects), their current values and
  the path of the loaded FMU.
  %\ebj{Add explanation of what the model descriptor looks like here and how the lifting works concretely!}
\end{example}

\begin{figure}[t]
\begin{smolframe}
FMO[in Double y, out Double x] prey = simulate("Prey.fmu");
FMO[in Double x, out Double y] predator = simulate("Predator.fmu");
Int i = 0;
while (i++ <= iterations) do
   prey.y = predator.y;
   predator.x = prey.x;
   prey.tick(1.0);
   predator.tick(1.0);
end
\end{smolframe}
\caption{\label{fig:lv}A co-simulation of a prey-predator system.}
\end{figure}

% The lifting of an FMU is the lifting of (a) the model description,
% and (b) the handling object, and follows the same structure as
% lifting a class and a normal object.

\subsubsection{Persistent State \& Garbage Collection}
% \newrev{While reflection in languages such as Java and Smalltalk
%   creates meta-representations of runtime stuctures, including objects
%   on the heap, these meta-objects can still be handled by a garbage
%   collector (although one needs to decide if objects that are only
%   accessible via reflection should be collected).}
Semantic reflection can retrieve objects that are no longer referenced
in a program state, as long as they exist on the heap. This means that
the result of a query to the semantically lifted program state may
depend on the garbage collector; i.e., there is a race between the
semantic lifting and runtime operations that are invisible to the
programmer.
% This prohibits any automatic garbage collection.
For example, consider the following program:\looseness=-1

\begin{smolframe}
main
    C c = new C(); 
    c = null; 
    List<C> l = access("SELECT ?x WHERE { ?x a prog:C });
    print(l); // non-null
end
\end{smolframe}

A tracing garbage collector could remove the object created on Line 2
before the query is executed on Line 3, depending on the timing of the
garbage collector. This renders the result of the query
non-deterministic. We see two possible approaches to render semantic
reflection deterministic in this context: disable garbage collection
\newrev{for objects reflected in the knowledge graph} or force garbage collection before semantic lifting. For simplicity in
\SMOL, we opted for the former and did not implement an automatic
garbage collector in the language interpreter so far. Instead, we have
introduced a simple form of
% However, the object is still lifted and would, as such, be retrieved
% by the query in the last statement.  An object may also not be
% barely retrieved by a query, but also be used for reasoning, so it
% is not syntactically possible to derive which objects can be removed
% without influencing any queries.  For this reason, \SMOL requires
manual memory management --- an object can be deallocated explicitly
using a \smol{destroy} statement. Note that the alternative, obtaining
deterministic behavior of queries to the semantic layer by forcing a
pass of the garbage collector before semantic lifting, can also be
realized by restricting the results of a query to the reachable
objects, thereby rendering the result of the query deterministic
independent of when the garbage collector is applied.%\ebj{Should we
  %mention Java's debugging interface and link to Anton's work here?}

\newrev{The race between semantic lifting and garbage collections manifests itself
  % Non-determinism arises as a problem
  if we consider objects only reachable through the knowledge graph as inaccessible. The underlying question is what we understand as inaccessible. In the example, the object is only accessible through the graph. One can reasonably argue that this can be considered inaccessible from the program's perspective (as it is a program object). One can equally argue that it is accessible, because the language provides a mechanism to access it through queries. Indeed, the first version of the geological case study relied on events being accessible through queries and did not store them in any variable.}

\begin{smolframe}
main
    C c = new C(); 
    destroy(c);
    c = null
    List<C> l = access("SELECT ?x WHERE { ?x a prog:C }); //empty
end
\end{smolframe}

\subsection{Applications}\label{sec:epplications}
Semantic lifting, and its realization in \SMOL, has been used and
validated in several applications, which we describe in this
section. Each of the applications is further detailed in the
referenced publications.  Throughout development, these applications
have influenced the design of \SMOL.  As discussed above, the
computational layer of semantic lifting (\cref{sec:cssa}) was removed
due to a lack of use and performance problems. Another removal was the
lifting of the process stack. Originally, \SMOL also lifted the full
function stack, including local variables and process
identifiers. However, this was removed to reduce the size of the
lifted state and because it was rarely needed in
applications.\looseness=-1

\subsubsection{Digital Twins}\label{sec:twins}
Semantic lifting has been used in case studies to enable a digital
twin to self-adapt to changes in a twinned system. In this line of
work, we exploited the ability to use graph queries on a knowledge
graph that contains information about \emph{both} the twinned system
% physical asset
and the controlling digital twin software. This has proven useful in
several scenarios, and we give only a short description of the main
idea here.  For a comprehensive overview over the use of semantic
lifting and reflection in digital twins, as well as more advanced
patterns, we refer to~\cite{tizianafest,edtconf}.

The first case study~\cite{isola1} considers a cyber-physical system
consisting of a (simplified) building as the twinned system (the
physical twin), and a \SMOL program as the digital twin.  The aim is
to ensure that as rooms are added and removed from the building, the
\SMOL program automatically detects these changes and adapts the
digital twin by removing or adding the objects that represent the
rooms. Each \SMOL object for a room contains an FMU that models its
temperature.

To self-adapt, the semantically lifted \SMOL program is compared
against
% is lifted, and the structure of the
% building,
a so-called asset model, represented as a knowledge graph.  The
\SMOL program that runs a \emph{defect query} over the combined
knowledge graph, where each such query expresses a relation between a
lifted \SMOL object and the part of the building that it models.  If
the defect query has a result, then it is either a \SMOL object that
must be removed (because the room it modeled was removed) or
information about how to create a \SMOL object to accommodate a new
room.\looseness=-1

This reflective digital twin architecture has been further applied and
generalized to a greenhouse in the \textsf{GreenhouseDT}
examplar~\cite{dtgreen}. Here, the physical twin is a greenhouse
containing pumps and plants, where plants are regularly replaced or
moved. In \textsf{GreenhouseDT}, \SMOL is used as an orchestrator
component for the digital twin: streams of sensor data and pump
controllers are realized as external components, while the \SMOL
component acts as the semantically reflected system orchestrator. The
reason for this decision is to separate data processing and numerical
operations from operations on the semantic structure to reconfigure
the digital twin components.

\newrev{
\paragraph*{Reclassification and Dynamic Background Knowledge}
The approach to semantic reflection as described here relies on the fact that the background knowledge is static, i.e., that the ontology provided to \SMOL does not change during program execution. Only the lifted knowledge graph changes. In Digital Twins, and in general when the background knowledge contains information about a running/existing system and not just general knowledge about universals, the background knowledge can evolve as well.}

\newrev{
Dynamic background knowledge is a problem in object-oriented programming, as this may cause that objects change their class after instantiation, which is known as \emph{dynamic object reclassification}~\cite{DBLP:conf/ecoop/DrossopoulouDDG01}. This is in particular important in the setting of self-adaptation, where objects need to be reclassified while retaining type safety. Sieve et al.~\cite{sieve25ecoop} discuss a reclassification extension of \SMOL, where an object changes its class, if the context it is linked to evolves. This declarative dynamic object reclassification has also been evaluated on the \textsf{GreenhouseDT} system and extended with type safety, but is so far not part of the main version of \SMOL.
}

\subsubsection{Simulation}
This case study exploits semantic reflection in \SMOL to facilitate
the interpretation of states in a simulator, using domain terminology.
The \emph{geological simulator} of Qu et al.~\cite{geo2} uses the
BFO-based GeoCore ontology~\cite{geocore}.  The motivating example in
\cref{sec:overview} is a simplified version of this simulator.  Using
complex triggers, the \SMOL program connects an
% in-use
advanced ontology with the simulation of geological process, without
the need to extend the ontology --- the new trigger concepts do not
refer to the \SMOL ontology.  The application is able to reproduce
results previously obtained manually by a team of geologists for the
Ekofisk geological formation in under 10 minutes.  As the so far
biggest case study with \SMOL, it also influences the design of the
language the most. To enhance performance, the \smol{hidden} keyword
was introduced and the use of guards in domain linkage proved very
useful to control the presence of kerogen depending on the object's
state.

\subsubsection{Semantic Lifting of JVM}
In contrast to the \SMOL-based applications above, the Java semantic
debugger (\textsf{sjdb}) of Hauber~\cite{tubiblio134113} implements the semantic lifting of the Java Virtual Machine (JVM). \newrev{It does not use \SMOL, but implements} semantic
lifting to a mainstream language. \newrev{Using such a language} introduces additional challenges
because the runtime state is not directly available or formalized.
Thus, \textsf{sjdb} uses the debugging interface of JVM as the basis
for semantic lifting --- the lifting does not serialize the runtime
state directly, but only the information exposed over this interface.

The \textsf{sjdb} tool consists of two parts. The \textsf{jvm2owl}
library that is used for semantic lifting, and \textsf{sjdb} itself,
which implements a debugger tool with breakpoints and a SPARQL
interface to examine the state. Here, the breakpoints can also be
semantic: execution is only halted if a certain query returns a
non-empty set. The \textsf{sjdb} tool does not support semantic
reflection, as it does not extend Java.  For this reason, the
programmer cannot control lifting without changing the code of
\textsf{jvm2owl} and exclude certain parts of the language.

% \subsection{Discussion}
% The above is not the only possibility to formalize semantic lifting
% and we give the reason for choosing a certain design over its
% alternative in the following.

% \paragraph*{The \SMOL Ontology}
% The \SMOL ontology models fields and variables in two ways: as
% \emph{objects} to express that a class has a field, and as
% \emph{properties} to express that an object has some value stored in
% the field. Thus, reasoners must use punning when handling the
% resulting knowledge base. However, we believe that this solution is
% significantly more readable and understandable to programmers than
% alternatives, such as the use of additional concepts and axioms to
% express that some object has stored something in a field. In
% particular, this would require property chains in OWL and blow up the
% size of even the simplest SPARQL queries concerned with storage.

% \paragraph*{The Modeling Bridge}
% The modeling bridge enables the user to program a part of the mapping
% directly in \SMOL.  An alternative design inside \SMOL is to use a
% special method to compute additional mappings, similar to the
% \smol{rule} methods introduced in the next section, but this would
% only allow to add single triples (computed from the return value of
% the method), while using guarded bridge expressions appears as more
% flexible for the user and can be done without detailed knowledge about
% the behavior of \SMOL methods --- it suffices to understand expressions
% and RDF.

%%% Local Variables: 
%%% mode: latex
%%% TeX-master: "../main"
%%% End: 

\section{Related Work}\label{sec:related}
Zhao \etal~\cite{DBLP:conf/ecoop/ZhaoCLS16} have proposed translating
programs, i.e., the static structure of types, variables, statements,
etc., to knowledge graphs in order to simplify and integrate static
analysis.  Similarly, ontologies for the static structure of Java
programs have been proposed by Kouneli
\etal~\cite{DBLP:conf/icwl/KouneliSPK12} and later Atzeni and
Atzori~\cite{DBLP:conf/semweb/AtzeniA17}.  Abstracting from concrete
programming languages, de~Aguiar \etal~\cite{DBLP:conf/er/AguiarFS19}
describe the OOC-O ontology that aims to give an integrated view for
multiple OO languages.  The knowledge graphs produced by these
approaches are similar to the static part of the \SMOL translation,
but runtime states are not considered.

The OPAL framework, proposed by Pattipati
\etal~\cite{DBLP:journals/spe/PattipatiNP20}, takes into account the
runtime semantics by representing the control flow graph and static
traces of C programs as triples. The mapping is based on a
\emph{static} analysis of the program and runtime states are still not
represented.  BOLD is an ontology-based log debugger for C programs
developed by Pattipati \etal~\cite{DBLP:journals/ase/PattipatiNP22},
building on OPAL.  In BOLD, programs are instrumented in order to
accumulate information about execution traces at runtime.  Debugging
then proceeds by querying this log information.  In contrast to \SMOL,
only a part of the execution state is captured, and only at selected
points in time, depending on the instrumentation.  The gathered
information cannot be accessed by the program, but it is available for
debugging, similar to ideas outlined in our prior
work~\cite{DBLP:conf/esws/KamburjanKSJG21}.  On the other hand, the
possibility to access information about a whole execution trace
instead of only the current state is of course particularly useful for
debugging, and this is not supported by \SMOL in its current state.  A
recent extension of semantic lifting to traces~\cite{fase} does not
consider semantic state access, but focuses on runtime enforcement.

%% \bigskip

%% From \cite{DBLP:journals/ase/PattipatiNP22} related work:
%% \begin{quote}
%%         PATO (Zhao et al. 2016) and CodeOntology (Atzeni and Atzori
%%         2017) frameworks proposed ontologies for C and Java languages
%%         respectively.

%%         The PATO framework (Zhao et al. 2016) converts C
%%         programs into triples and performs static analysis using these
%%         triples.

%%         The OPAL framework (Pattipati et al. 2020) represents
%%         control flow graph and static trace as triples and uses them
%%         for static analysis.

%%         \mg{The following have to do with programs and ontologies, but I think they are not really relevant to our work.}
%%         The SmartAPI (Eberhart and Argawal 2004)
%%         approach uses ontologies for automatic code generation. They
%%         expand libraries with domain ontology which assist easy
%%         location of code for a given goal.
%%         Ontologies are used to
%%         publish meta-information of software (Malone et
%%         al. 2014). This caters to the needs of a broad range of
%%         stakeholders.
%%         Ontologies are used for code maintenance
%%         (Devanbu et al. 1990) and teaching programming languages
%%         (Ganapathi et al. 2011; Sosnovsky and Gavrilova 2006). The
%%         applications of ontologies in software engineering are
%%         discussed by Happel et al. (2006).
%% \end{quote}

\medskip

%\paragraph{Transition Systems over Knowledge Bases.}
Connections between imperative programming languages and transition
systems over knowledge graphs have been investigated in multiple lines
of work, where the idea is to define languages that can operate
directly on knowledge graphs through atomic actions.
An early proposal is Golog~\cite{DBLP:journals/jlp/LevesqueRLLS97}, a
language based on McCarthy’s situation
calculus~\cite{McCarthy69sitcal}, that uses first-order logic guards
to examine and pick elements from its own state.
Around the same time, Fagin \etal proposed to make explicit the
distinction between what is the case in the world and what is known to
a program, by means of epistemic (multi-)modal operators $K_i$,
leading to the concept of \emph{knowledge-based}
programs~\cite{DBLP:journals/dc/FaginHMV97}.
This line of work is particularly interesting in multi-agent
scenarios, where programs benefit not only from access to their own
knowledge but also from reasoning about that of other programs.
Zarrieß and Claßen~\cite{ZarriessC15} expand on this line of work
% pick up the idea of Golog and
by integrating description logic into a concurrent extension of
Golog. They show how to verify CTL~\cite{DBLP:conf/lop/ClarkeE81}
properties with description logic assertions. In contrast to our work,
knowledge is managed explicitly and programs do not reflect upon
themselves.
% we note that such assertions are easily realised in \SMOL using \smol{assert}.

\medskip

A challenge for programs that can directly modify a knowledge graph
is that an ABox may change in such a way that it violates the TBox,
meaning that the system’s state becomes inconsistent.
Calvanese \etal~\cite{Calvanese11} propose two operations
$\mathtt{ASK}$ and $\mathtt{TELL}$ for transition systems defined
\emph{explicitly} over knowledge graphs. The $\mathtt{ASK}$ operator
corresponds roughly to our \smol{access}, while $\mathtt{TELL}$
performs a required action on the knowledge graph.
% The details of
This operation is based on the theory of knowledge base revision
\cite{DBLP:conf/kr/KatsunoM91}, in particular for DL-Lite knowledge
bases \cite{DBLP:conf/aaai/GiacomoLPR06,DBLP:conf/aaai/GiacomoLPR07}.

In contrast to this line of work, the transition system of \SMOL is
\emph{implicit}, following the semantics of a fairly standard
object-oriented language. The advantage of the $\mathtt{TELL}$
operator is clearly that state updates, like state access, closely
match the semantic view.  On the other hand, the advantage of our
approach with \SMOL is that well-established principles from
programming languages carry over, avoiding to reinvestigate concepts
such as modularity, runtime semantic structure and control flow for
knowledge graphs.
While all changes to the knowledge graph are global in the work of
Calvanese \etal~\cite{Calvanese11}, global changes in \SMOL only
happen in the part of the knowledge graph inferred from user-provided
axioms; the part inferred from the mapping only changes locally.
% after a single step in \SMOL.

More generally, the effects of combining rule systems with description
logics, and how to accommodate the differences in semantics, have been
the object of much study.  In particular, the work of Eiter
\etal~\cite{DBLP:journals/ai/EiterILST08} concentrates on using rule
systems as a programming language; technically, answer set programming
is enhanced with access to knowledge graphs.  The rules are similar to
what is commonly used in non-monotonic logic programs, but rule bodies
may also contain queries to the knowledge graph, possibly under
default negation.  In contrast to this line of work on rules and
description logics, our work with \SMOL has
% does not aim to manipulate the
% knowledge graphs directly; instead, we have
concentrated on the semantic lifting of a more ‘mainstream’
object-oriented language.

\medskip

In contrast to the previously discussed work, we have not targeted
% tried to define
a language that operates directly on description logic interpretations
or knowledge graphs.  Instead, we aim is to enhance a language similar
to mainstream programming languages by semantic technologies.

Closest to our approach in this respect is the work on
\emph{ontology-mediated} programming of Dubslaff, Koopmann and Turhan
\cite{DubslaffKT19,DubslaffKT20}.  Instead of operating on a knowledge
graph, they define the concept of an ‘interface’ between the program
and the knowledge graph.  Technically, the interface defines explicit
mappings from program states to the description logic, and vice versa,
where the interaction happens through a number of designated variables
and ‘hooks.’  As underlying programming language, the authors use a
stochastic guarded command language similar to PRISM \cite{PRISM4.0},
such that it is possible to perform probabilistic model checking on
the ‘ontologized programs.’  In contrast to our work, the programming
language provides neither
% itself is not so interesting for us, providing neither
semantic reflection nor typing.  However, it is interesting to compare
the interface mechanism itself to our work in more detail.

\SMOL uses an implicit ‘direct’ mapping for semantic lifting, and does
not require that the correspondence between program and knowledge
graph is described in two directions.  From the point of view of the
program, the variables $\mathit{Var}_{\mathbf{O}}$ and hooks
$H_{\mathbf{O}}$ of an interface can be compared to the variables used
in \smol{access} queries.  \newrev{The variables $\mathit{Var}_{\mathbf{O}}$ are those variables whose values are lifted into the knowledge graph, while $H_{\mathbf{O}}$ are concepts in the ontologies that are explicitly used inside the program.} From the point of view of the knowledge
graph, \SMOL reflects the complete program state (including all
variables) into the knowledge graph, but our implementation of
virtualization ensures that only those parts needed for semantic state
access are actually generated.  We believe that the semantic lifting
of \SMOL subsumes the language concepts of ontology-mediated
programming in terms of expressivity.

%In \SMOL, every symbol not defined in the program but described in the
%user ontology is analogous to the variables part of the interface of
%Dubslaff \etal.  thus, the language concepts of ontology-mediated
%programming are thus subsumed by programming with semantical lifted
%states.

\medskip

% \paragraph{Semantic Approaches to Programming Languages.}
Ontologies have also been explored in the context of type systems for
programming languages.
% There is work on making an
Leinberger \etal~\cite{DBLP:conf/esop/LeinbergerLS17} study DL concept
expressions
% of a description logic
as static types in a $\lambda$-calculus, such that terms can be
type-checked using
% .  Type checking is based on
SHACL constraints~\cite{DBLP:conf/semweb/LeinbergerSSLS19}.  Existing
programming languages can be integrated with
% aware of
RDF data
% to integrate it with
using the type systems of Paar and
Vrandecic~\cite{DBLP:conf/esws/PaarV11} and Leinberger
\etal~\cite{DBLP:conf/semweb/LeinbergerSLSTV14}. It is interesting to
observe that the difference between ontologies and regular types is
not just about taste: (a) concepts allow more expressive structure
than type hierarchies and (b) classes in programming languages are
designed by the user to fit the needs of its application, while the
concepts of the domain are designed to accommodate the needs of a
general domain.
% While, other systems use static types to connect ontologies and
% programming languages.
% \SMOL is dynamically typed and the concepts of the domain and
% mapping in \SMOL are disjoint and need to be connected using
% additional axioms.
The connection to types has also been investigated through
mappings~\cite{DBLP:conf/seke/KalyanpurPBP04} and code
generation~\cite{DBLP:conf/caise/StevensonD11}.
While this line of work attempts to unify two tools made for different
tasks, our approach with \SMOL is to propose a sensible interface.

\newrev{Thimmaiah et al.~\cite{DBLP:conf/icse/ThimmaiahLR024} implement a simple (compared to sjdb) form of semantic lifting of Featherweight Java into Neo4j, but focus on efficiency of the subsequent data access instead of modeling or the connection between behavioral and ontological modeling. It provides no form of static analysis or type checking. To the best of our knowledge, their approach has not been applied to projects beyond performance evaluations.}
% \MG{I propose to remove the remaining references here.  There is a
% mountain of literature about rules on knowledge bases (most of the
% RuleML conference is about that) and the interaction of rules with
% DL semantics.  Of course, part of the point is to use the rules to
% program, but we don’t want to cite all that.  I added text earlier
% that tries to explain that all this work exists but is not very
% relevant here.}

% K{\"a}fer and Harth~\cite{DBLP:conf/www/KaferH18} perform actions on
% RDF files in the semantic web using linked data, operating on a set
% of user-input rules for an abstract state machine.  Horne \etal
% define an operational semantics for SPARQL
% updates~\cite{DBLP:conf/aswc/HorneSG11} and a system that
% internalizes queries into a process
% algebra~\cite{DBLP:journals/corr/abs-1108-0229}.

%%% Local Variables: 
%%% mode: latex
%%% TeX-master: "../main"
%%% End: 

\section{Conclusion}\label{sec:conclusion}
Semantic reflection opens new perspectives on how semantic
technologies and programming languages can be combined.  Our work with
\SMOL introduces a clear separation of concerns between
\emph{computations} (in the \newrev{imperative} programming language) and \emph{domain
  description} (in the ontology), and provides a clear interface
between these concerns (queries and domain linkage). This way, we are
able to reuse standard technologies and, thus, reduce the need to
learn a new formalism for users. Furthermore, we provide basic tool
and analysis support for this interface through a type
system. \looseness=-1

We believe that this research direction, at the intersection of
semantic technologies and programming languages, opens up interesting
possibilities for integrating domain knowledge and behavioral
modeling.  Complementing the foundational study presented here, we
have also described first applications and case studies.  Together,
this work demonstrates the versatility and robustness of semantical
reflection. \looseness=-1

\newrev{Beyond the technical level of knowledge graphs and imperative programs, semantic lifting also provides an opportunity to study the relation between knowledge evolution ~\cite{DBLP:journals/tgdk/PolleresPBDDDEF23} and software evolution. Other kinds of interfaces between data and software artifacts, such as object-relational mappings, are known to be challenging to maintain~\cite{DBLP:conf/icse/ChenSJHNF14,DBLP:conf/msr/ChenSYHGNF16}, but the interface between graph data and software has received less attention so far.}\looseness=-1

A possible direction for future work at the foundational level, is the
extension of the type checking towards more dynamic queries, e.g.,
queries assembled at runtime using string operations, investigate
the connection between semantic lifting and knowledge graph
construction pipelines, \newrev{and investigate semantic reflection for non-imperative programming languages, e.g., functional languages.}. \looseness=-1

%%% Local Variables: 
%%% mode: latex
%%% TeX-master: "../main"
%%% End: 

\bibliography{ref}
\pagebreak

% \appendix
\begin{appendices}
\section{Full Runtime Semantics}\label{app:language}
  The full
% complete grammar of the
syntax of \SMOL is given by the grammar in \cref{fig:app:syntax}, including
% , with all
the primitives for
semantic reflection.
% , is as shown in Fig.~\ref{fig:app:syntax},
The runtime configurations (cf. \cref{def:conf}) are defined by
the  grammar
$$\begin{array}{l@{\,::=\,}l@{\qquad}l@{\,::=\,}l@{\qquad}l@{\,::=\,}l}
    \conf & \classtable~\mathsf{obs}~\mathsf{prs} 
    &\mathsf{rs} & \mathsf{Stmt} \ssep \mathsf{Loc} \leftarrow
                   \xsmol{stack}\xsmol{;}~\mathsf{Stmt}
           & \mathsf{Cl} & \xsmol{C} \ssep \xsmol{List<C>} \\
    \mathsf{obs} & \many{(\mathsf{Cl},\rho)_\mathtt{X}}
          & \mathsf{prs} & \many{(\xsmol{m},\mathtt{X},\mathsf{rs},\sigma)}.
\end{array}$$

\noindent
Here, $\sigma$ ranges over local stores, i.e., maps from variables to
DEs, $\rho$ over object stores, i.e., maps from fields to DEs,
$\mathsf{CT}$ over class tables, and $\mathtt{X}$ over object
identifiers. The remaining terms are defined in \cref{def:syntax}.

\begin{figure}[t]
    \begin{minipage}{\textwidth}
    \begin{align*}
    \mathsf{Prog} ::=\;& \many{\mathsf{Class}}~\xsmol{main}~\mathsf{Stmt}~\xsmol{end}&&\text{Programs}\\
    \mathsf{Class} ::=\;& \xsmol{class C}\big[\xsmol{extends C} \big]
    (\many{\mathsf{Field}})~[\mathsf{Linkage}]~\many{\mathsf{Met}}~\xsmol{end}&&\text{Classes}\\
    {\mathsf{Type} ::=\;}& \xsmol{t} \ssep \xsmol{C} \ssep \xsmol{List<C>} &&{\text{Types}}\\
    \mathsf{Field} ::=\;& \big[\xsmol{hidden} \ssep \xsmol{domain}\big]~{\mathsf{Type}}~\xsmol{f}&&\text{Fields}\\
    \mathsf{Linkage} ::=\;& \many{\xsmol{links(}\mathsf{Expr}\xsmol{) le;}}~\xsmol{links le;}&&\text{Domain linkage}\\
    \mathsf{Met} ::=\;& 
    {\mathsf{Type}}~\xsmol{m}(\many{{\mathsf{Type}}~\xsmol{v}})~{\mathsf{Stmt}}~\xsmol{end}&&\text{Methods}\\
    \mathsf{Stmt} ::=\;& 
    \mathsf{Loc}~\xsmol{=}~\mathsf{RHS} \xsmol{;} 
        \ssep\xsmol{if}~\mathsf{Expr}~\xsmol{then}~\mathsf{Stmt}~\xsmol{else}~\mathsf{Stmt}~\xsmol{end} 
    &&\text{Statements}\\
    &
    \ssep \mathsf{Expr}\xsmol{.m}(\many{\mathsf{Expr}})\xsmol{;} \ssep \xsmol{skip;} 
    \ssep \xsmol{while}~\mathsf{Expr}~\xsmol{do}~\mathsf{Stmt}~\xsmol{end} \\
    &\ssep\mathsf{Type}~\xsmol{v}~\xsmol{=}~\mathsf{RHS}\xsmol{;} \ssep \mathsf{Stmt~Stmt} \ssep \xsmol{return}~\mathsf{Expr} \xsmol{;} \\
    \mathsf{RHS} ::=\; & 
    \xsmol{new C}(\many{\mathsf{Expr}})~[\mathsf{Linkage}]
    \ssep \mathsf{Expr}.\xsmol{m}(\many{\mathsf{Expr}}) \ssep \mathsf{Expr}  &&\\
    &\ssep \xsmol{access}(\mathtt{sparql},\many{\xsmol{Expr}}) \ssep \xsmol{member}(\mathtt{owl}) \ssep \xsmol{validate}(\xsmol{shacl})  
    &&\text{RHS expressions}\\
    \mathsf{Expr} ::=\;& \xsmol{this} \ssep \xsmol{null} \ssep \mathsf{Loc} \ssep \xsmol{a}
                     \ssep \mathsf{Expr}~\mathit{op}~\mathsf{Expr} \ssep \mathsf{Expr}~\xsmol{==}~\mathsf{Expr} \ssep \mathsf{Expr}~\xsmol{!=}~\mathsf{Expr} &&\text{Expressions}\\
      \mathsf{Loc} ::=\;& \mathsf{Expr}\xsmol{.f} \ssep \xsmol{v} 
    &&\text{Locations}
    \end{align*}
    \end{minipage}
\caption{Full Syntax of \SMOL\label{fig:app:syntax}
}
\end{figure}

In the following, we present a Structural Operation Semantics
(SOS)~\cite{Plotkin} for \SMOL as a set of rules that defines
transitions between runtime configurations. These rules formally
define the relation $\rightarrow_\mathsf{er}^\mathcal{K}$ from
\cref{ssec:reflect} and have \cref{def:validate,def:member,def:access}
as special cases in rules \rulename{validate}, \rulename{member}, and
\rulename{access}. Expressions $\mathsf{Expr}$ are evaluated using an
evaluation function
$\sem{\mathsf{Expr}}_{\mathtt{X}}^{\sigma,\mathsf{obs}}$ with respect
to an object identifier $\mathtt{X}$ to resolve the expression
\smol{this}, a local store $\sigma$ to resolve local variables and a
set of objects $\mathsf{obs}$ to resolve field accesses.  To simplify
the presentation, we will also allow object identifiers as the
right-hand-sides of assignments; object identifiers simply evaluate to
ehemselves.
% The evaluation of expressions is the identity on object identifiers.

We say that an object identifier is \emph{fresh} if it does not appear
in a runtime configuration.  We group the rules into three parts: rules with global effect, rules for semantic reflection, and rules with local effect.

\newcommand{\ertransition}{\rightarrow_\mathsf{er}^\mathcal{K} }
\begin{description}
\item[Rules with global effect.]  The rules in \cref{fig:app:methods}
  have a global effect; they either create new objects or manipulate
  the call stack.
  \begin{itemize}
  \item Rule \rulename{new} allocates a new object in the runtime
    configuration, initializing it and reduces to an assignment ---
    thus, we do not need several rules for all forms of locations. The
    rule's first premise assigns to each field in the object memory
    $\rho$ the evaluation of the corresponding parameter. The second
    premise creates a fresh object identifier $\mathtt{X}$. The third
    premise ensures that the number of parameters is the same as the
    number of fields.
  \item Rule \rulename{call} deals with method calls. The rule is
    analogous to rule \rulename{new}, but modifies the stack instead of
    the runtime configuration.  The rule's premises evaluate the
    target expression $\mathsf{Expr}$ to an object identifier,
    retrieves the class of this object from the runtime configuration,
    checks that the number of arguments is correct by a lookup in the
    class table (using $\mathsf{vars}$).  The rule creates an initial
    store $\sigma'$ by evaluating the parameters, and adds a new
    process to the configuration.  The old process has its active
    statement replaced by the waiting statement
    $\mathsf{Loc}~ \leftarrow \mathtt{stack}\xsmol{;}$ that records
    where the return value will be stored once the called method
    terminate.
  \item Rule \rulename{return} deals with \smol{return} statements and
    removes one process from the stack, but also modifies the calling
    process by replacing its waiting statement
    $\mathsf{Loc}~ \leftarrow \mathtt{stack}\xsmol{;}$ by an
    assignment of the return value to the stored location
    $\mathsf{Loc}$.
  \end{itemize}

\item[Rules for semantic reflection.] The rules in
  \cref{fig:app:semantic} correspond directly to
  \cref{def:validate,def:member,def:access}.  There are three
  analogous rules in the case that the target location is a declared
  variable.

\item[Local effect without lifting.] Finally, the rules in
  \cref{fig:app:local} are standard programming constructs.
\begin{itemize}
\item Rules \rulename{iftrue} and \rulename{iffalse} reduce a
  branching statement to the first or second branch, depending on the
  evaluation of the guarding expression.
\item The rule \rulename{loop1} unrolls the loop body once if the loop
  guard evaluates to true, while \rulename{loop2} removes the loop
  from the active statement and continues with the next statement.
\item The rule \rulename{assign1} deals with assignment of side effect
  free expressions (and object identifiers) to fields. It evaluates
  $\mathsf{Expr}$ to an object identifier $\mathtt{Y}$ and updates its
  memory. Rules \rulename{assign2} and \rulename{assign3} update the
  local memory of the stack frame for new declared and
\item Finally, \rulename{skip} merely removes a \smol{skip} statement
  without further effect and \rulename{callIn} introduces a fresh
  variable to handle method calls without a target location.
\end{itemize}
\end{description}

\begin{figure}[t]
\input{appendix/app_global}
\caption{Rules with global effect: Object creation, method call,
  method return.\label{fig:app:methods}}
%   \end{figure}  
% \begin{figure}[t]
\input{newrules}
\caption{Rules for semantic reflection.\label{fig:app:semantic}}
\end{figure}

\begin{figure}[t]
\input{appendix/app_local}
\caption{Rules with a local effect that do not depend on semantic
  lifting: branching, iteration, assignment, variable declarations and
  calls without target variable.\label{fig:app:local}}
\end{figure}

%%% Local Variables: 
%%% mode: latex
%%% TeX-master: "../main"
%%% End: 

% %%%
% \section{Full Axioms for \smolkb}\label{app:axioms}
% \input{app_axioms}
\clearpage
% %%%
\section{The Type System}\label{app:type}
We use a normalized form of \SMOL in this section, in order to make
the formalization of the type system more simple.  In particular, (1)
all expressions with side-effects target variable declarations, and
(2) all methods end with a return statement. We further consider a
Java-style type hierarchy that distinguishes basic and class types.
Given a program $\mathsf{Prog}$, we denote by
$\mathtt{C}_1 ~\mathit{extends}~ \mathtt{C}_2$ that $\mathtt{C}_1$ is
declared to extend class $\mathtt{C}_2$ in $\mathsf{Prog}$.

\begin{definition}[Type Hierarchy \& Subtyping]
  Given a program $\mathsf{Prog}$, let $\mathcal{T}$ denote the
  associated type hierarchy, defined as follows:
\begin{itemize}
\item
  $\xsmol{Object}, \xsmol{Unit}, \xsmol{Int}, \xsmol{Bool}, \top, \bot
  \in \mathcal{T}$ and
\item $\xsmol{C} \in \mathcal{T}$ and
  $\xsmol{List<C>} \in \mathcal{T}$ for all classes \smol{C}.
\end{itemize}
Subtyping is defined as the minimal partial order $\preceq$ over
$\mathcal{T}$, satisfying the following conditions:
\begin{itemize}
\item $\forall T \in \mathcal{T}.~T \preceq \top$,
\item $\forall T \in \mathcal{T}.~\bot \preceq T$,
\item $\forall \xsmol{C}.~\xsmol{C} \preceq \xsmol{Object}$,
\item $\xsmol{List<C>} \preceq \xsmol{Object}$,
\item $\xsmol{C}_1 \preceq \xsmol{C}_2$ if $\xsmol{C}_1$
  $\mathit{extends}$ $\xsmol{C}_2$,
\item $\xsmol{List<C}_1\xsmol{>} \preceq \xsmol{List<C}_2\xsmol{>}$ if
  $\xsmol{C}_1 \preceq \xsmol{C}_2$, and
\item $\xsmol{Int},\xsmol{Unit},\xsmol{Bool} \not\preceq \xsmol{Object}$
\end{itemize}
\end{definition}

If there is no such partial order because of cycles in the $\mathit{extends}$ relation, then type checking immediately fails.
We write $\mathsf{defines}(\classtable, \mathsf{Type})$ if type $\mathsf{Type}$ is either a basic type, or defined in the program, i.e., $\mathsf{Type} \in \mathbf{dom}(\classtable)$.

\subsection{Typing Surface Syntax}
We first describe the typing judgements and rules for
% the first three layers of syntax:
programs, classes and methods, given in \cref{fig:app:type:global}.
The typing hierarchy and class table are implicitly given, to avoid
syntactic clutter.
% \looseness=-1

\begin{description}
\item[Program Layer] The type judgement
  $\vdash^{\mathcal{K}}_\mathsf{er} \Prgm$ holds if the program
  $\Prgm$ is type-safe with respect to knowledge base $\mathcal{K}$
  and entailment regime $\mathsf{er}$. In practice, we always use the
  \SMOL ontology and the user provided domain knowledge here, i.e.,
  $\mathcal{K} = \smolkb \cup \userkb$.  The rule \rulename{prog}
  expresses simply that all classes and the main block must be
  well-typed with respect to the class table of the program and the
  given knowledge base and entailment regime.
\item[Class Layer] The type judgement
  $ \vdash^{\mathcal{K}}_\mathsf{er} \mathsf{Class}$ holds if the
  class $\mathsf{Class}$ is type-safe with respect to knowledge base
  $\mathcal{K}$, and entailment regime $\mathsf{er}$.  The rule
  \rulename{class} checks that the extended class exists, that no
  declared field exists in a superclass, that all field types are
  defined and then type checks all methods.
\item[Method Layer] The type judgement
  $\Gamma\vdash^{\mathcal{K}}_\mathsf{er} \mathsf{Met}$ holds if the
  method $ \mathsf{Met}$ is type-safe with respect to knowledge base
  $\mathcal{K}$ and entailment regime $\mathsf{er}$. Here, the typing
  environment $\Gamma$ captures the additional context for the type
  judgment, this typing environment consists of types for the fields of the
  surrounding class.  The rule \rulename{method} again checks that all
  used types are defined and type checks the method body.
\end{description}

\begin{figure}[t]
\noindent\resizebox{\textwidth}{!}{
\begin{minipage}{1.1\textwidth}
\TINFER{prog}{
\emptyset \vdash^\mathcal{K}_\mathsf{er} \mathsf{Stmt}  : \xsmol{Unit}\vartriangleright \Gamma
\qquad\forall i \leq n.~ \vdash^\mathcal{K}_\mathsf{er} \mathsf{Class}_i
}{
 \vdash^\mathcal{K}_\mathsf{er} \mathsf{Class}_1 \dots \mathsf{Class}_n~\xsmol{main}~\mathsf{Stmt}~\xsmol{end}
}

\TTINFER{class}{
\forall i\leq m.~ \{\xsmol{f}_1 \mapsto \mathsf{Type}_1,\dots,\xsmol{f}_n \mapsto \mathsf{Type}_n, \xsmol{this} \mapsto \xsmol{C}\} \vdash^\mathcal{K}_\mathsf{er}  \mathsf{Met}_i
\qquad\xsmol{D} \in \mathbf{dom}(\classtable)
}{ \forall \xsmol{E}.~\big(\xsmol{C} \prec \xsmol{E} \rightarrow \not\exists T.~T~\xsmol{f}_i \in \mathsf{fields}(\classtable,\xsmol{E})\big)\Big)
\qquad\forall i\leq n.~\mathtt{defines}(\classtable, \mathsf{Type}_i)
}{
 \classtable \vdash^\mathcal{K}_\mathsf{er} \xsmol{class C}~\xsmol{extends D}
 (\mathsf{Type}_1~\xsmol{f}_1, \dots,\mathsf{Type}_n~\xsmol{f}_n)~
 \mathsf{Met}_1\dots\mathsf{Met}_m~\xsmol{end}
}

\TTINFER{method}{
\mathtt{defines}(\classtable, \mathsf{Type}) \qquad \forall i\leq n.~\mathtt{defines}(\classtable, \mathsf{Type}_i)
}{
    \Gamma \cup \{\xsmol{v}_1 \mapsto \mathsf{Type}_1,\dots,\xsmol{v}_n \mapsto \mathsf{Type}_n\} \vdash^\mathcal{K}_\mathsf{er} \mathsf{Stmt}~\xsmol{return}~\mathsf{Expr}\xsmol{;} :\mathsf{Type}
}{
\Gamma \vdash^\mathcal{K}_\mathsf{er} \mathsf{Type}~\xsmol{m}(\mathsf{Type}_1~\xsmol{v}_1, \dots,\mathsf{Type}_n~\xsmol{v}_n)~\mathsf{Stmt}~\xsmol{return}~\mathsf{Expr}\xsmol{;}~\xsmol{end}
}
\end{minipage}
}
\caption{Typing Rules for Programs, Classes and Methods.}
\label{fig:app:type:global}
\end{figure}

The type judgement for statements is given as
$\Gamma\vdash^\mathcal{K}_\mathsf{er} \mathsf{Stmt} : \mathsf{Type}
\vartriangleright \Gamma'$, and expresses that the statement
$\mathsf{Stmt}$ returns a value of type $\mathsf{Type}$ under
environment $\Gamma$ and updates the environment to $\Gamma'$ (this
update of the typing environmentis needed for variable
declarations).\footnote{By slight abuse of notation, we consider
  \xsmol{return} a statement here.}  The type $\xsmol{Unit}$ is used
for statements that do not return, but are still well-typed in the
sense of containing no substatements that could cause a runtime error,
as discussed.

\begin{figure}[t]
\begin{minipage}{0.5\textwidth}
\TTINFER{T-fresh}{   
    \xsmol{v} \not\in \mathbf{dom}~\Gamma\qquad
\Gamma' \vdash^\mathcal{K}_\mathsf{er} ~\xsmol{v :=}~\mathsf{RHS}\xsmol{;} : \xsmol{Unit}
}{
  \Gamma' = \Gamma[\xsmol{v} \mapsto \mathsf{Type}]{}\qquad  \Gamma' \vdash^\mathcal{K}_\mathsf{er} \xsmol{Stmt} : \mathsf{Type}
}{
\Gamma \vdash^\mathcal{K}_\mathsf{er} ~\xsmol{Type v :=}~\mathsf{RHS}\xsmol{;}~ \xsmol{Stmt}~ : \xsmol{Unit}
}
\end{minipage}~
\begin{minipage}{0.49\textwidth}
\TTINFER{T-wh}{ 
\Gamma \vdash \xsmol{Expr} : \xsmol{Bool}
}{
\Gamma \vdash \xsmol{Stmt; skip;}:\mathsf{Type}
}{
\Gamma \vdash^\mathcal{K}_\mathsf{er} \xsmol{while Expr do Stmt end}:\mathsf{Unit}
}
\end{minipage}

\medskip

\begin{minipage}{0.5\textwidth}
\TTTINFER{T-if}{
\Gamma \vdash \xsmol{Expr} : \xsmol{Bool}
}{
\Gamma \vdash \xsmol{Stmt}_1\xsmol{; skip;}  : \mathsf{Type}
}{
\Gamma \vdash \xsmol{Stmt}_2\xsmol{; skip;}  : \mathsf{Type}
}{
\Gamma \vdash^\mathcal{K}_\mathsf{er} \xsmol{if Expr then Stmt}_1~\xsmol{else Stmt}_2~\xsmol{end} : \mathsf{Type}
}
\end{minipage}
\qquad
\begin{minipage}{0.45\textwidth}
\TTINFER{T-sequence}{
\Gamma \vdash^\mathcal{K}_\mathsf{er} \xsmol{Stmt}_1 : \xsmol{Unit}
}{
\Gamma \vdash^\mathcal{K}_\mathsf{er} \xsmol{Stmt}_2 : \mathsf{Type}
}{
\Gamma \vdash^\mathcal{K}_\mathsf{er} \xsmol{Stmt}_1~\xsmol{Stmt}_2 : \mathsf{Type} 
}
\end{minipage}
% \caption{Typing Rules for Statements (I).}
% \label{fig:app:type:stmt1}
% \end{figure}

\bigskip

% \begin{figure}
\begin{minipage}{.5\textwidth}    
\TINFER{T-skip}{ 
}{
\Gamma \vdash^\mathcal{K}_\mathsf{er} \xsmol{skip;} : \xsmol{Unit}
}

\TINFER{T-return}{ 
    \Gamma \vdash^\mathcal{K}_\mathsf{er} \xsmol{Expr} : \mathsf{Type}
}{
\Gamma \vdash^\mathcal{K}_\mathsf{er} \xsmol{return Expr;} : \mathsf{Type}
}
\end{minipage}
\begin{minipage}{.5\textwidth}    
\TTINFER{T-assign}{
\Gamma \vdash \mathsf{Loc} : \mathsf{Type} 
}{
\Gamma \vdash \mathsf{RHS} : \mathsf{Type}
}{
\Gamma \vdash^\mathcal{K}_\mathsf{er} \mathsf{Loc}~\xsmol{:=}~\mathsf{RHS}\xsmol{;} : \xsmol{Unit}
}
% \TTINFER{T-fresh}{   
% \xsmol{v} \not\in \mathbf{dom}~\Gamma
% \qquad \Gamma' = \Gamma[\xsmol{v} \mapsto \mathsf{Type}]
% }{
% \Gamma' \vdash^\mathcal{K}_\mathsf{er} ~\xsmol{v :=}~\mathsf{RHS}\xsmol{;} : \xsmol{Unit}
% }{
% \Gamma \vdash^\mathcal{K}_\mathsf{er} ~\xsmol{Type v :=}~\mathsf{RHS}\xsmol{;} : \xsmol{Unit}
% }
\end{minipage}

% \TTINFER{T-declare}{   
% \Gamma \vdash \mathsf{Expr} : \mathsf{Type}_1
% }{
% \mathsf{Type} \preceq \mathsf{Type}_1
% \qquad \xsmol{v} \not\in \mathbf{dom}~\Gamma
% \qquad \Gamma' = \Gamma[\xsmol{v} \mapsto \mathsf{Type}]
% }{
% \Gamma \vdash^\mathcal{K}_\mathsf{er} \mathsf{Type}~\xsmol{v :=}~\mathsf{Expr}\xsmol{;} : \xsmol{Unit} \vartriangleright \Gamma'
% }

\caption{Typing Rules for Statements.\label{fig:app:type:stmt}}
\end{figure}

\Cref{fig:app:type:stmt} gives the rules for
% composed
statements. We first consider composed statements.  Sequential
composition is handled by two rules. Rule \rulename{T-fresh} updates
the typing environment and continues with type checking the remainder of
the program.  Rule \rulename{T-sequence} first type checks the first
and then the second statement in the sequential composition.  The typing
environment is not updated, because there is no rule for a variable
declaration except \rulename{T-fresh} --- if the first statement
introduces a new variable, then rule \rulename{T-sequence} is
\emph{not} applicable.  Rule \rulename{T-if} handles branching. The
guard expression is typed as a Boolean, and both branches must have
the same type. Note that the environment is not relevant --- variables
declared in one branch are not available afterward.  Rule
\rulename{T-wh} is analogous for loops.
% \Cref{fig:app:type:stmt2} gives the rules for non-composed,
% non-semantic statements.
% We then consider non-composed statements.
Rule \rulename{T-skip} always types the \xsmol{skip} statement with
\xsmol{Unit}.  Rule \rulename{T-return} types the return statement
with the type of the returned expression.
% returned expression with the required type.
Rule \rulename{T-assign} is for assignments that do not declare new
local variables.  It, thus, does not update the environment. It
type-checks the right-hand side expression. Assignability is ensured
through subtyping, for which we have a special rule \rulename{subtype}
for expressions (see below).
% Rule \rulename{T-declare} is for assignments that declare a new
% local variable.  It type checks the right-hand side expression and
% checks that its type is a supertype of the declared type.
% Furthermore it modifies the environment: \smol{v} must not be
% declared (i.e., not be in the domain of $\Gamma$) and is added with
% its declared type.

The remaining rules consider the typing of expressions and right-hand sides.
% are different right-hand side expressions.
\Cref{fig:app:type:stmt3} gives the rules for the right-hand side
expressions handling semantic state access.  Rule
\rulename{T-validate} always returns a Boolean.  The reasoning behind
\rulename{T-acc} is discussed in \cref{sec:typesystem}.
% given in the main part of this article.
Rule \rulename{T-member} uses an analogous check on the given DL
concept to ensure that the returned objects can be represented at
runtime.
% Rule \rulename{T-weak} allows us to forget additionally declared
% variables and is needed for technical reasons.

\begin{figure}[t]
  \begin{minipage}{0.5\textwidth}
\TINFER{T-validate}{
\quad
}{
\Gamma \vdash^\mathcal{K}_\mathsf{er}  \xsmol{validate(shacl)}
:\xsmol{Bool}
}
\end{minipage}
\begin{minipage}{0.5\textwidth}
\TINFER{T-member}{
\xsmol{dl}~\sqsubseteq_{\mathsf{er}}^{\mathcal K}~\progpre{\xsmol{C}}
}{
\Gamma \vdash^\mathcal{K}_\mathsf{er}  \xsmol{member(dl)}
:\xsmol{List<C>}
}
\end{minipage}

\bigskip

\TTINFER{T-acc}{
\forall i \leq n .~\Gamma \vdash \mathsf{Expr}_i : \mathsf{Type}_i
}{
    \mathtt{SELECT~?obj~\{P.~?v_1~a~\mu(\mathsf{Type}_1).~\dots.~?v_n~a~\mu(\mathsf{Type}_n)\}}
   \subseteq_{\mathsf{er}}^{\mathcal K} \mathtt{SELECT~?obj}\{\mathtt{?obj~a}~\progpre{\xsmol{C}}\}
}{
\Gamma \vdash^\mathcal{K}_\mathsf{er}  \xsmol{access("SELECT ?obj}\{\xsmol{P}\}\xsmol{"},\mathsf{Expr}_1\xsmol{,}\dots\xsmol{,}\mathsf{Expr}_n\xsmol{)} 
:\xsmol{List<C>}
}

% \TTINFER{T-weak}{
% \Gamma \vdash^\mathcal{K}_\mathsf{er}:\mathsf{Type}\vartriangleright \Gamma''
% }{
% \mathbf{dom}~\Gamma' \subseteq \mathbf{dom}~\Gamma'' \qquad \forall \xsmol{v} \in \mathbf{dom}~\Gamma'.~\Gamma'(\xsmol{v}) = \Gamma''(\xsmol{v})
% }{
% \Gamma \vdash^\mathcal{K}_\mathsf{er}:\mathsf{Type}\vartriangleright \Gamma'
% }

\caption{Typing Rules for Right-Hand Sides.\label{fig:app:type:stmt3}}

% \end{figure}

\bigskip

% \begin{figure}
\begin{minipage}{0.3\textwidth}
\TINFER{literal-int}{
%    \mathsf{Type} \preceq \mathtt{Int}
}{
\Gamma\vdash \mathtt{n} : \mathsf{Int}
}
  \end{minipage}\quad
\begin{minipage}{0.3\textwidth}
\TINFER{this}{
\Gamma(\xsmol{this}) = \mathsf{Type}
}{
\Gamma\vdash \xsmol{this} : \mathsf{Type}
}
\end{minipage}\quad
\begin{minipage}{0.3\textwidth}
\TINFER{var}{
\Gamma(\mathtt{v}) = \mathsf{Type}
}{
\Gamma \vdash \mathtt{v} : \mathsf{Type}
}
\end{minipage}

\bigskip

\begin{minipage}{0.3\textwidth}
\TINFER{literal-null}{ 
    \mathsf{Type} \preceq \mathtt{Object}
}{
\Gamma\vdash \mathtt{null} : \mathsf{Type}
}
\end{minipage}\quad
\begin{minipage}{0.3\textwidth}
\TTINFER{compose}{ 
    \Gamma \vdash \mathsf{Expr} : \mathtt{C}
}{
    \mathsf{Type}~\xsmol{f} \in\mathsf{fields}_\classtable(\mathtt{C})
}{
\Gamma \vdash \mathtt{Expr.f} : \mathsf{Type}
}
\end{minipage}\quad
\begin{minipage}{0.3\textwidth}
\TTINFER{add}{
\Gamma \vdash \xsmol{Expr}_1 : \xsmol{Int}
}{
\Gamma \vdash \xsmol{Expr}_2 : \xsmol{Int}
}{
\Gamma \vdash \xsmol{Expr}_1~\mathtt{+}~\xsmol{Expr}_2 : \xsmol{Int}
}
\end{minipage}

\bigskip

\begin{minipage}{0.3\textwidth}
\TTINFER{subtype}{
\Gamma\vdash \mathsf{Expr} : \mathsf{Type_1}}
{    \mathsf{Type_1} \preceq \mathtt{Type_2}
}{
\Gamma\vdash \mathsf{Expr} : \mathsf{Type_2}
}
\end{minipage}\qquad
\begin{minipage}{0.6\textwidth}
\TTINFER{T-call}{
\Gamma \vdash \mathsf{Expr}:\xsmol{C}\qquad
\Gamma \vdash \mathsf{Expr}_i : \mathsf{Type}_i ~\text{for all}~ i \leq n
}{
\mathsf{Type}~\xsmol{m}(\mathsf{Type}_1~\xsmol{f}_1,\dots,\mathsf{Type}_n~\xsmol{f}_n) \in \mathsf{methods}_\classtable(\xsmol{C}) 
}{
\Gamma \vdash^\mathcal{K}_\mathsf{er} \mathsf{Expr}\xsmol{.m(}\mathsf{Expr}_1\xsmol{,}\dots\xsmol{,}\mathsf{Expr}_n\xsmol{)} 
: \mathsf{Type}
}

\TTINFER{T-new}{
\mathsf{fields}_\classtable(\xsmol{C}) = \{\mathsf{Type}_1~\xsmol{f}_1,\dots,\mathsf{Type}_n~\xsmol{f}_n\}
}{
\xsmol{C} \in \mathbf{dom}(\classtable) \qquad \Gamma \vdash \mathsf{Expr}_i : \mathsf{Type}_i ~\text{for all}~ i \leq n
}{
\Gamma \vdash^\mathcal{K}_\mathsf{er}  \mathsf{Type}~\xsmol{new C(}\mathsf{Expr}_1\xsmol{,}\dots\xsmol{,}\mathsf{Expr}_n\xsmol{)} 
:\xsmol{C}
}
\end{minipage}
\caption{Typing rules for expressions and right hand sides\label{fig:app:type:expr}}
\end{figure}

The remaining typing rules for expressions are given in
\cref{fig:app:type:expr}.  The typing judgement has the form
$\Gamma \vdash \mathsf{Expr} : \mathsf{Type}$, with the intuitive
meaning that under typing environment $\Gamma$ the expression
$\mathsf{Expr}$ has type $\mathsf{Type}$.  Rule \rulename{this} types
the \smol{this} literal with the carried type $\mathsf{self}$.  Rule
\rulename{literal-null} types the \smol{null} literal with any subtype
of $\xsmol{Object}$. %\ektodo{add the type hierarchy in the beginning}
Rule \rulename{literal-int} is representative for all typing rules for
literals. It types every integer literal with \smol{Int}.  Rule
\rulename{var} looks up the type of A variable in the typing
environment.  Rule \rulename{compose} types the sub-expression with
some class \smol{C}, and looks up the type of the field in the class
table %.\ektodo{Check the role of $G$ when we do this.}
Rule \rulename{add} is representative for the underspecified set of
expressions with operators.  It types both sub-expressions with
\smol{Int}, and types the overall expression with \smol{Int} as well.
Rule \rulename{T-call} handles method calls, and its type is the
return type of the method. The parameters require a more elaborate
check: First, the target expression is typed with a class, then the
class table is accessed to retrieve the abstract method parameters.
Each of the expressions is then checked against the type of the
corresponding abstract parameter. Furthermore, the number of concrete
and abstract parameters must be the same.  Rule \rulename{T-new} is
analogous for object allocation.

\subsection{Typing Runtime Syntax}

We now consider the typing of runtime configurations, which amounts to
% Typing the runtime syntax is reduced to
typing all statements in the process stack and class table, and
checking consistency constraints; e.g., every runtime return statement
must have a corresponding statement to continue in the next lower
process on the stack. Also, we need to generate the corresponding
typing environments.

\Cref{fig:runtime:global} gives the typing rules for runtime
configurations.  Rule \rulename{R-conf} checks a whole configuration.
The first premise type-checks all classes in the class table, the
second the objects, assuming a well-typed class table, and the last
checks the processes, using the well-typed class table and
objects. %\todo{A bit strange to type-check the class table for every
  %runtime configuration, since it is static and implicit. Couldn't we
  %just assume that is is well-typed from compile time?}  
  As we will see, the first premise is  not required if the configuration is reached from an initial configuration. Rule
\rulename{R-obs-1} states that an empty set of objects is well typed,
and rule \rulename{R-obs-2} decomposes checking a set of objects into
checking each object in isolation.  Rule \rulename{R-obs-3} checks a
single object. Each field of the given class must have a value
assigned in the memory, and the type of the value (checked as an
expression) must be of the declared type.  Rule \rulename{R-prs-1}
states that an empty process stack is well-typed, and rule
\rulename{R-prs-2} handles all processes on the stack with a
non-return statement on top.  It reduces to check the statement in the
context generated from the heap and local memory with
\begin{align*}
\Gamma(\sigma,\rho,\mathtt{C}, \mathsf{obs}) =& 
\phantom{\cup} \{ \xsmol{v} \mapsto \mathsf{Type} \ssep \xsmol{v} \in \mathbf{dom}~\sigma, \emptyset \vdash \sigma(\xsmol{v}) : \mathsf{Type}  \text{ or } (\mathsf{Type}, \rho')_{\sigma(\xsmol{v})} \in \mathsf{obs} \} \\
&\cup
\{ \xsmol{f} \mapsto \mathsf{Type} \ssep  \xsmol{f} \in \mathbf{dom}~\rho, \emptyset \vdash \rho(\xsmol{f}) : \mathsf{Type} \text{ or } (\mathsf{Type}, \rho')_{\rho(\xsmol{f})} \in \mathsf{obs}\} \\
&\cup \{\xsmol{this} \mapsto \mathtt{C}\}
\end{align*}
Finally, rule \rulename{R-prs-3} is concerned with process stacks
where the next statement to be executed is a return; here, the next
lower process must start with a continuation.  Note that any process
stack not matching these rules, is ill-typed.

\begin{figure}[t]
\begin{minipage}{0.33\textwidth}
\TTINFER{R-conf}{
\forall i \leq n.~  \vdash^\mathcal{K}_\mathsf{er} \mathsf{Class}_i
}{
 \vdash \mathsf{obs}
\qquad \mathsf{obs} \vdash \mathsf{prs}
}{
\vdash \mathsf{obs}~\mathsf{prs}
}
\end{minipage}
\begin{minipage}{0.33\textwidth}
\TINFER{R-obs-1}{
\quad
}{
\vdash \emptyset
}
\end{minipage}
\begin{minipage}{0.33\textwidth}
\TINFER{R-obs-2}{
\forall i \leq n.~\classtable \vdash\mathsf{ob}_i
}{
 \vdash \{\mathsf{ob}_1\dots\mathsf{ob}_n\}
}
\end{minipage}

\vspace{-1mm}

\TTINFER{R-obs-3}{
\mathsf{fields}_\classtable(\xsmol{C}) = \{\mathsf{Type}_\xsmol{f}~\xsmol{f}\}
}{
\mathbf{dom}~\rho = \{\xsmol{f}~|~\mathsf{Type}_\xsmol{f}~\xsmol{f} \in \mathsf{fields}(\xsmol{C})\} 
\qquad
\forall \xsmol{f} \in \mathbf{dom}~\rho.~ \emptyset\vdash \rho(\xsmol{f}) : \mathsf{Type}_\xsmol{f}
}{
\classtable \vdash (\xsmol{C}, \rho)_\mathtt{X}
}

\vspace{-1mm}

\noindent\resizebox{\textwidth}{!}{
\begin{minipage}{1.15\textwidth}
\begin{minipage}{0.25\textwidth}
\TINFER{R-prs-1}{
\quad
}{
\mathsf{obs} \vdash \emptyset
}
\end{minipage}\hspace{5mm}
\begin{minipage}{0.5\textwidth}
\TTTINFER{R-prs-2}{
\mathsf{obs} \vdash \mathsf{prs}
\qquad
\Gamma(\sigma, \rho, \mathtt{C}, \mathsf{obs}) \vdash \mathsf{Stmt} : \mathsf{Type} \vartriangleright \Gamma'
}{
    \mathsf{Type}~\xsmol{m}(\dots) \in \mathsf{methods}_\classtable(\xsmol{C})\qquad
(\mathtt{C}, \rho)_\mathtt{X} \in \mathsf{obs} 
}{
\mathsf{Stmt} \neq \mathsf{Stmt}'\xsmol{;}~\xsmol{return}~\mathsf{Expr}\xsmol{;}
\qquad
    \mathsf{Stmt} \neq \mathsf{Loc}\leftarrow\xsmol{stack;}\mathsf{Stmt}\xsmol{'}
}{
\mathsf{obs} \vdash \mathsf{prs}, (\mathtt{m}, \mathtt{X}, \mathsf{Stmt}, \sigma)
}
\end{minipage}
\end{minipage}
}

\vspace{2mm}

% \noindent\resizebox{\textwidth}{!}{
% \begin{minipage}{1.3\textwidth}
\TTTINFER{R-prs-3}{
(\mathtt{C}_1, \rho_1)_{\mathtt{X}_1} \in \mathsf{obs} 
\qquad 
(\mathtt{C}_2, \rho_2)_{\mathtt{X}_2} \in \mathsf{obs} 
}{
\mathsf{obs} \vdash \mathsf{prs}
\qquad
\Gamma(\sigma_2, \rho_2, \mathtt{C}_2, \mathsf{obs})\vdash \mathsf{Expr} : \mathsf{Type}(\mathsf{Loc})
}{
\Gamma(\sigma_1, \rho_1, \mathtt{C}_1, \mathsf{obs}) \vdash \mathsf{Stmt}_1 : \mathsf{Type}_1 \vartriangleright \Gamma_1
\qquad 
\Gamma(\sigma_2, \rho_2, \mathtt{C}_2, \mathsf{obs}) \vdash \mathsf{Stmt}_2 : \mathsf{Type}_2 \vartriangleright \Gamma_2
}{
\mathsf{obs} \vdash \mathsf{prs}, (\mathtt{m}_1, \mathtt{X}_1, \mathsf{Loc}\leftarrow\xsmol{stack;}\mathsf{Stmt}_1, \sigma_1),~
(\mathtt{m}_2, \mathtt{X}_2, \mathsf{Stmt}_2\xsmol{;}~\xsmol{return}~\mathsf{Expr}\xsmol{;}, \sigma_2)
}
% \end{minipage}
% }

\caption{Typing rules for runtime configurations.}
\label{fig:runtime:global}
\end{figure}

\subsection{Soundness}

\begin{lemma}[Initial State]\label{thm:initial}
  The initial configuration of a well-typed program is well-typed:
  \[ \vdash\mathsf{Prog} \quad\Rightarrow\quad
    \vdash\mathsf{init}_\mathsf{Prog}\ .\]
\end{lemma}
\begin{proof}
We need to show that we can construct a proof tree to type check $\mathsf{init}_\mathsf{Prog}$ with the rule \rulename{R-conf} as its root, given 
a tree for $\mathsf{Prog}$ with \rulename{prog} as its root.
\begin{description}
\item[First premise:] This follows from the second premise of rule \rulename{prog}, except the class \smol{Entry}, 
which is well-typed if the main statement is well-typed, which is the first premise of \rulename{prog}.
\item[Second premise:] By definition there is only one object that is already created, which is of class \smol{Entry}. Ergo, we must type is with rule \rulename{R-obs-3}.
This class has no fields, thus the first premise trivially holds. The generated heap is empty, thus the second and third premises also hold.
\item[Third premise:] By definition there is only one process, 
which we must type with \rulename{R-prs-2}.
As the main block is well-typed with $\mathsf{Unit}$, which is the type of the generated \smol{entry} method, the first two premises hold.
The third premise holds trivially by generation. The forth one is guaranteed to hold as the identifier and class name are fixed for all initial configurations. The last premises hold as return and the waiting statement are not allowed in the main block.\qedhere
\end{description}
\end{proof}

Before we show that every the lifting of well-typed configuration is
consistent, we give an auxiliary structures as an intermediate steps.

\begin{lemma}\label{lem:aux:ct}
  The lifting of a class table of every well-typed program is
  consistent:
\[
  \vdash \mathsf{Prog} \quad \Rightarrow \quad \bigcup_{\mathtt{C} \in
    \classtable_\mathsf{Prog}} \mu(\mathtt{C}) \cup \mathcal{K}_\SMOL
  \text{ is consistent.}
\]
\end{lemma}
\begin{proof}
Let $|\mathsf{Prog}|$ be the number of classes in a program.
Let $|\mathsf{Class}|$ be the number of fields and methods within a class.
We prove the theorem by induction on $n = |\mathsf{Prog}|$.
\begin{description}
\item[Induction hypothesis 1:]
\(
\forall n.~|\mathsf{Prog}| = n \Rightarrow \big(\vdash \mathsf{Prog}  \Rightarrow  \bigcup_{\mathtt{C} \in \classtable_\mathsf{Prog}} \mu(\mathtt{C}) \cup \mathcal{K}_\SMOL\text{ is consistent}\big)
\)
\item[Induction base $n=0$:]
In this case, the class table is empty and the lifting consist only of $\mathcal{K}_\SMOL$. 
This set of axioms is consistent, as is easily shown by inputing it into a DL reasoner. 
\item[Induction step $n>0$:] 
Before we continue, we observe the structure of the lifted class table: Besides additional axioms stemming from the subclass relation, it is saturated, i.e., no new axioms can be derived. Furthermore, it contains no counting axioms, as the only axioms that limit the number of members of a concept are in $\mathit{close}$.
Thus, there are only 4 ways to make the knowledge graph inconsistent: (1) violating a domain axiom, (2) violating a range axiom, (3) violating a disjointness axiom, and (4) violating a set axiom from $\mathit{close}$. Note that all of these inconsistencies do involve more than one axiom, but each inconsistency must involve (at least) one of the above.

We remind that the axioms generated from lifting a class are as
follows.
\begin{owlframe}
Individual: $\progpre{\mathtt{C}}$ 
  Facts: a $\smolpre{\mathtt{Class}}$, $\smolpre{\mathtt{hasName}}$ "C",
  [$\smolpre{\mathtt{subClass}}$ $\progpre{\mathtt{D}}$],     // if $\mathtt{C}$ extends $\mathtt{D}$
  [$\smolpre{\mathtt{subClass}}$ $\smolpre{\mathtt{Any}}$],    // otherwise
  $\smolpre{\mathtt{hasMethod}}$ $\progpre{\mathtt{m}_1}$,  $\dots$, $\smolpre{\mathtt{hasMethod}}$ $\progpre{\mathtt{m}_n}$,   
  $\smolpre{\mathtt{hasField}}$ $\progpre{\mathtt{f}_1}$,  $\dots$, $\smolpre{\mathtt{hasField}}$ $\progpre{\mathtt{f}_k}$   

Individual: $\progpre{\mathtt{m}_1}$ Facts: a $\smolpre{\mathtt{Method}}$, $\smolpre{\mathtt{hasName}}$ "$\mathtt{m}_1$"
$\dots$
Individual: $\progpre{\mathtt{f}_1}$ Facts: a $\smolpre{\mathtt{Field}}$, $\smolpre{\mathtt{hasName}}$ "$\mathtt{f}_1$"
$\dots$
\end{owlframe}

Let $\mathsf{Class}$ be any class in $\mathsf{Prog}$, such that (by induction hypothesis 1) the other $n$ classes are lifted to a consistent knowledge graph.
We now proceed with an induction on $m = |\mathsf{Class}|$.
\begin{description}
\item[Induction hypothesis 2:]
\(
\forall m.~|\mathsf{Class}| = m \Rightarrow \big(\mu(\mathsf{Class}) \cup \mathcal{K}_\SMOL\text{ is consistent}\big)
\)
\item[Induction base $m=0$:]
In this case the class \smol{C} has no fields or methods. 
\begin{itemize}
\item
The first axiom generated connects the name to the IRI of the class using $\smolpre{\mathtt{hasName}}$.
The only interactions this axiom have is with (1) the domain axiom for this property, which is observed, because the lifting of \classtable states that 
$\progpre{\mathtt{C}}~\mathtt{a}~\smolpre{\mathtt{Class}}$ explicitly, and (2) the range axiom, which is observed, as the name has the right data type, namely a \texttt{xsd\!:\!String}.
\item 
The second group of axioms model the subtyping relation and the disjointness. 
These can only interact with each other, but all 
the subtyping relation axioms are consistent, as they form a tree by definition of the class system which excludes cycles, and the program is well-typed by the assumption of this theorem.
% The disjointness axioms are consistent as follows: The first $n$ classes are consistent per induction hypothesis. The class in question
\item
The last axiom is the set axiom for $\smolpre{\mathtt{Class}}$, which cannot cause an inconsistency as the membership of $\progpre{\mathtt{C}}$ is state explicitly.
\end{itemize}
\item[Induction step $m>0$:]
We distinguish the cases for fields and methods.
\begin{description}
\item[Fields:]
The range and domain axioms cannot cause inconsistencies because the field is never used. The axioms for $\smolpre{\mathtt{hasField}}$ and $\smolpre{\mathtt{hasName}}$
again fulfill their range and domain axioms by construction and the last axiom $\progpre{\mathtt{f}}~\mathtt{a}~\smolpre{\mathtt{Field}}$.
% The consistency of the set axiom for $\smolpre{\mathtt{Field}}$ added by $\mathit{close}$ is shown analogously to the case for classes.
\item[Methods:]
Analogous to fields. \qedhere
\end{description}
\end{description}
\end{description}
\end{proof}

For our main theorem, we consider the following property that connects
typability of runtime configurations and consistency of the
corresponding knowledge bases.

\begin{theorem}[Connection]\label{thm:connect}
The lifting of every well-typed configuration is consistent:
\[
  \vdash\!\mathsf{conf} \Rightarrow \mu(\mathsf{conf})\text{ is
    consistent.}
\]
\end{theorem}

\begin{proof}
We have 
\[
\mu(\mathtt{conf}) = 
\bigcup_{\xsmol{C} \in \mathbf{dom}(\classtable)} \mu(\mathtt{C}) \cup \bigcup_{1 \leq \mathtt{X}\leq n}\big(\mu(\mathsf{ob}_i) \cup \mathsf{links}(\mathtt{X})[\runpre{\mathtt{X}},\mathtt{conf}] \big)  \cup \mathit{close}
\]
and, by \Cref{lem:aux:ct}, the lifted class table
($\bigcup_{\xsmol{C} \in \mathbf{dom}(\classtable)} \mu(\mathtt{C})
\cup \mathcal{K}_\SMOL$) in itself is consistent.  We show that (1)
the lifting of objects is consistent, (2) that the lifting of objects
is consistent with the lifted class table, and (3) both liftings are
consistent with the closure axioms.
\begin{itemize}
    \item We show that the following is consistent:
    \[
    \bigcup_{1 \leq \mathtt{X}\leq n}\big(\mu(\mathsf{ob}_i) \cup \mathsf{links}(\mathtt{X})[\runpre{\mathtt{X}},\mathtt{conf}] \big) \cup \mathcal{K}_\SMOL\ .
    \]
    Following the structure of \cref{lem:aux:ct}, we consider each
    added axiom in isolation, and compare it with the range and domain
    axioms of the respective property.  It is easy to see that the
    lifting follows domain and range, with the only critical point
    being the distinction between data and object property.  The
    linkage is consistent due to being a conservative
    extension.
    \item We show that the union of of lifted class table and lifted objects is consistent:
    \[
    \bigcup_{\xsmol{C} \in \mathbf{dom}(\classtable)} \mu(\mathtt{C}) \cup \bigcup_{1 \leq \mathtt{X}\leq n}\big(\mu(\mathsf{ob}_i)\cup \mathsf{links}(\mathtt{X})[\runpre{\mathtt{X}},\mathtt{conf}] \big) \cup \mathcal{K}_\SMOL\ .
    \]
    The two axioms sets, which are consistent in themselves, interact only on class and field individuals. It is easy to see that the use of the $\smolpre{\mathtt{entryOf}}$ and $\smolpre{\mathtt{implements}}$
    adhere to the domain and range axioms of the ontology.
    \item It remains to show that the union of the above with $\mathit{close}$ is consistent. This is straigthforward: this set of axioms only defines classes in terms of sets of individuals.
    All these sets are disjoint, and there are no relevant subclass relations. There are no counting axioms in the SMOL ontology that could limit the number of individuals, and that the domain knowledge is a conservative extension and cannot introduce such restrictions other. Thus, there are no axioms that could be used to derive a contradiction.\qedhere
\end{itemize}
\end{proof}

As our evaluation of side-effect and non-semantic expressions is
underspecified, we impose the following restriction on it, which is
standard and independent of semantic state access, as these constructs
are handled by evaluation of statements, not expressions.  We require
the following property to connect typability of expression with their
evaluation. As we underspecify all expressions except the ones related
to retrieve objects, we only state this assumption for objects.

\begin{assumption}\label{lem:aux:obj}
If $\Gamma \vdash \mathsf{Expr} : \mathtt{C}$, $\classtable \vdash \mathsf{obs}$ and $(\mathtt{C},\rho)_\mathtt{X} \in \mathsf{obs}$,
then either (1) $\llbracket\mathsf{Expr}\rrbracket^{\sigma,\mathsf{obs}}_\mathtt{X} = \xsmol{null}$ or (2) $\llbracket\mathsf{Expr}\rrbracket^{\sigma,\mathsf{obs}}_\mathtt{X} = \mathtt{Y}$, such that $(\mathtt{D},\rho)_\mathtt{Y} \in \mathsf{obs}$ and $\mathtt{D} \preceq \mathtt{C}$\ .
\end{assumption}

We now prove the subject reduction theorem and show that being well-typed is an invariant at runtime.  We refrain from giving full formal details because besides the case for \smol{access}, the system is a standard object-oriented language.
\begin{lemma}[Subject Reduction]\label{thm:subject}
Every transition from a well-typed configuration results in a well-typed configuration.
\[
\vdash\mathsf{conf} \wedge 
\mathsf{conf} \rightarrow^\mathcal{K}_\mathsf{er} \mathsf{conf'}
\quad\Rightarrow\quad \vdash\mathsf{conf'}\ .
\]
\end{lemma}
\begin{proof}
Case distinction on the rule used to make the transition.
\begin{itemize}
\item \textbf{Rule \rulename{iftrue}:}
We must show that for all $\Gamma$ if
\[ \Gamma \vdash
\mathsf{CT}~\mathsf{obs}~\mathsf{prs}, (\xsmol{m},\,\mathtt{X},\,\xsmol{if}~\mathsf{Expr}~\xsmol{then}~\mathsf{Stmt}_1~\xsmol{else}~\mathsf{Stmt}_2~\xsmol{end}~\mathsf{Stmt},\,\sigma)
\]
then
\[
\ertransition \mathsf{CT}~\mathsf{obs}~\mathsf{prs}, (\xsmol{m},\,\mathtt{X},\,\mathsf{Stmt}_1\,\mathsf{Stmt}, \,\sigma)\ .
\]

First, we observe that the type trees differ only in the statement, as the rule does not change the environment, objects or process. Thus, we only need to prove that if
\[
\Gamma \vdash^\mathcal{K}_\mathsf{er} \xsmol{if}~\mathsf{Expr}~\xsmol{then}~\mathsf{Stmt}_1~\xsmol{else}~\mathsf{Stmt}_2~\xsmol{end}~\mathsf{Stmt}:~\mathsf{Type}
\]
then 
\[
\Gamma \vdash^\mathcal{K}_\mathsf{er} \mathsf{Stmt}_1~\mathsf{Stmt}:~\mathsf{Type}\ .
\]

By assumption, we have the following derivation tree:

\bigskip

%\resizebox{\textwidth}{!}{
% \begin{minipage}{1.3\textwidth}
% \begin{prooftree}
% \AxiomC{$(\ast)$}
% \end{prooftree}
% \end{minipage}
%}

\bigskip

% \resizebox{\textwidth}{!}{
% \begin{minipage}{1.1\textwidth}
\begin{prooftree}
\AxiomC{$(1)$}
\UnaryInfC{$
\Gamma \vdash^\mathcal{K}_\mathsf{er} \mathsf{Stmt}_2:~\mathsf{Type}
$}
\AxiomC{$(2)$}
\UnaryInfC{$
\Gamma \vdash^\mathcal{K}_\mathsf{er} \mathsf{Stmt}_1:~\mathsf{Type}
$}
\RightLabel{\rulename{T-if}}
\BinaryInfC{$
\Gamma \vdash^\mathcal{K}_\mathsf{er} \xsmol{if}~\mathsf{Expr}~\xsmol{then}~\mathsf{Stmt}_1~\xsmol{else}~\mathsf{Stmt}_2~\xsmol{end}:~\mathsf{Type}
$}
\AxiomC{$(3)$}
\UnaryInfC{$
\Gamma_2 \vdash^\mathcal{K}_\mathsf{er} \mathsf{Stmt}:~\mathsf{Type}
$}
\RightLabel{\rulename{T-sequence}}
\BinaryInfC{$
\Gamma \vdash^\mathcal{K}_\mathsf{er} \xsmol{if}~\mathsf{Expr}~\xsmol{then}~\mathsf{Stmt}_1~\xsmol{else}~\mathsf{Stmt}_2~\xsmol{end}~\mathsf{Stmt}:~\mathsf{Type}
$}
\end{prooftree}
% \end{minipage}
% }

\bigskip

Here $(1), (2), (3)$ are closed derivation trees,  $\mathbf{dom}\Gamma_3 \supseteq \mathbf{dom}\Gamma_2$, and $\Gamma_3$ and $\Gamma_2$ do agree in their image on the domain of $\Gamma_2$.
It is easy to see that if a statement can be typed with $\Gamma_2$, then it can also be typed with $\Gamma_3$.
Thus, we can construct the following derivation tree for the target judgement:

% \resizebox{\textwidth}{!}{
% \begin{minipage}{1.2\textwidth}
\begin{prooftree}
\AxiomC{$(3)$}
\UnaryInfC{$
\Gamma \vdash^\mathcal{K}_\mathsf{er} \mathsf{Stmt}_1:~\mathsf{Type}
$}
\RightLabel{\rulename{T-weak}}
\UnaryInfC{$
\Gamma \vdash^\mathcal{K}_\mathsf{er} \mathsf{Stmt}_1:~\mathsf{Type}
$}
\AxiomC{$(2)$}
\UnaryInfC{$
\Gamma_2 \vdash^\mathcal{K}_\mathsf{er} \mathsf{Stmt}:~\mathsf{Type}
$}
\RightLabel{\rulename{T-sequence}}
\BinaryInfC{$
\Gamma \vdash^\mathcal{K}_\mathsf{er} \mathsf{Stmt}_1~\mathsf{Stmt}:~\mathsf{Type}
$}
\end{prooftree}
% \end{minipage}
% }

\medskip

\item \textbf{Rules \rulename{iffalse}, \rulename{loop1}, \rulename{loop2}:}
Analogously to \rulename{iftrue}, we just copy over the right subtrees in the rewritting.

\item \textbf{Rule \rulename{assign1}:}
We must show that, under the conditions described by the premises, if 
\[
\Gamma \vdash^\mathcal{K}_\mathsf{er} \mathsf{Expr}.{f :=}~\mathsf{Expr'}\xsmol{;}~\mathsf{Stmt}:~\mathsf{Type}
\]
and 
\[\classtable \vdash (\mathtt{C},\rho)_\mathtt{Y}\]
then 
\[
\Gamma \vdash^\mathcal{K}_\mathsf{er} \mathsf{Stmt}:~\mathsf{Type}
\]
and 
\[
\classtable \vdash (\mathtt{C},\rho[\xsmol{f} \mapsto \mathtt{e}])_\mathtt{Y}\ .
\]
By assumption, we have the following (slightly simplified) derivation tree

% \begin{prooftree}
% \AxiomC{$
% (\ast)
% $}

% \end{prooftree}

% \medskip

\begin{prooftree}
\AxiomC{$(3)$}
\UnaryInfC{$
\Gamma \vdash^\mathcal{K}_\mathsf{er} \mathsf{Expr'} : \mathsf{Type_2}\vartriangleright \Gamma'
$}
\AxiomC{$(2)$}
\UnaryInfC{$
\Gamma \vdash^\mathcal{K}_\mathsf{er} \mathsf{Expr}:~\mathtt{C}
$}
\UnaryInfC{$
\Gamma \vdash^\mathcal{K}_\mathsf{er} \mathsf{Expr}.{f}:~\mathsf{Type_1}
$}
\BinaryInfC{$
\Gamma \vdash^\mathcal{K}_\mathsf{er} \mathsf{Expr}.{f :=}~\mathsf{Expr'}\xsmol{;} : \mathsf{Unit}
$}\AxiomC{$(1)$}
\UnaryInfC{$
\Gamma'' \vdash^\mathcal{K}_\mathsf{er} \mathsf{Stmt}:~\mathsf{Type}
$}
\RightLabel{\rulename{T-sequence}}
\BinaryInfC{$
\Gamma \vdash^\mathcal{K}_\mathsf{er} \mathsf{Expr}.{f :=}~\mathsf{Expr'}\xsmol{;}~\mathsf{Stmt}:~\mathsf{Type}
$}
\end{prooftree}

\noindent
Obviously $(1)$ is a derivation tree for $\Gamma \vdash^\mathcal{K}_\mathsf{er} \mathsf{Stmt}:~\mathsf{Type}$.
For the second statement, we must show that $\mathtt{e} : \mathsf{Type}_\xsmol{f}$, which follows from \cref{lem:aux:obj}. 

\item \textbf{Rules \rulename{assign2}, \rulename{assign3}]:}
Analogous to \rulename{assign1}.

\item \textbf{Rule \rulename{skip}:}
We must show that if 
\[
\Gamma \vdash^\mathcal{K}_\mathsf{er} \xsmol{skip;}~\mathsf{Stmt}:~\mathsf{Type}
\]
then 
\[
\Gamma \vdash^\mathcal{K}_\mathsf{er} \mathsf{Stmt}:~\mathsf{Type}
\]
By assumption, we have the following derivation tree

\medskip

\begin{prooftree}
\AxiomC{$ $}
\RightLabel{\rulename{T-skip}}
\UnaryInfC{$
\Gamma \vdash^\mathcal{K}_\mathsf{er} \xsmol{skip;}:~\mathsf{Unit}
$}
\AxiomC{$(1)$}
\UnaryInfC{$
\Gamma \vdash^\mathcal{K}_\mathsf{er} \mathsf{Stmt}:~\mathsf{Type}
$}
\RightLabel{\rulename{T-Sequence}}
\BinaryInfC{$
\Gamma \vdash^\mathcal{K}_\mathsf{er} \xsmol{skip;}~\mathsf{Stmt}:~\mathsf{Type}
$}
\end{prooftree}

\medskip

\noindent
Thus, there is a derivation tree for $(1)$, which is exactly the statement we need to show.

\item \textbf{Rule \rulename{callIn}:} This rule is not applicable in
  the small language, but it is trivial to see that the explicit check
  for the return type ensures the applicatbility of
  \rulename{T-call}. %\todo{Shouldn't we exclude thise case, if it is
   % not in the grammar we consider?}
\item \textbf{Rule \rulename{new}:}
We need to show if the whole configuration is well-typed and, thus, there derivation tree for the creation statement, rooted in \rulename{T-new},
then there is one for the declaration for the declaration, rooted in \rulename{T-declare}, and that the newly created object is well-typed.

For the derivation tree we have, by construction, $\mathtt{Y} : \mathtt{C}$, and all subtrees can be copied over directly.
For the newly created object, we must show that all evaluated values respect the type of the field they are assigned to.
This is the same argument as for storing a single value in a field in \rulename{assign1} using \cref{lem:aux:obj}.
%: the expression is typed with the type of the field,
%by Lemma~\ref{lem:aux:obj} this means that the object identity the expression evaluates to respect the type of the expression.
%For non-object types, the evaluation is underspecified, but the required generalization is trivial to achieve.
\item \textbf{Rule \rulename{call}:}
  % The only part that is not analogous to the above rules handling
  % the statement change, is the type check for the extended process
  % stack.
  We must ensure that rule \rulename{R-prs-3} is applicable after the
  transition. Thus,
  % and the only change is that
  we must ensure the required syntactic form, as the rest follows from
  the typability of the prior configuration (such as typability of the
  lower process stack).  For this it suffices to observe that
  $\mathsf{Stmt}'$ ends in a \smol{return} statement, as required by
  the rule.
\item \textbf{Rule \rulename{return}:} This removes exactly one pair
  of runtime stack statements, we must show that the resulting value
  is respecting the type of the expression, which is analogous to the
  above cases, as it is checked explicitly by \rulename{T-return}, and
  this derivation subtree can be reused by \rulename{T-assign}.  Note
  that $\mathsf{Stmt}$ always ends in a \smol{return} by definition,
  as it is always a method body, and, thus, we do not need to reason
  about its exact structure.
\item \textbf{Rule \rulename{validate}:}
We must show that applicability of \rulename{T-validate} implies applicability of \rulename{T-declare} after the transition.
By definition, $\mathsf{Sha}$ returns a Boolean literal, the other subtrees carry over directly.
\item \textbf{Rule \rulename{member}:}
We must show that applicability of \rulename{T-member} implies applicability of \rulename{T-declare} after the transition.
For this, we must show that we can listify the results of the membership query into a list of \smol{C} elements.

Only objects of \smol{C} and its subclasses are described by
$\progpre{C}$, the additional premise, explicitly checks that the
membership query is on a concept that is a subconcept of
$\progpre{C}$, i.e., a subset of objects of \smol{C} and its
subclasses. By \cref{thm:connect} the original configuration is
consistent, so the reasoner indeed does so.  We remind the reader that
we include the \textit{close} axioms to ensure that no further
individuals (that cannot be represented at runtime) can be added by
the domain knowledge.Thus, every subconcept of $\progpre{C}$ is a
subset of the individuals of representable objects of the required
class.
\item \textbf{Rule \rulename{access}:}
Analogous to \rulename{member}, except that the argument is over query containment instead of concept subclassing.\qedhere
\end{itemize}
\end{proof}

We obtain that for a well-typed program, every reachable state is well-typed.
\begin{lemma}[Well-Typed Reachable States]\label{thm:reachtype}
\[
\vdash\mathsf{Prog} 
\wedge
\mathsf{init}_{\mathsf{Prog}} \rightsquigarrow^{\userkb}_\mathsf{er} \mathsf{conf}
\quad\Rightarrow\quad\vdash \mathsf{conf}
\]
\end{lemma}
\begin{proof}
Follows directly from \cref{thm:initial,thm:subject}.
\end{proof}

%\ektodoin{We do not need this, strictly spoken, so we may remove it to slim down the technical load}
% Let $\mathsf{term}(\mathsf{conf})$ denote that $\mathsf{conf}$ is successfully terminated.
% \begin{lemma}[Progress]\label{thm:progress}
% From a well-typed configuration, either a step is possible, or the configuration is successfully terminated.
% \[\forall \mathsf{conf}.~ 
% \big((\vdash\!\mathsf{conf} \Rightarrow  
% \exists\mathsf{conf'}.~\mathsf{conf} \rightarrow^\mathcal{K}_\mathsf{er} \mathsf{conf'} 
% \vee\mathsf{term}(\mathsf{conf})\big)
% \]
% \end{lemma}
% \begin{proof}
% \ektodoin{todo}
% \end{proof}

We can now prove \cref{thm:safety}.
\safety*
% \begin{quote}
%   Let $\mathsf{Prog}$ be a program that is well-typed with respect to
%   $\vdash^{\userkb}_\mathsf{er}$,
% %  $\vdash^{\smolkb \cup \userkb}_\mathsf{er}$,
%   where $\userkb$ is a conservative extension of $\smolkb \cup \mu(\classtable_\mathsf{Prog})$.
% % \begin{enumerate}
% % \item
%  Every reachable configuration of $\mathsf{Prog}$ can be lifted
%   to a consistent knowledge graph:
%   \[\forall \mathsf{conf}.~\mathsf{init}_{\mathsf{Prog}}
%     \rightsquigarrow^{\userkb}_\mathsf{er} \mathsf{conf}~
%     ~\rightarrow~ ~\mu(\mathsf{conf}) \cup \smolkb \cup \userkb \text{
%       is consistent}\]
% %     \item If $\mathsf{Prog}$ terminates, then it terminates successfully:
% % \[\forall \mathsf{conf}.~\mathsf{init}_{\mathsf{Prog}} \Downarrow^{\userkb}_\mathsf{er} \mathsf{conf}~ ~\rightarrow~ \mathsf{conf} \text{ is successfully terminated}\]
% % \end{enumerate}
% \end{quote}
%\todo{Should we restate the theorem? There is some latex package for that...}

\begin{proof}
% For 1.: 
  By \cref{thm:reachtype}, every reachable configuration is well-typed
  and by \cref{thm:connect} every well-typed configuration is lifted
  to a consistent knowledge graph.
% For 2.: Follows directly from \cref{thm:progress}.
\end{proof}

%%% Local Variables: 
%%% mode: latex
%%% TeX-master: "../main"
%%% End: 

\end{appendices}

% \clearpage
%  \tableofcontents
\end{document}